\newif\ifjournal
\newif\ifdoublecol

\journalfalse
\doublecolfalse

\ifdoublecol
\documentclass[10pt,twocolumn,twoside]{IEEEtran}
\else	\documentclass[11pt,draftclsnofoot,onecolumn,twoside,romanappendices]{IEEEtran}
\fi

\usepackage{amsfonts,amsmath,theorem}
\usepackage{color}
\usepackage[table]{xcolor}
\usepackage{bm}
\usepackage[sort,nospace,compress]{cite}
\usepackage{algorithm,algorithmic}
\usepackage{graphicx}
\usepackage[caption=false]{subfig}
\usepackage{fixltx2e}
\usepackage{afterpage}
\graphicspath{{figures/}}

\def\figwidth{0.3\textwidth}
\def\figwidth2{0.2\textwidth}

\ifjournal
\def\boundsfigheight{0.16\textwidth}
\newcommand\forinitialsubmission[2]{#2}
\else
\def\boundsfigheight{0.32\textwidth}
\newcommand\forinitialsubmission[2]{#1}
\fi

\def\hpsnrfigheight{0.35\columnwidth}
\def\perfheight{0.35\columnwidth}

\newcommand{\makered}[1]{}

\def\argmax{\mathop{\rm argmax}}
\def\argmin{\mathop{\rm argmin}}
\newcommand{\trieq}{\stackrel{\bigtriangleup}{=}}
\newcommand{\expec}[1]{\mathbb{E}\left[#1\right]}
\newcommand{\expecY}[1]{\mathbb{E}_{\vY(T)}\left[#1\right]}
\newcommand{\expecYY}[1]{\mathbb{E}_{\vY(T-1)}\left[#1\right]}
\newcommand{\expecy}[1]{\mathbb{E}_{\vy(T)|\vY(T-1)}\left[#1\right]}

\newcommand{\expecgen}[2]{\mathbb{E}_{#1}\left[#2\right]}
\newcommand{\expecN}[1]{\expecgen{\vN}{#1}}
\newcommand{\expecNN}[1]{\expecgen{\vN\setminus N_1|N_1}{#1}}
\newcommand{\expecNone}[1]{\expecgen{N_1}{#1}}
\newcommand{\vlam}{{\bm \lambda}}

\newcommand{\olami}{{\overline{\lambda_i}}}
\newcommand{\vY}{{\bm Y}}
\newcommand{\vX}{{\bm X}}
\newcommand{\vy}{{\bm y}}
\newcommand{\vC}{{\bm C}}
\newcommand{\vu}{{\bm u}}
\newcommand{\vN}{{\bm N}}

\newcommand{\Iic}[1]{I_i^{(#1)}}
\newcommand{\pic}[1]{p_i^{(#1)}}
\newcommand{\sic}[1]{\sigma_i^{(#1)}}

\newcommand{\sictsqd}[2]{(\sic{#1}(#2))^2}
\newcommand{\si}{\sigma_i}
\newcommand{\hatXic}[1]{\hat{X}_i^{(#1)}}
\newcommand{\zi}{z_i}
\newcommand{\vxi}{\bm \xi}
\newcommand{\calC}{{\cal C}}
\newcommand{\set}[1]{\left\{#1\right\}}
\newcommand{\ignore}[1]{}
\newcommand{\pt}{\tilde{p}}
\newcommand{\pbar}{\bar{p}_1}

\newcommand{\firstmom}{m_{h,1}}
\newcommand{\secondmom}{m_{h,2}}
\newcommand{\firstmomsq}{\firstmom^2}

\definecolor{DarkGreen}{rgb}{0,0.5,0}
\definecolor{DarkRed}{rgb}{0.75,0,0}
\definecolor{blue}{rgb}{0.0,0,0.5}

\theoremstyle{plain} \theorembodyfont{\upshape}
\newtheorem{theorem}{\noindent Theorem}
\newtheorem{prop}[theorem]{\indent Proposition}
\newtheorem{lemma}{\noindent Lemma}
\newtheorem{proof}{\noindent Proof}
\newtheorem{assumption}{\noindent Assumption}

\ifjournal
\title{\forinitialsubmission{Importance-weighted adaptive search for multi-class targets}{\huge Importance-weighted adaptive search for multi-class targets}}
\else
\title{Resource-Constrained Adaptive Search for Sparse Multi-Class Targets with Varying Importance}
\fi
\author{Gregory~E.~Newstadt,~\IEEEmembership{Member,~IEEE,~}Beipeng~Mu,~\IEEEmembership{Member,~IEEE,~}Dennis~Wei,~\IEEEmembership{Member,~IEEE,~}\\Jonathan~P.~How~\IEEEmembership{Senior Member,~IEEE,~}and~Alfred~O.~Hero~III,~\IEEEmembership{Fellow,~IEEE}
\thanks{Gregory Newstadt is with Google Pittsburgh, Pittsburgh, PA 15206, E-mail: gregoryn@google.com}
\thanks{Beipeng Mu and Jonathan How are with Dept. of Aeronautics and Astronautics, Massachusetts Institute of Technology, Cambridge, MA 02139, E-mail: (\{mubp\},\{jhow\})@mit.edu}
\thanks{Dennis Wei is with the Thomas J. Watson Research Center, IBM Research, Yorktown Heights, NY 10598, E-mail: dwei@us.ibm.com.}
\thanks{Alfred Hero is with the Dept. of Electrical Engineering and Computer Science, University of Michigan, Ann
Arbor, E-mail: hero@umich.edu.}
\thanks{The research in this paper was partially supported by Army Research Office MURI grant number W911NF-11-1-0391.}
}

\begin{document}

\maketitle

\begin{abstract}
In sparse target inference problems it has been shown that significant gains can be achieved by adaptive sensing using convex criteria. We generalize previous work on adaptive sensing to (a) include multiple classes of targets with different levels of importance and (b) accommodate 
multiple sensor models. New optimization policies are developed to allocate a limited resource budget to simultaneously locate, classify and estimate a sparse number of targets embedded in a large space. Upper and lower bounds on the performance of the proposed policies are derived by analyzing a baseline policy, which allocates resources uniformly across the scene, and an oracle policy which has a priori knowledge of the target locations/classes.  These bounds quantify analytically the potential benefit of adaptive sensing as a function of target frequency and importance. 
Numerical results indicate that the proposed policies perform close to the oracle bound 
when signal quality is sufficiently high (e.g.~performance within 3 dB for SNR above 15 dB). Moreover, the proposed policies improve on previous policies in terms of reducing estimation error, reducing misclassification probability, and increasing expected return. To account for sensors with different levels of agility, three sensor models are considered: 
global adaptive (GA), which can allocate different amounts of resource to each location in the space; global uniform (GU), which can allocate resources uniformly across the scene; and local adaptive (LA), which can allocate fixed units to 
a subset of locations. Policies that use a mixture of GU and LA sensors are shown to perform similarly to those that use GA sensors while being more easily implementable.  
\end{abstract}

\section{Introduction}
This work considers localization, classification, and estimation of targets from observations taken sequentially and adaptively.  In particular, we focus on the regime where (a) targets are sparse compared to the size of the scene, and (b) some of the targets have higher importance than 
others. For example, in search-and-rescue missions, detection of survivors has significantly higher mission importance than detection of other features in the environment. 
Similarly, a radar operator may be more interested in detecting/tracking a tank rather than a car, though both might sparsely populate the scene.

Viewing the targets as sparse signals, adaptive localization and estimation use past observations to shape future measurements of the scene 
\cite{Castro-04-coarse-to-fine,Castro-05-faster-rate-reg-via-act-learn, Bashan08_TSP, Bashan11_TSP, Hitchings_Castanon_AdaptiveSensing2010, haupt2011distilled, haupt2012sequentially, Wei13_TSP,malloy2012sequential,MalloyAdaptiveCompressed13},
which can result in stronger signal-to-noise ratios. The performance gain occurs by adaptively focusing the majority of sensing resources only on the positions that contain targets. Applications where adaptive sensing has been used include image \forinitialsubmission{acquisition/compression}{processing}, spectrum sensing, agile radars, and medical imaging\cite{Wei13_TSP,haupt2011distilled,Averbuch12_Image,Indyk11_sparsity,Bashan11_TSP,Tajer12_TIT}. 

\ignore{Previous work has shown that adaptive sensing problems can be formulated as partially observable Markov decision processes (POMDPs) \cite{Chong09_sensorPOMDP,Natarajan12_ICDSC,Krishnamurthy05_TAES,Capitan12_icra}. 
Applications include surveillance \cite{Chong09_sensorPOMDP,Natarajan12_ICDSC} and robot coordination \cite{Capitan12_icra}. However, the complexity of POMDP solutions grows exponentially with the number of targets and sensors, making them generally intractable for the size of problems considered here. The scaling is further complicated by facts that sensing result of one sensor can affect the result of another; signals are continuous thus must be discretized to apply POMDP, and sensors have various models. This work provides approximate solutions that are both tractable and also perform very well in comparison to baseline policies (lower bound) and oracle policies (upper bound).}

Previous work on adaptive sensing for sparse targets \cite{Bashan08_TSP, Bashan11_TSP, Hitchings_Castanon_AdaptiveSensing2010, haupt2011distilled, haupt2012sequentially, Wei13_TSP,malloy2012sequential,MalloyAdaptiveCompressed13,Averbuch12_Image,Indyk11_sparsity,Tajer12_TIT,Wei-Hero-guarantees-adaptive-estimation-sparse-signals}
considered only the two-class detection problem, where the target is either absent or present.  In many applications, such as surveillance and search-and-rescue missions, targets may have different classes with varying importance to the mission. In this setting, detection-based methods may waste resources on unimportant targets and thus suffer performance losses for the important ones. This work  explicitly accounts for multi-class targets with different mission importance.

Previous work in adaptive sensing \cite{Bashan08_TSP,Wei13_TSP,haupt2011distilled} also assumed the availability of an agile sensor that can allocate sensing resources to any combination of locations in the scene at the same time and with potentially different effort  (i.e. dwell time or energy).  In some cases we may only have access to sensors with restricted agility in terms of allocating different quantities 
of resources to different locations.  For example, a global sensor (e.g. in wide-area surveillance) may only be able to allocate resources uniformly across the scene; a local sensor (e.g. on an unmanned aerial vehicle) may only be able to allocate resources to a small subset of the scene.
This paper provides resource allocation policies for when a globally adaptive sensor is available, as well as when the agility of the sensor is limited.  Moreover, the differences in performance of these policies are analyzed and the regimes where each of the policies perform well are identified.

There is relatively limited work for planning with multiple types of sensors and tasking agents, particularly when the size of the scene is large.  Previous work  \cite{Mu13_GNC,Bertuccelli11_JACIC} proposed a planning algorithm for a team of heterogeneous sensors with the goal of maximizing mission importance. This algorithm showed that tasking agents achieved significantly better performance gains when target exploration included mission importance in the planning stage, as compared to an exploration policy that only depended on target uncertainty.
This resource allocation problem included many additional parameters (e.g. timing and delays), but was limited to the situation where the number and location of targets were assumed to be known a priori.  In this work, we relax this assumption and simultaneously enumerate and localize targets along with sensor planning.


We provide a Bayesian formulation and objective function that generalize our previous work
\cite{Bashan08_TSP,Wei13_TSP} to include targets with different mission importance.  
 Subsequently, upper bounds on the performance of adaptive resource allocation policies are derived through analysis of a full oracle policy with complete knowledge of the target locations and mission importances and a location-only oracle with knowledge only of target locations.  Similarly, a lower bound on performance is derived from analysis of a baseline policy which uniformly allocates resources everywhere in the scene.  These bounds indicate that there are regimes in which incorporating mission importance can lead to significant gains over both the baseline policy, as well as adaptive policies which do not incorporate mission importance.  Moreover, the bounds yield a simple method for mathematically predicting the benefit of adaptive sensing as a function of the system and target parameters.

This paper also provides resource allocation policies that are implementable (in contrast to the oracle policies) and approximately optimize the objective function in multiple scenarios, depending on the agility of the available sensors.  The performance of these policies is compared numerically to baseline policies, oracle policies and previously proposed policies, and the regimes (i.e., conditions on model parameters) in which the new policies perform well are identified.  Results indicate that the proposed policies perform significantly better than policies which do not include mission importance.  Moreover, in the high-resource regime, 
the proposed approach performs nearly as well as the oracle policies.  In scenarios where a globally adaptive sensor is not available, we further identify the tradeoffs between performance and sensor complexity (e.g., number of local sensors and/or inclusion of a global uniform sensor).  Finally, while the objective function is chosen for its convexity and ease of optimization, it is shown numerically that our proposed approach performs well over many performance metrics, including reducing mean squared estimation error and misclassification probability.

The rest of the paper is organized as follows: Section \ref{sec:model} presents the target and observation models, as well as the objective function; Section \ref{sec:performance-bounds} derives and compares the performance bounds; Section \ref{sec:policy} provides approximate optimization methods for multiple sensing scenarios; Section \ref{sec:sim} compares performance of the proposed policies; and the paper concludes in Section \ref{sec:conclusion}.

\section{Model}\label{sec:model}
Consider a scene with $N$ discrete locations, which contains a small number of targets of interest, for example, vehicles in wide area imagery. The $i$-th location has an unknown class label $C_i\in\calC = \set{1,2,\dots}$, where $C_i=1$ indicates the absence of a target. Define $\set{p_c}_{c\in\calC}$ as the prior distribution of the target class at a location, where $p_c = \Pr(C_i=c)$ and $\sum_{c\in\calC} p_c=1$.  Thus, the few-target (i.e. sparsity) assumption can be restated as $p_1\approx 1$.  
The prior distribution is assumed to be known. 
Moreover, this formulation can be easily extended so that the prior probabilities also depend on location.  

Correct classification of a location to class $c$ yields reward $h(c)$ to the mission (hereafter referred to as the mission importance).  Here,  $h:\calC\rightarrow[0,\infty)$ is a known reward function that maps the target class to its mission importance.  Moreover, it is assumed that the value of the no-target class is always zero, so that $h(1)=0$.  An example of a 3-class problem includes the no-target class ($c=1$), a low-value target class ($c=2$, e.g. a car) and high-value target class ($c=3$, e.g. a tank) where
\begin{align}\label{eq:hc}
h(c)=\left\{
\begin{array}{cc}
0, & c=1\\
1, & c=2 \\
10, & c=3\\
\end{array}\right. .
\end{align}
Without loss of generality, $h(c)$ is also assumed to be monotonically increasing in $c$.

Associated with each location $i$ is an amplitude $X_i$, called the signal and modeled as follows: 
\begin{align}\label{eq:xi_given_ci}
X_i|C_i=c \sim \left\{
\begin{array}{ll}
\forinitialsubmission{\mathrm{Normal}}{\mathcal{N}}(\mu_c,\sigma_c^2),& c> 1 \\
0,	& c=1 
\end{array}\right.
\end{align}
In words, conditioned on the event $C_i=c$ for $c > 1$, $X_i$ is distributed as a Gaussian random variable with known mean $\mu_c$ and variance $\sigma_c^2$, while $X_i$ is equal to zero with probability one if $c=1$. 

Observations of the signals are collected over $T$ stages with resource allocations $\lambda_i(t)$, which are a function of both location $i$ and stage index $t=0, \cdots, T-1$.  The resource $\lambda_i(t)$ can depend on observations up to and including stage $t$. Examples of resource  $\lambda_i(t)$ include integration time, number of samples, or transmitted power.  Given  $\lambda_i(t-1)>0$, the next 
observation $y_i(t)$ takes the form:
\begin{align}\label{eq:yt}
y_i(t)= X_i + \frac{n_i(t)}{\sqrt{\lambda_i(t-1)}}, ~i=1,\dots,N,\ t=1,\dots,T
\end{align}
where $\set{n_i(t)}_{i,t}$ is i.i.d. zero-mean Gaussian noise with variance $\nu^2$. If $\lambda_i(t-1)=0$,  no observation is taken.  A key point of this model is that the observation quality increases with sensing effort (i.e., $\lambda_i(t-1)$). 
The resource allocations are constrained with the following total resource budget:
\begin{align}
\label{eq:total-budget-constraint}
\sum_{t=0}^{T-1}\sum_{i=1}^N \lambda_i(t)=\sum_{t=0}^{T-1} \Lambda(t)=\Lambda,
\end{align}
where $\Lambda(t)$ are the per-stage budgets.  
For convenience, we write $\vlam(t)=[\lambda_1(t)\ \lambda_2(t)\dots \lambda_N(t)]^T$, $\vy(t)=[y_1(t)\ y_2(t)\dots y_N(t)]^T$ (similarly for other indexed quantities) and define $\vY(t)=\set{\vy(1),\vy(2),\dots,\vy(t)}$. 
The sequence of effort allocations $\vlam = \set{\vlam(t)}_{t=0}^{T-1}$ is called the allocation policy, where $\vlam(t)$ is a mapping from $\vY(t)$ to $[0,\Lambda(t)]^N$.

A straightforward multi-class extension of results in \cite{Wei13_TSP} yields the following posterior model for targets and their associated signals. 
Conditioned on the measurements up until time $t$ and the class $C_i$, we have 
\begin{equation}
\label{eq:xi_given_Y_ci}
X_i \big| \vY(t), C_i=c \sim \forinitialsubmission{\mathrm{Normal}}{\mathcal{N}}(\hatXic{c}(t), \sictsqd{c}{t}),
\end{equation} 
where the posterior mean (conditional mean estimator) and variance are defined as 
\begin{align}
\label{eq:Xi-cme}
\hatXic{c}(t)&=\expec{X_i\big|\vY(t),C_i=c}\\
\label{eq:posterior-variance-def}
\sictsqd{c}{t}&=\mathrm{var}\left[X_i\big|\vY(t),C_i=c\right]
\end{align} 
with $\hatXic{c}(0) = \mu_{c}$ and $\sictsqd{c}{0} = \sigma_{c}^{2}$. 
The posterior probability for a target at location $i$ having class $c$ is
\begin{align}
\label{equ:pic-def}
\pic{c}(t)&=\Pr(C_i=c|\vY(t))
\end{align} 
with $\pic{c}(0) = p_{c}$. 
When the observation $y_{i}(t)$ is taken, \eqref{eq:Xi-cme}-\eqref{equ:pic-def} satisfy the update equations
\begin{align}
\label{eq:var-update}& {\sictsqd{c}{t}} = \nu^2\left[\frac{\nu^2}{\sictsqd{c}{t-1}} + \lambda_i(t-1)\right]^{-1} \\
\label{eq:mean-update}& \hatXic{c}(t) = \sictsqd{c}{t}\left(\frac{\hatXic{c}(t-1)}{\sictsqd{c}{t-1}}+\frac{y_i(t)\lambda_i(t-1)}{\nu^2} \right) \\
\label{eq:prob-update}
&\pic{c}(t) = \frac{\pic{c}(t-1)f(y_i(t)|\vY(t-1),C_i=c)}{\sum_{c'\in\calC}\pic{c'}(t-1) f(y_i(t)|\vY(t-1),C_i=c')}
\end{align}
where $y_i(t+1)|\vY(t),C_i=c$
\begin{equation}
\label{equ:y_t-prob}
\sim
\begin{cases}
\forinitialsubmission{\mathrm{Normal}}{\mathcal{N}}\left(0,\nu^2/\lambda_i(t)\right)& c=1,\\
\forinitialsubmission{\mathrm{Normal}}{\mathcal{N}}\left(\hatXic{c}(t),\sictsqd{c}{t}+\nu^2/\lambda_i(t)\right), & c>1,
\end{cases}
\end{equation}
If $y_{i}(t)$ is not taken, the posterior parameters do not change. 

\subsection{Objective function for resource allocations}

We consider a generalization of the resource allocation objective functions in \cite{Bashan08_TSP,Wei13_TSP}. In analogy with \cite{Bashan08_TSP,Wei13_TSP}, define 
$\Iic{c}$ to be the following indicator function
\begin{equation}
\Iic{c}=\begin{cases}
1,&C_i=c\\
0,&C_i\neq c
\end{cases}.
\end{equation}
The proposed objective function can then be expressed as 
\begin{equation}
\label{eq:multiclass-objective-function1}
J_T(\vlam)= \expecgen{\vY(T),\vC,\vX}{\sum\limits_{i=1}^N\sum\limits_{c=2}^{|\cal{C}|}\Iic{c} h(c)\left(X_i-\hatXic{c}(T)\right)^2},
\end{equation}
where $\expecgen{\vY(T),\vC,\vX}{\cdot}$ is the expectation operator over $\vY(T),\vC,$ and $\vX$.
The inner sum corresponds to the 
mean-squared error (MSE) in estimating $X_i$ given all observations $\vY(T)$ and class $C_{i}$, weighted by mission importance $h(C_{i})$. 
Note that we have already substituted the posterior mean estimator $\hatXic{c}(T)$ in \eqref{eq:multiclass-objective-function1}, which minimizes the MSE. 

The two-class problem considered in \cite{Bashan08_TSP,Wei13_TSP} is a special case of \eqref{eq:multiclass-objective-function1} with $\calC=\set{1,2}$, $h(1)=0$ and $h(2)=1$.  
These previous works derived optimal resource allocation policies that minimize (\ref{eq:multiclass-objective-function1}) when there is only a single target class with nonzero mission importance. The generalization to multiple nonzero target classes in \eqref{eq:multiclass-objective-function1} penalizes the policy for incorrect target classification in addition to incorrect detection. 

To explicitly reflect the fact that the target classes are random and not known exactly, we may rewrite \eqref{eq:multiclass-objective-function1} in terms of the posterior class probabilities $\pic{c}(t)$ instead of the indicators $\Iic{c}$, as shown below.

\begin{prop} 
\label{prop:cost-MSE-equivalence}
The objective function (\ref{eq:multiclass-objective-function1}) is equivalent  to
\begin{equation}\label{eq:final-cost-form}
\ignore{\begin{split}
J_T(\vlam)&=\nu^2\expecYY{\sum\limits_{i=1}^N\frac{ \zi(T-1)}{\nu^2/\si^2(T-1)+ \lambda_i(T-1)}},\\
&=\nu^2\expecYY{\sum\limits_{i=1}^N\frac{ \zi(T-1)}{\nu^2/\sigma^2_0+\olami}},
\end{split}
}
\begin{split}
J_T(\vlam)&=\nu^2\expecYY{\sum\limits_{i=1}^N\sum_{c=2}^{|\calC|}\frac{ \pic{c}(T-1)h(c)}{\nu^2/\sictsqd{c}{T-1}+ \lambda_i(T-1)}},\\
&=\nu^2\expecYY{\sum\limits_{i=1}^N\sum_{c=2}^{|\calC|}\frac{ \pic{c}(T-1)h(c)}{\nu^2/\sigma_c^2+ \olami}},
\end{split}
\end{equation}
where 
\ignore{
\begin{align}
\zi(T-1) &= \sum_{c\in\cal{C}}h(c)\pic{c}(T-1) \label{equ:Zi}\\
}
$\olami=\sum\limits_{t=0}^{T-1}\lambda_i(t)$.
\end{prop}
\begin{IEEEproof}
\ifjournal
\else
	The proof is given in Appendix \ref{app:cost-MSE-equivalence}.
\fi
\end{IEEEproof}
The alternative form of the objective function in \eqref{eq:final-cost-form} also shows its explicit dependence on the allocated resources $\vlam = \set{\vlam(t)}_{t=0}^{T-1}$, which will be useful in the sequel. 

While it is apparent from 
\eqref{eq:multiclass-objective-function1} 
that the objective function explicitly penalizes estimation errors, classification errors are implicitly penalized as well. More specifically, \eqref{eq:multiclass-objective-function1} encourages the allocation of more resources to reduce the MSE in locations where targets have high value $h(C_i)$.  However, in order to selectively concentrate resources in this way, targets must also be correctly classified. Proposition \ref{prop:misclass-prob} confirms the intuition that allocating resource to a location 
reduces the corresponding mis-classification error. 
We consider 
a modified Maximum Posterior (MAP) class estimator: 
\begin{equation}\label{equ:estimated-C}
\hat{C}_i(t) = \left\{ \begin{array}{ll}
\arg\max_{c\in\calC}~\pic{c}(t) & \text{when } y_i(t)\geq \epsilon \\
1					& \text{when } y_i(t)<\epsilon
\end{array}\right.
\end{equation}
where $\epsilon$ is a user-chosen threshold.  Define misclassification probability of location $i$ with true class $C_i$, given measurements up until stage $t$ as
\begin{align}\label{equ:missP-def}
q_i(t) &=\Pr\left(\hat{C}_i(t)\neq C_i\big| \vY(t-1)\right)
\end{align}
We define $\sigma_0^2 \trieq \max\limits_{c>1} \sigma_c^2$ and introduce the following assumption of equal prior target variances: 
\begin{assumption}\label{ass:equal-variance}
$\sigma_c^2 = \sigma_0^2$ 
for all $c>1$.
\end{assumption}
This assumption leads to 
upper bounds on both the mis-classification error as well as the objective function (\ref{eq:final-cost-form}) since target variances smaller than the maximum would make estimation and classification easier.  

\begin{prop} 
\label{prop:misclass-prob}
Under Assumption \ref{ass:equal-variance}, when $\hatXic{c}(t-1) > 0$ for $c>1$ and $\epsilon=0$, $q_i(t)$ is monotonically decreasing with $\lambda_i(t-1)$.
\end{prop}
Assumption \ref{ass:equal-variance} and the condition $\hatXic{c}(t-1) > 0$ for $c>1$ simplify the proof of Proposition \ref{prop:misclass-prob} given in Appendix \ref{app:missclassification-prob}. If $\mu_c > 0$ for $c > 1$ and the sensor noise variance $\nu^2$ is small, then $\hat{X}_i^{(c)}(t-1)$ will be positive with high probability. A result analogous to Proposition \ref{prop:misclass-prob} holds if instead $\hatXic{c}(t-1) < 0$ for $c > 1$. It is conjectured but not proven in this paper that $q_{i}(t)$ is monotonically decreasing in $\lambda_{i}(t-1)$ under much weaker conditions than stated in Proposition \ref{prop:misclass-prob}.

\forinitialsubmission{Bounds on the objective function proposed in this section will be given in the next section.}{}

\section{Performance bounds}
\label{sec:performance-bounds}

This section presents bounds on the performance gain that can be achieved using adaptive resource allocation policies.  These bounds are used to quantify the benefit of adaptive sensing, particularly in the presence of targets with varying mission importance. Lower bounds on the objective function are obtained by analyzing the performance of an oracle policy, which are optimal allocations that depend on exact knowledge of the locations and (in some cases) the classes of the targets.  An upper bound is obtained by analyzing the performance of a baseline non-adaptive policy that allocates resources uniformly over the scene, called the uniform policy.

Assumption \ref{ass:equal-variance} on the equality of prior variances $\sigma_{c}^{2}$ is used in this section in order to derive bounds in Propositions \ref{prop:full-oracle-cost-upper-bound}--\ref{prop:uniform-cost} that are closed-form and facilitate interpretation. 
Exact expressions that do not require Assumption \ref{ass:equal-variance} are provided in 
\ifjournal
\else
Appendix~\ref{app:proof-prop-oracle-cost-general}. A comparison of the two approaches in Fig.~\ref{fig:senstvt-equ-var} shows that the main conclusions of this section remain valid without Assumption \ref{ass:equal-variance}.
\fi

In all of the policies considered in this section,  $\vlam$ is a deterministic function of the target classes $\vC$. For example, in the oracle policy, the target class is known, and the same allocation is given to targets of the same class, while in the uniform policy, the same allocation is given to targets of all classes. In this case, the objective function can be further simplified, as given below:

\begin{lemma}{
\label{lemma:cost-determ-lambda}
When $\vlam$ is a deterministic (non-random) function of $\vC$, 
\begin{equation}
\label{eq:oracle-cost-function}
\begin{split}
J_T(\vlam) &= \nu^2
\expecgen{\vC}{\sum\limits_{i=1}^N\frac{h(C_i)}{\nu^2/\sigma_c^2+\olami}}.
\end{split}
\end{equation}
}
\end{lemma}
\begin{IEEEproof}
\ifjournal
\else
The proof is given in Appendix \ref{app:cost-determ-lambda}.
\fi
\end{IEEEproof}

\subsection{Full oracle policy}
The full oracle policy has exact knowledge of both the locations and classes of targets within the scene. In other words, $\pic{c}(t)=\Iic{c}$ for all $i=1,2,\dots,N$, $c\in \calC$, and $t=0,1,\dots,T-1$. The optimal full-oracle policy 
is given by
\ifjournal
\else
\begin{equation}
\label{eq:oracle-allocation}
{\overline{\lambda_{\pi(i)}^{o}}}=\begin{cases}
\dfrac{\left(\Lambda+\nu^2\sum\limits_{j=1}^{k^*}\sigma_{C_{\pi(j)}}^{-2}\right)\sqrt{h(C_{\pi(i)})}}{\sum_{j=1}^{k^*}\sqrt{h(C_{\pi(j)})}}-\dfrac{\nu^2}{\sigma_{C_{\pi(i)}}^2}, & i=1,\dots,k^*\\
0, & \mathrm{otherwise}\end{cases}.
\end{equation}
\fi
where the number of non-zero allocations $k^*$ depends on SNR, 
and $\pi$ is an index permutation that sorts $\sqrt{h(C_i)}\sigma_{C_i}^2$ in non-increasing order so that
\begin{equation}
\label{eq:index-permutation-definition}
\sqrt{h(C_{\pi(1)})}\sigma_{C_{\pi(1)}}^2 \geq 
\ifjournal
\else
\sqrt{h(C_{\pi(2)})}\sigma_{C_{\pi(2)}}^2 \geq 
\fi
\cdots \geq \sqrt{h(C_{\pi(N)})}\sigma_{C_{\pi(N)}}^2.
\end{equation}
\ifjournal
\else
The derivation of the optimal full-oracle policy is given in Appendix \ref{app:full-oracle-deriv}.  
\fi

To analyze the performance of the full oracle policy, we make the following definitions and assumptions. 
Denote the number of targets with class $c$ as
\begin{equation}
N_c = |\set{i:C_i = c}|
\end{equation}
Define the first and second moments of $\sqrt{h(C_i)}$ conditioned on $C_i>1$ (so that $h(C_i)>0$):
\begin{align}
\label{eq:first-moment-sqrt-h}
\firstmom &\trieq \expecgen{C_i>1}{\sqrt{h(C_i)}} = \frac{1}{\pbar}\sum\limits_{c=2}^{|\calC|} p_c\sqrt{h(c)}\\
\label{eq:second-moment-sqrt-h}
\secondmom &\trieq \expecgen{C_i>1}{h(C_i)} = \frac{1}{\pbar}\sum\limits_{c=2}^{|\calC|} p_ch(c)
\end{align}
where $\pbar = 1-p_1$. We invoke Assumption \ref{ass:equal-variance} and the following:
\begin{assumption}
\label{ass:high-snr-assumption-oracle1}
The total resource budget is greater than $\Lambda_{\mathrm{min}}$,
\begin{equation}
\Lambda_{\mathrm{min}} \geq c_0 \max\set{\Lambda_0,\Lambda_1}, \label{eq:lambda-min-oracle1}
\end{equation}
where $c_0 = \nu^2/\sigma_0^2$ and 
\begin{align}
\Lambda_0 &= \sum\limits_{c=2}^{|\calC|} N_c\sqrt{h(c)/h(2)}-(N-N_1)\label{eqn:Lambda0}\\
\Lambda_1 &= \left(\secondmom/\firstmomsq\right)-1
\end{align}
\end{assumption}
The condition $\Lambda_\mathrm{min}\geq c_0\Lambda_0$ ensures that at least some effort is distributed to all classes with non-zero mission importance (i.e. $h(c)>0$).
This allows \eqref{eq:oracle-allocation} to be rewritten as 
\begin{equation}
\label{eq:oracle-allocation-high-SNR}
{\overline{\lambda_{i}^{o}}}=
\dfrac{\left(\Lambda+(N-N_1)c_0\right)\sqrt{h(C_{i})}}{\sum_{c\in\calC}N_c\sqrt{h(c)}}-c_0,
\end{equation}
when $C_i>1$ and 0 otherwise. The second condition $\Lambda_\mathrm{min} \geq c_0\Lambda_1$ implies a certain functional convexity that facilitates bounding expectations. 
Note that $\Lambda_{0}$ in \eqref{eqn:Lambda0} depends on the histogram of classes $\vN = \set{N_c}_{c=1,2,\dots,|C|}$, which is random.  In 
\ifjournal
\else
Appendix \ref{app:snr-condition-prob}, 
\fi
we show
that 
if $\Lambda_{0}$ in \eqref{eq:lambda-min-oracle1} is replaced by the deterministic quantity 
\begin{equation}\label{eqn:Lambda0'}
\Lambda_{0}' = \alpha N \pbar \left( \frac{\firstmom}{\sqrt{h(2)}} - 1 \right),
\end{equation}
where $\alpha > 1$ is an arbitrary constant, then the original condition $\Lambda_\mathrm{min}\geq c_0\Lambda_0$ is satisfied with probability 
converging exponentially to $1$ as $N$ increases. 

\ignore{Moreover, we generally have $\Lambda_0>\Lambda_1$.  For example, using the simulation parameters in Section \ref{sec:sim} and evaluating $\Lambda_0$ at $\expec{\vN}$, we have $\Lambda_0 = 99.8$ and $\Lambda_1 = 0.27$.}

\begin{prop}
\label{prop:full-oracle-cost-upper-bound}
Given Assumptions \ref{ass:equal-variance} and \ref{ass:high-snr-assumption-oracle1}, the cost \eqref{eq:oracle-cost-function} of the full oracle policy (\ref{eq:oracle-allocation})
is bounded from below $J_T(\vlam^o)\geq L_T(\vlam^o)$ with 
\begin{equation}
\label{eq:expected-cost-over-realizations}
L_T(\vlam^o)\trieq\frac{\nu^2N\pbar}{\Lambda+N\pbar c_0}(\secondmom+(N\pbar-1)\firstmomsq),
\end{equation}
and bounded from above $J_T(\vlam^o)\leq U_T(\vlam^o)$ with
\begin{equation}
\label{eq:oracle-cost-upper-bound}
\begin{split}
&U_T(\vlam^o)\trieq L_T(\vlam^o) + \nu^2\frac{((\Lambda+c_0)\firstmomsq - c_0 \secondmom)Np_1\pbar}{\Lambda^2}\\
&=\frac{\nu^2N\pbar}{\Lambda+N\pbar c_0}\left(\secondmom+(N\pbar-1 + O(1))\firstmomsq
\right)
\end{split}
\end{equation}
\end{prop}
\begin{IEEEproof}
The proof is given in Appendix \ref{app:proof-prop-oracle-cost}.
\end{IEEEproof}

\subsection{Location-only oracle policy}
A location-only oracle policy has knowledge of the locations of targets with $h(C_i)>0$, but not the classes themselves.  The location-only oracle policy is 
\begin{equation}
\label{eq:location-oracle-allocation}
{\overline{\lambda_{i}^{lo}}}=\begin{cases}
\Lambda/(N-N_1), & h(C_i)>0\\
0, & \mathrm{otherwise}\end{cases}
\end{equation}
This policy was studied in previous work \cite{Bashan08_TSP,Wei-Hero-guarantees-adaptive-estimation-sparse-signals} for the case $|\calC|=2$.  The proposition below gives a lower bound \eqref{eq:expected-cost-over-realizations-location-only} on the cost of the policy that generalizes a lower bound in \cite[Sec.~III-B]{Wei-Hero-guarantees-adaptive-estimation-sparse-signals} for the $2$-class case, and additionally provides an upper bound \eqref{eq:upper-bound-location-only} that converges to the lower bound for large $N$. 
\begin{prop}
\label{prop:location-oracle-cost-upper-bound}
Given Assumption \ref{ass:equal-variance}, the cost of the location-only oracle policy is bounded from below with $J_T(\vlam^{lo})\geq L_T(\vlam^{lo})$ with
\begin{equation}
\label{eq:expected-cost-over-realizations-location-only}
L_T(\vlam^{lo})\trieq\frac{\nu^2N^2\pbar^2\secondmom}{\Lambda+N\pbar c_0}.
\end{equation}
and bounded from above with $J_T(\vlam^{lo})\leq U_T(\vlam^{lo})$ with
\begin{equation}
\label{eq:upper-bound-location-only}
\begin{split}
U_T(\vlam^{lo})&\trieq L_T(\vlam^{lo}) + \frac{Np_1\pbar}{\Lambda}(\nu^2\secondmom)\\
&=\frac{\nu^2N\pbar}{\Lambda+N\pbar c_0}\left(N\pbar+O(1)\right)\secondmom
\end{split}
\end{equation}
\end{prop}
\begin{IEEEproof}
\ifjournal
\else
The proof is given in Appendix \ref{app:proof-prop-location-oracle-cost}.
\fi
\end{IEEEproof}

\subsection{Uniform (baseline) policy}
The uniform policy uniformly allocates resources everywhere in the scene, \forinitialsubmission{:
\begin{equation}
\label{eq:uniform-allocation-onestage}
\lambda_i^u(t) = \Lambda(t)/N
\end{equation}
}
{$\lambda_i^u(t) = \Lambda(t)/N$}, 
and the allocations across all stages are
\begin{equation}
\label{eq:uniform-allocation-allstages}
{\overline{\lambda_i^u}} = \sum_{t=0}^{T-1}\lambda_i^u(t) = \Lambda/N
\end{equation}
\begin{prop}
\label{prop:uniform-cost}
Given Assumption \ref{ass:equal-variance}, the cost of the uniform policy is 
\begin{equation}
\label{eq:expected-cost-uniform}
J_T(\vlam^{u})=\frac{\nu^2N\pbar\secondmom}{c_0+\Lambda/N}
\end{equation}
\end{prop}
\begin{IEEEproof}
\ifjournal
\else
The proof is given in Appendix \ref{app:proof-prop-uniform-cost}.
\fi
\end{IEEEproof}

\subsection{Comparisons of bounds}

\ignore{{\color{red} another case where I disagree with the change} These bounds yield insight into the regimes where there might be significant gains in comparison to the baseline (uniform) policy $\vlam^u$, the full oracle policy $\vlam^o$, and the location-only oracle policy $\vlam^{lo}$, which only has knowledge of location, but not classification of targets.}

The bounds in Propositions \ref{prop:full-oracle-cost-upper-bound}--\ref{prop:uniform-cost} are compared to 
yield insight into the regimes where one might expect significant gains in comparison to the non-adaptive baseline (uniform) policy $\vlam^u$, as well as the regimes in which incorporating mission importance (through the full oracle policy $\vlam^o$) yields significant gains over a detection-only policy (i.e., the location-only oracle $\vlam^{lo}$).  The model parameters of interest are the total effort budget $\Lambda$ or equivalently SNR\footnote{SNR is defined as a mapping from the total budget $\Lambda$: $\Lambda=10^{\mathrm{SNR}/10}N$.}, the prior distribution of targets $\set{p_c}_{c\in\calC}$, and the mission importance $\set{h(c)}_{c\in\calC}$.  Note that $\set{p_c,h(c)}_{c\in\calC}$ determine the moments $\firstmom$ and $\secondmom$ on which the bounds depend directly.

Define $G(\vlam^0,\vlam^1)$ as the gain of policy $\vlam^1$ with respect to another policy $\vlam^0$: 
\begin{equation}\label{eq:gain-definition}
G(\vlam^0,\vlam^1) = J_T(\vlam^0)/J_T(\vlam^1).
\end{equation}
From Propositions \ref{prop:full-oracle-cost-upper-bound}--\ref{prop:uniform-cost}, 
we obtain the following bounds on the performance gains of the oracle policies with respect to the uniform policy:
\begin{align}
\label{eq:gain-full-oracle}
G(\vlam^u,\vlam^o) &\leq \left(\frac{\Lambda+N\pbar c_0}{Nc_0+\Lambda}\right)g_1(N,\pbar,\firstmom,\secondmom)\\
\label{eq:gain-location-only-oracle}
G(\vlam^u,\vlam^{lo}) &\leq \left(\frac{\Lambda+N\pbar c_0}{Nc_0+\Lambda}\right)g_2(\pbar)
\end{align}
where the functions $g_1$ and $g_2$ are defined as
\begin{align}
g_1(N,\pbar,\firstmom,\secondmom) &= \frac{N\secondmom}{\secondmom+(N\pbar -1)\firstmomsq}\\
g_2(\pbar) &= {(\pbar)}^{-1}
\end{align}
The location-only oracle bound \eqref{eq:gain-location-only-oracle} coincides with the oracle bound \cite[eq.~(16)]{Wei-Hero-guarantees-adaptive-estimation-sparse-signals} in the 2-class case. 

The first term in both (\ref{eq:gain-full-oracle}) and (\ref{eq:gain-location-only-oracle}) approaches $1$ as the SNR $\Lambda\gg 1$. In this regime, 
the gain of the location-only oracle depends only on the proportion of non-zero targets through $\pbar$.  
To further interpret the full oracle bound \eqref{eq:gain-full-oracle}, we observe that $\firstmom$ is the first moment of the random variable 
$\sqrt{h(C_i)}$ 
conditioned on $C_i>1$, while $\secondmom$ is the corresponding second moment.  Then,
since the second moment is greater than the first moment squared, $\firstmomsq\leq\secondmom$, we have 
\begin{equation}
\begin{split}
g_1(N,\pbar,\firstmom,\secondmom) &\leq \frac{N\secondmom}{\firstmomsq+(N\pbar-1)\firstmomsq}\\
&= \frac{\secondmom}{\pbar\firstmomsq},
\end{split}
\end{equation}
where equality holds asymptotically as $N \to \infty$.  The ratio $\secondmom/\firstmomsq$ can be interpreted as $1 + \mathrm{CV}_{C_i>1}(\sqrt{h(C_i)})$, where the latter is the coefficient of variation (CV, a normalized measure of variability) of the mission importance $\sqrt{h(C_i)}$ conditioned on $C_i>1$.  

\begin{figure}[ht]
\centering
\hspace{-0.1in}
\subfloat[\small (\ref{eq:gain-full-oracle}) vs. ($\pbar$,SNR)]{
\label{fig:gains-p1-SNR-oracle-uniform}
\includegraphics[height=\boundsfigheight]{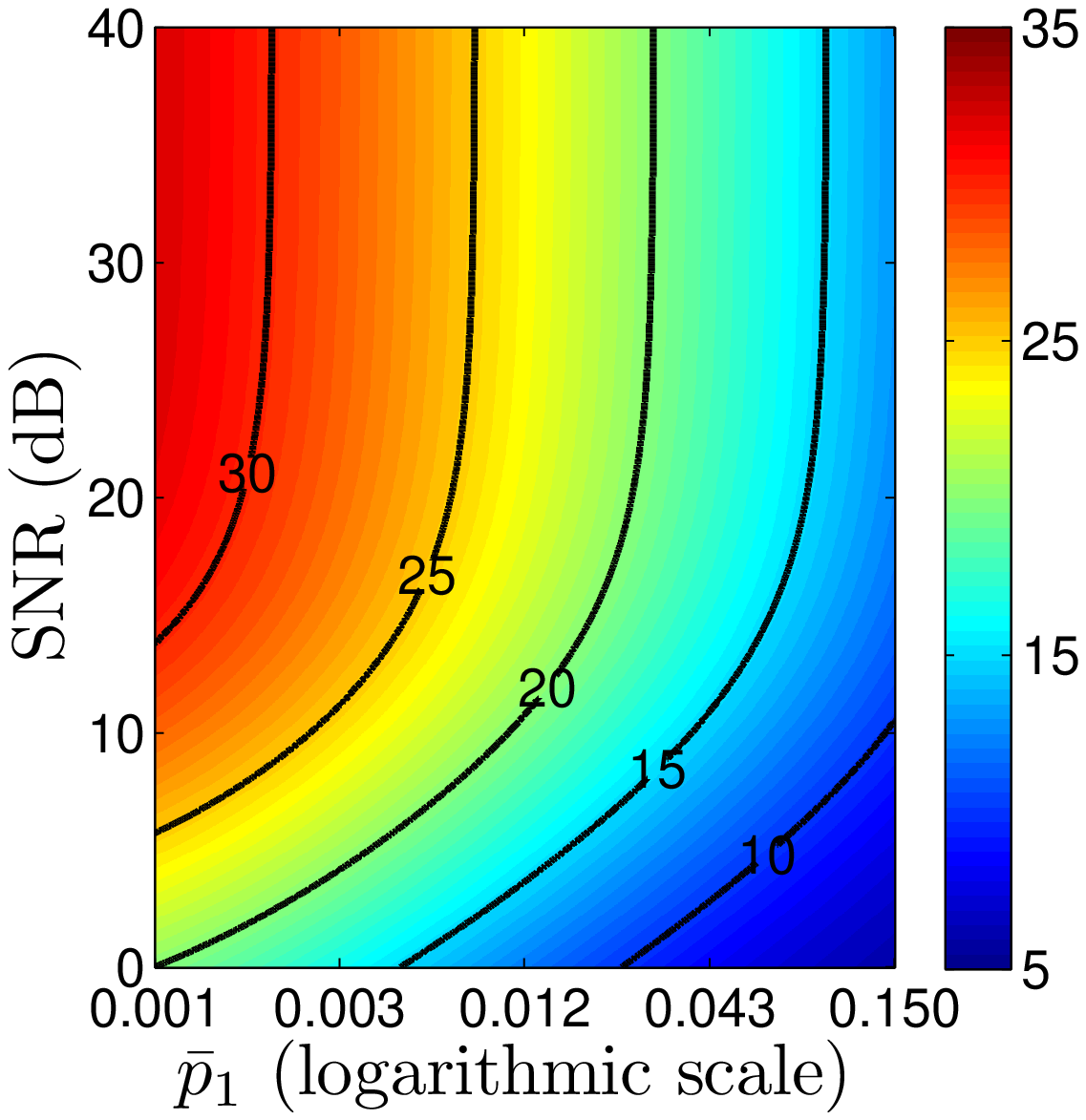}
}
\subfloat[(\ref{eq:gain-full-oracle}) vs.
 $\Bigl(\frac{\secondmom}{\firstmomsq},\pbar\Bigr)$]{
\label{fig:gains-p1-m2m1ratio-oracle-uniform}
\includegraphics[height=\boundsfigheight]{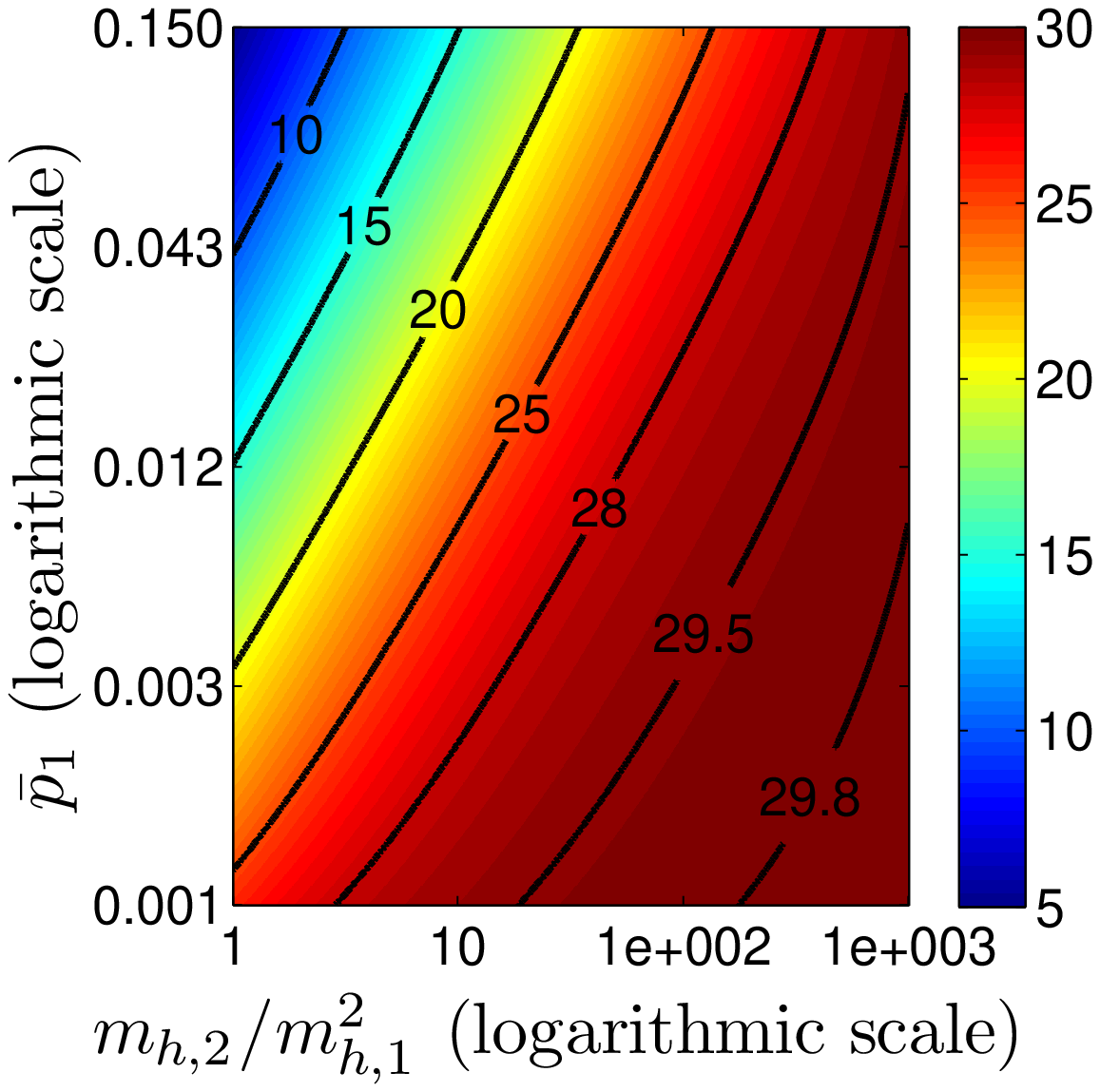}
}
\subfloat[(\ref{eq:gain-full-oracle}) {\small vs.}  ($N$,SNR)]{
\label{fig:gains-N-SNR-oracle-uniform}
\includegraphics[height=\boundsfigheight]{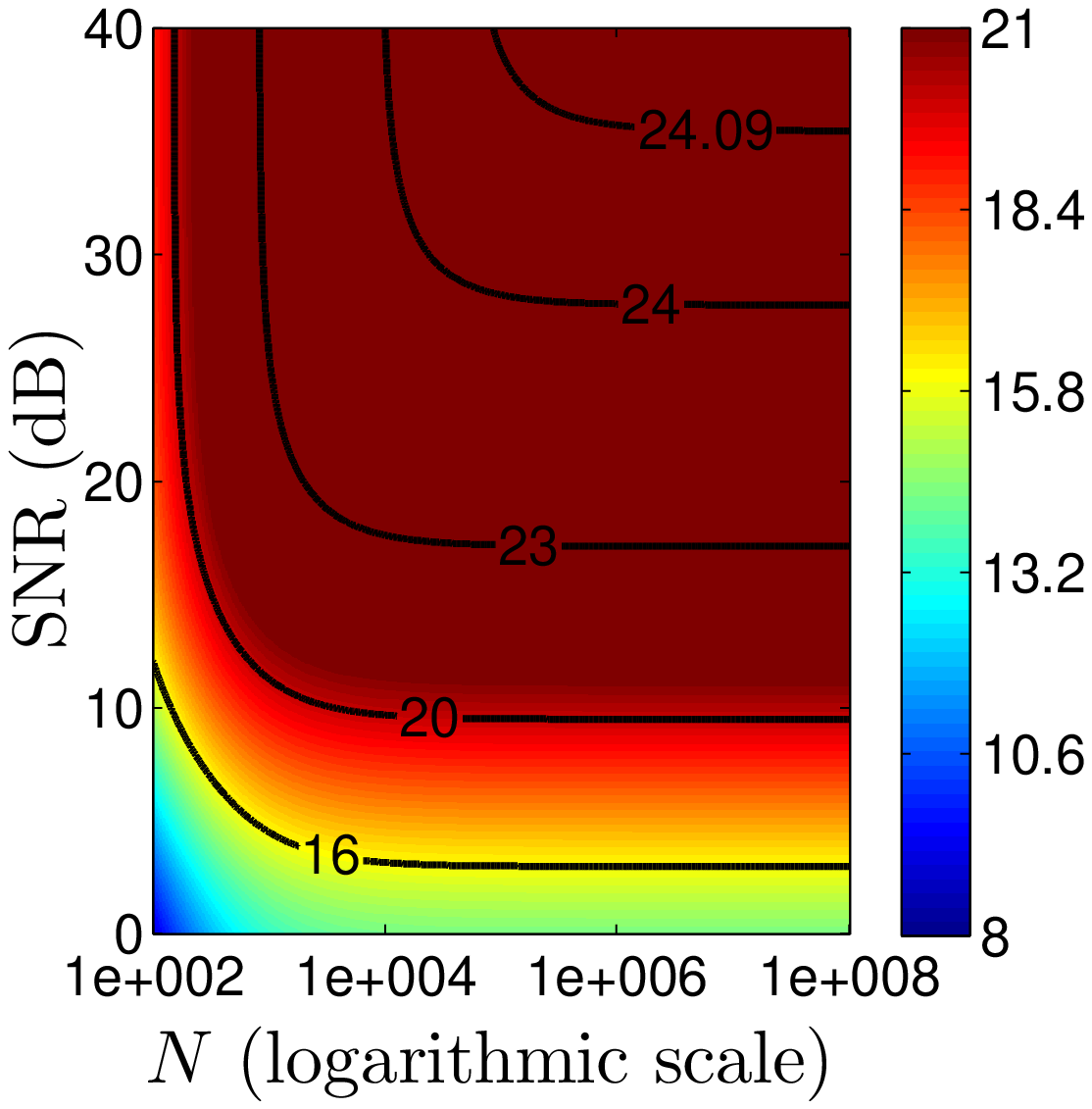}
}
\ignore{\subfloat[Bound (\ref{eq:gain-full-oracle}) vs. ($\secondmom/\firstmomsq$,$\firstmom$)]{
\label{fig:gains-m2m1ratio-m1-oracle-uniform}
\includegraphics[height=\boundsfigheight]{m1_ratio_oracle_gain}
}}
\caption{
Bound (\ref{eq:gain-full-oracle}) on the gain $G(\vlam^u,\vlam^o)$ of 
the full oracle policy w.r.t. the uniform policy across the model parameters.  The bound is plotted in units of dB ($10\log_{10}$).  Decreasing $\pbar$ (leading to fewer overall targets) and increasing SNR, $N$ and the ratio $(\secondmom/\firstmomsq)$ lead to larger gains. 
}
\label{fig:oracle-gains-over-uniform}
\end{figure}

Figure \ref{fig:oracle-gains-over-uniform} evaluates the bound \eqref{eq:gain-full-oracle} on 
the gain of the full oracle policy with respect to the uniform policy $G(\vlam^u,\vlam^o)$ 
across the model parameters.  Unless otherwise stated, the simulation parameters are given in Table \ref{table:simulation-parameters}.  It is seen that decreasing $\pbar$ (leading to fewer overall targets) and increasing SNR, $N$ and the ratio $(\secondmom/\firstmomsq)$ all lead to larger gains.  Moreover, the gains incurred by increasing $N$ and/or SNR approach a finite value (i.e., diminishing returns given sufficiently high $N$ or SNR), while they continue to grow with decreasing $\pbar$ 
or increasing $(\secondmom/\firstmomsq)$.  

\begin{table}
\centering
\caption{Simulation parameters}
\label{table:simulation-parameters}
\begin{tabular}{|l|c|}
\hline
Parameter & Value\\
\hline
Number of locations, $N$ & 2500\\
Number of classes, $|\calC|$ & 3\\
Class probabilities, $\set{p_c}_{c\in\calC}$ & $\set{0.95,0.049,0.001}$\\
Class importance, $\set{h(c)}_{c\in\calC}$ & $\set{0,1,2500}$\\
Target prior means, $\set{\mu_c}_{c\in\calC}$ & $\set{0,3,1.5}$\\
Target prior variances, $\set{\sigma_c^2}_{c\in\calC}$ & $\set{0,1/16,1/16}$\\
Noise variance, $\nu^2$ & 1\\
Number of sensors for LA policy, $M$ & 400\\
Number of sensors for GU/LA policy, $M$ & 50\\
\hline
\end{tabular}
\end{table}

Next, the bounds on the full and location-only oracles are compared directly to obtain 
\begin{equation}
\label{eq:gain-diff-oracles-upper}
\begin{split}
G(\vlam^{lo},\vlam^{o}) &\leq \frac{(N\pbar+O(1))\secondmom}{\secondmom+(N\pbar-1)\firstmomsq}\\
&\stackrel{N\to\infty}{\to} \frac{\secondmom}{\firstmomsq} = 1+\mathrm{CV}_{C_i>1}(\sqrt{h(C_i)}).
\end{split}
\end{equation}
Similarly, we have
\begin{align}
\label{eq:gain-diff-oracles-lower}
G(\vlam^{lo},\vlam^{o}) &\geq \frac{(N\pbar)\secondmom}{\secondmom+(N\pbar-1 + O(1))\firstmomsq}\\
&\stackrel{N\to\infty}{\to} \frac{\secondmom}{\firstmomsq} = 1+\mathrm{CV}_{C_i>1}(\sqrt{h(C_i)}).
\notag
\end{align}
Thus, when $N$ is large, the upper and lower bounds converge to the same value (i.e., 1 plus the CV).  
The CV is equal to zero in the 2-class case or when the target classes have the same importance. In these cases, the full and location-only oracles coincide. 
Otherwise, the CV is positive and the asymptotic gain is greater than 1, indicating that the full-oracle 
performs better than the location-only oracle.  Hence the CV can be seen as representing the value of knowledge of target importance.

\begin{figure}[ht]
\centering
\subfloat[\eqref{eq:gain-diff-oracles-upper} vs. ($N$,SNR)]{
\label{fig:gains-N-SNR-oracle-lo}
\includegraphics[height=\boundsfigheight]{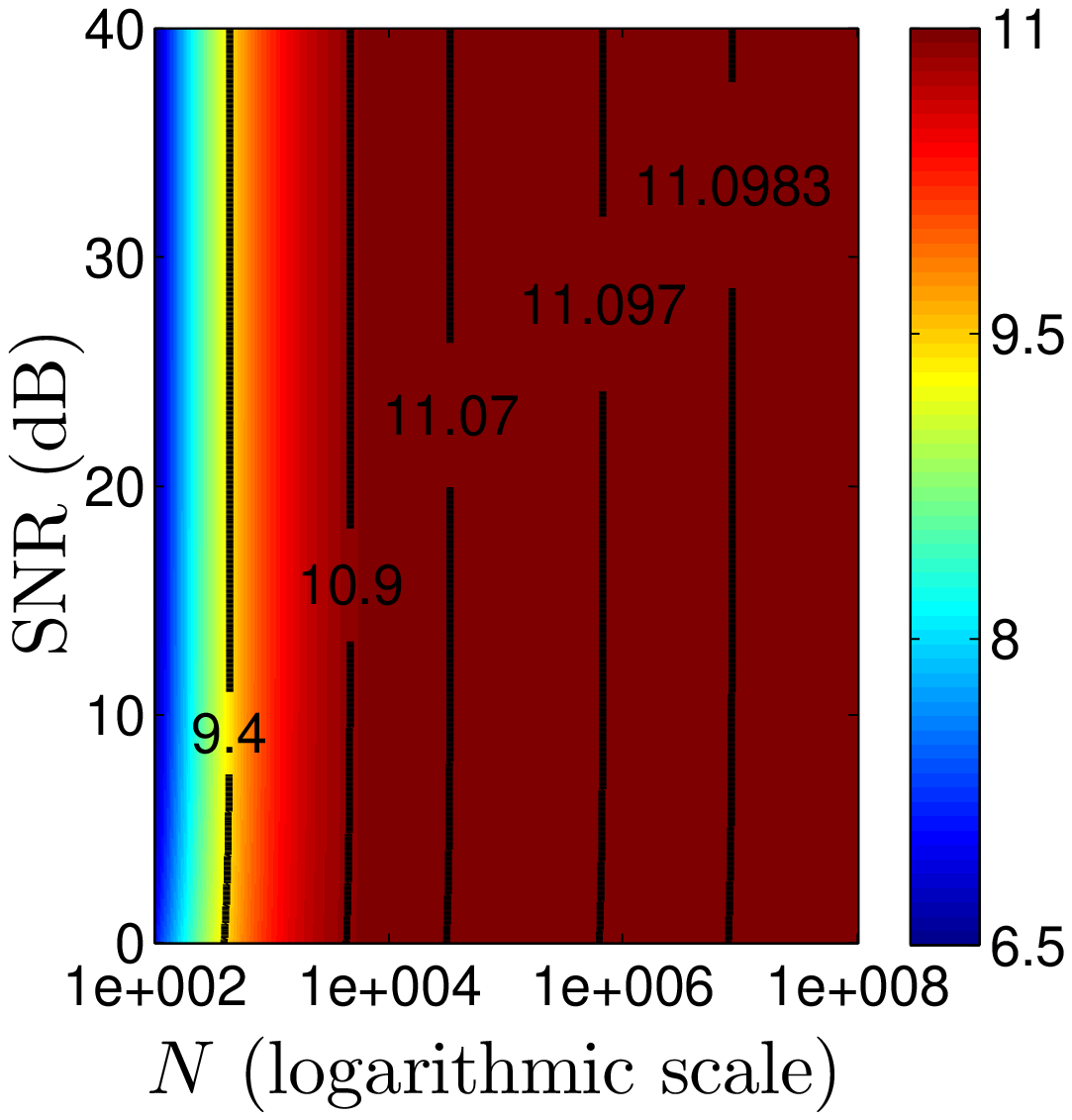}
}
\hspace*{0.1in}\subfloat[\eqref{eq:gain-diff-oracles-upper} vs. $\Bigl(\frac{\secondmom}{\firstmomsq},\pbar\Bigr)$]{
\label{fig:gains-p1-m2m1ratio-oracle-lo}
\includegraphics[height=\boundsfigheight]{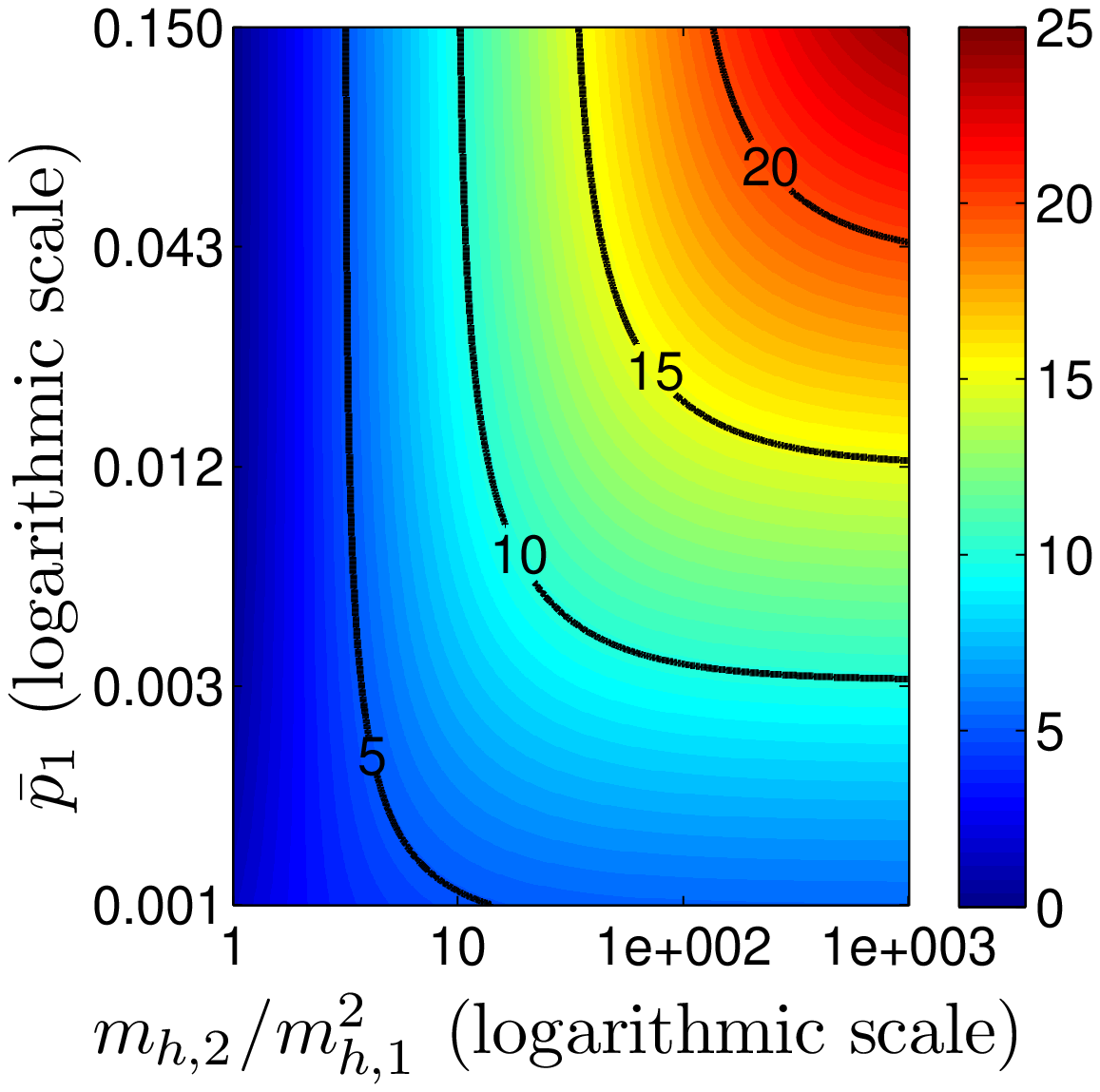}
}
\caption{
Bound (\eqref{eq:gain-diff-oracles-upper}, in dB) on the gain $G(\vlam^{lo},\vlam^o)$ of the full oracle policy with respect to the location-only oracle policy.  Increasing the ratio $(\secondmom/\firstmomsq)$ leads to larger gains as in Fig. \ref{fig:oracle-gains-over-uniform}(b).  Increasing $N$ also leads to marginally larger gains.  However, increasing $\pbar$ 
(i.e., increasing the number of targets) increases the gains between the oracle policies, which is the opposite pattern as compared to Fig. \ref{fig:oracle-gains-over-uniform}(b).}
\label{fig:oracle-gains-over-location-only}
\end{figure}

Figure \ref{fig:oracle-gains-over-location-only} evaluates 
the bound \eqref{eq:gain-diff-oracles-upper} on the gain of the full oracle policy with respect to the location-only oracle policy $G(\vlam^{lo},\vlam^o)$ across model parameters.  Increasing the ratio $(\secondmom/\firstmomsq)$ leads to larger gains, as expected from \eqref{eq:gain-diff-oracles-upper}, \eqref{eq:gain-diff-oracles-lower} and similar to Fig.~\ref{fig:oracle-gains-over-uniform}(b).   Increasing $N$ also leads to marginally larger gains.  However, increasing $\pbar$ 
(i.e., increasing the number of targets) increases the gains between the oracle policies, which is the opposite pattern as compared to Fig. \ref{fig:oracle-gains-over-uniform}(b). This suggests that when there is either a large spread in mission importance or many targets of interest, the full oracle policy has large gains over the location-only oracle.  

\begin{figure}[ht]
\centering
\includegraphics[height=\boundsfigheight]{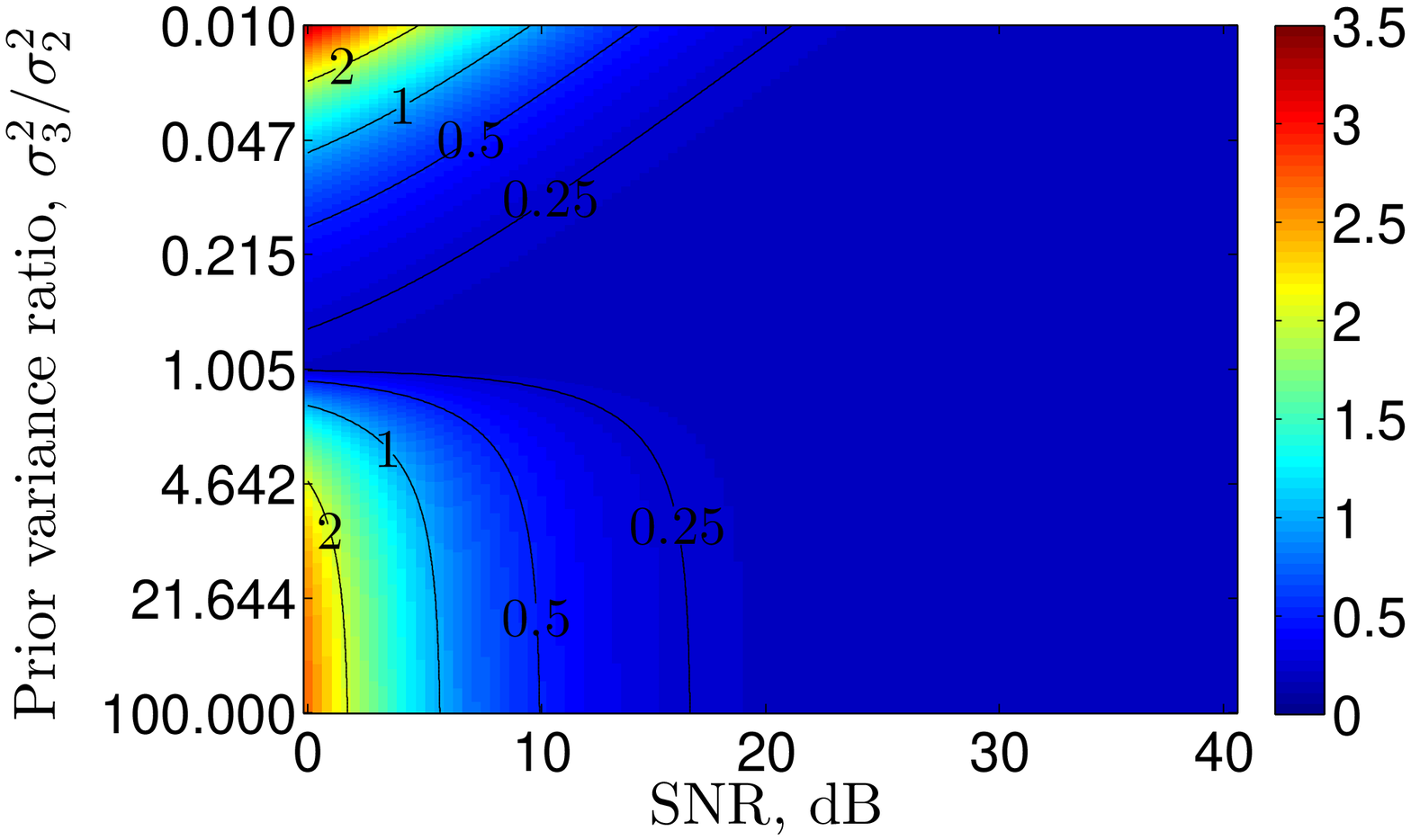}
\caption{Ratio (in dB) of the upper bound \eqref{eq:oracle-cost-upper-bound} on the full oracle cost to its exact value computed without Assumption \ref{ass:equal-variance}. There is no significant difference when $\sigma_2^{2}$ is within a factor of 100 of $\sigma_3^{2}$ or SNR is moderate to high.}
\label{fig:senstvt-equ-var}
\end{figure}

Finally we numerically evaluate 
the impact of Assumption \ref{ass:equal-variance} on the analysis of the full oracle policy. Figure \ref{fig:senstvt-equ-var} shows the ratio of the analytical upper bound \eqref{eq:oracle-cost-upper-bound} on the full oracle cost 
to the exact expression (\ref{eq:oracle-cost-function}) without Assumption \ref{ass:equal-variance}.  
The ratio is always greater than one (above 0 dB) since Assumption \ref{ass:equal-variance} gives an upper bound on the actual cost. 
It is seen that when (a) $\sigma_2^{2}$ is either 100 times larger or smaller than $\sigma_3^{2}$, and (b) SNR is low, there are gaps but the ratio never exceeds 2 (3 dB).  

\section{Search Policies}\label{sec:policy}
This section develops implementable policies that adaptively allocate resources to locations with targets. Recall a policy $\set{\vlam(t)}_{t=0}^{T-1}$ is a sequence of effort allocations, which are mappings from previous observations $\mathbf Y(t)$ to $[0,\Lambda(t)]^N$. Denote $\mathcal{U}(t)$ 
as the set of locations $i$ for which $\lambda_i(t)>0$. For example, $\mathcal{U}(t) 
=\{1,\cdots, N\}$ for the uniform policy and $\mathcal{U}(t) 
=\{i|C_i>1\}$ for the full oracle policy.

We introduce several models to emulate realistic types of sensors. Sensors typically vary in sensing agility, e.g.~field of view and accuracy, and in cost. Specifically, three types of sensors are considered: the global uniform (GU) sensor, the global adaptive (GA) sensor, and the local adaptive (LA) sensor. The global adaptive sensor can choose any subset of the scene and allocate different amounts of sensing resource to the selected locations.  These sensors provide the highest agility, but may be difficult or very expensive to realize. 
The global uniform sensor explores all locations with equal sensing resource in each location; e.g., a large field of view radar.  Therefore, it is likely to spend resources in locations without targets and suffer performance degradation compared to the global adaptive sensors.
Finally, the local adaptive sensor can only explore a small number of locations within the scene, albeit with high resolution, e.g., unmanned aerial vehicles (UAVs).  These sensors may be easily available in practice and are simpler to deploy as compared to global adaptive sensors.
Table \ref{table:sensor-types} compares these sensor types with regard to their allocations, ease of implementation, and ability to adapt. 

\begin{table}
\centering
\caption{Comparison of sensor types and allocations}
\label{table:sensor-types}\begin{tabular}{|p{.55in}|p{1.0in}|c|c|}
\hline
Sensor Type & Allocation & Adaptivity & {Implementation} \\
\hline
\rule{0pt}{8pt}Global Adaptive & $\lambda_i(t) \in [0,\Lambda/T]$, $\mathcal{U}(t)$ any subset 
& Full 
& Difficult\\
\hline
\rule{0pt}{8pt}Global Uniform & $\lambda_i(t) = \Lambda/(NT)$, $\mathcal{U}(t) = \{1,\ldots,N\}$ 
& None & Easy\\
\hline
\rule{0pt}{8pt}Local Adaptive  & $\lambda_i(t) = k\Lambda/(MT)$, $k = \set{0,1,\dots,M}$, $\lvert\mathcal{U}(t)\rvert \leq M$ 
& Limited
& Medium\\
\hline
\end{tabular}
\end{table}


\subsection{Global adaptive (GA) policy}
\label{subsec:policyGA}
As discussed in previous work on similar problems \cite{Wei13_TSP}, it is possible in principle to use dynamic programming (DP) to globally optimize $J_T(\vlam)$ with respect to the entire allocation policy $\vlam$ subject to the resource constraint \eqref{eq:total-budget-constraint}. 
However, for $T>2$, this exact solution is computationally intractable.  
As an alternative we present a myopic method that determines 
each $\vlam(t)$ independently for $t=0,1,\dots,T-1$ as the solution to 
\begin{align}\label{eq:greedy-problem}
\vlam^{ga}(t) = \argmin \limits_{\vlam(t)} ~ {K(t;\vlam(t))}\quad \mathrm{s.t.} ~\sum_{i=1}^N \lambda_i(t)=\Lambda(t),
\end{align}
where the myopic 
cost at stage $t$ given $\vY(t)$ is defined as
\begin{equation}
\label{eq:myopic-cost}
K(t;\vlam(t)) = \sum_{i=1}^N
\sum_{c=2}^{|\calC|}\frac{ \pic{c}(t) h(c)}{\nu^2/\sictsqd{c}{t}+ \lambda_i(t)},
\end{equation}
and $\Lambda(t)$ is some fraction of the total budget $\Lambda$. In this work, $\Lambda$ is equally allocated across stages, $\Lambda(t)=\Lambda/T$.
The myopic cost \eqref{eq:myopic-cost} is inspired by the exact cost function \eqref{eq:final-cost-form} and is equivalent to \eqref{eq:final-cost-form} when $t = T-1$.  However, \eqref{eq:myopic-cost} requires no expectation because we condition on (i.e.~have collected and incorporated) measurements up to stage $t$. 

The myopic allocation problem \eqref{eq:greedy-problem} is a convex optimization because the cost function is a sum of convex functions of the form $a / (b + \lambda_{i}(t))$, $a, b \geq 0$, and the constraints are linear (recall that $\lambda_{i}(t)$ is also constrained to be non-negative).  Hence it may be solved efficiently using a variety of iterative algorithms.  Furthermore, the optimal solution $\vlam^{ga}(t)$ satisfies the following precedence property: if 
\begin{equation}\label{eqn:precedence}
\sum_{c=2}^{|\calC|} \pic{c}(t) h(c) \bigl( \sigma_{i}^{(c)}(t) \bigr)^{4} \leq \sum_{c=2}^{|\calC|} p_{j}^{(c)}(t) h(c) \bigl( \sigma_{j}^{(c)}(t) \bigr)^{4},
\end{equation}
then we cannot have $\lambda_{i}^{ga}(t) > 0$ and $\lambda_{j}^{ga}(t) = 0$.  This property may be derived in a manner similar to \cite[App.~C]{Wei13_TSP} by invoking the optimality condition for \eqref{eq:greedy-problem}. 
The property implies that if $\vlam^{ga}(t)$ has $k$ nonzero components, they must correspond to the $k$ largest of the quantities appearing in \eqref{eqn:precedence}.

Under Assumption \ref{ass:equal-variance}, the posterior variances do not depend on class, $\sictsqd{c}{t} = \si^{2}(t)$.  In this case, by defining 
\begin{equation}\label{equ:Zi}
z_{i}(t) = \sum_{c=2}^{|\calC|} \pic{c}(t) h(c),
\end{equation}
the optimal solution $\vlam^{ga}(t)$ can be expressed analytically as in \cite{Wei13_TSP}.  We refer the reader to eqs.~(21)--(24) of \cite{Wei13_TSP} with $p_{i}(t)$ replaced by $z_{i}(t)$.

\ignore{
The optimal solution follows \cite{Wei13_TSP} and begins by defining $\pi$ to be an index permutation that sorts  $\sqrt{z_i(t)}\sigma_i^2(t)$ in non-increasing order:
\begin{equation}\label{equ:sort-Zi}
\sqrt{z_{\pi(1)}(t)}\sigma_{\pi(1)}^2(t)\geq \cdots \geq \sqrt{z_{\pi(N)}(t)}\sigma_{\pi(N)}^2(t).
\end{equation}
Let $c_i(t)=\nu^2/\sigma_i^2(t)$ and define $g(k)$ to be the monotonically non-decreasing function of $k=0,\dots,N$ with $g(0)=0$, 
\begin{equation}
\label{eq:g-func}
g(k) = \frac{c_{\pi(k+1)}(t)}{\sqrt{z_{\pi(k+1)}(t)}} \sum\limits_{i=1}^k\sqrt{z_{\pi(i)}(t)}- \sum\limits_{i=1}^kc_{\pi(i)}(t),
\end{equation}
for $k=1,\dots,N-1$, and $g(N)=\infty$. Define $k^*$ by the interval $(g(k-1),g(k)]$ to which the budget parameter $\Lambda(t)$ belongs.  Since $g(k)$ is monotonic, the mapping from $\Lambda(t)$ to $k^*$ is one-to-one. Then 
$\mathcal{U}(t) = \{\pi(1)\cdots,\pi(k^*)\}$ and the solution to (\ref{eq:greedy-problem}) is
\begin{equation}
\label{eq:greedy-allocation}
\lambda_{i}^{ga}(t)=
\left(\Lambda(t)+\sum\limits_{j\in\mathcal{U}(t)}
c_{j}(t)\right)\dfrac{\sqrt{z_i(t)}}{\sum_{j\in\mathcal{U}(t)}
\sqrt{z_{j}(t)}}-c_{i}(t),
\end{equation}
when $i \in \mathcal{U}(t)$,
and zero else.
}  

The GA policy including resource allocation, measurement, and posterior update steps is summarized in Algorithm \ref{alg:global-adaptive} with inputs $\Lambda$ and prior belief state $\vxi(0)$, where $\vxi(t)$ is defined as 
\begin{equation}
{\bm \xi}(t) = \set{\pic{c}(t),\hatXic{c}(t),\sictsqd{c}{t}}_{i=1,\dots,N,c\in\calC}
\end{equation}

\begin{algorithm}[t]
\begin{algorithmic}
\ignore{
	\STATE \textbf{Inputs}:
	\STATE \quad Total budget, $\Lambda$
	\STATE \quad Prior distribution, ${\bm \xi}(0) = \set{\sigma_i^2(0),\pic{c}(0),\hatXic{c}(0)}_{i=1,\dots,N,c\in\calC}$
\ignore{	\STATE \qquad prior variance $\sigma_i(0)$, $i=1,\cdots,N$
	\STATE \qquad prior class probability $p^{(c)}(0)$, $c\in\calC$, $i=1,\cdots,N$
	\STATE \qquad prior signal mean $\hatXic{c}(0)$, $i=1,\cdots,N$
}
	\STATE \textbf{Outputs}: 
	\STATE \quad Allocations $\vlam^{ga}(t)=\set{\lambda_i(t)}_{i=1,\dots,N}$ for $t=0,1,\dots,T-1$.
\STATE \quad Measurements $\vy(t)=\set{y_i(t)}_{i=1,\dots,N}$ for $t=1,\dots,T$.
\STATE \quad Posterior distribution, ${\bm \xi}(T)$.
\newline
}
	\STATE 0. Set $\Lambda(t)=\Lambda/T$ for $t=0,1,\dots,T-1$.
	\FOR{$t=1,2,\dots,T$}
	\STATE 1. Calculate $\lambda_i^{ga}(t-1)$ through convex optimization \eqref{eq:greedy-problem}.
	\STATE 2. Take measurements $y_i(t)$ as in (\ref{eq:yt}).
	\STATE 3. Update posterior dist. $\vxi(t)$ with (\ref{eq:var-update}),(\ref{eq:mean-update}),(\ref{eq:prob-update}).
	\ENDFOR
\end{algorithmic}
\caption{$\set{\vlam^{ga}(t-1),\vxi(t)}_{t=1}^{T}=$GlobalAdapt$(\Lambda,\vxi(0);T)$}
\label{alg:global-adaptive}
\end{algorithm}

\begin{algorithm}[t]
\begin{algorithmic}
	\STATE 0. Set $\Lambda(t)=\Lambda/T$ for $t=0,1,\dots,T-1$.
	\FOR{$t=1,2,\dots,T$}
	\STATE 1. Set $\hat{u}_i(t-1)=0$ for $i=1,2,\dots,N$.
	\STATE 2. Calculate $\Delta_i(t-1)$ for $i=1,\dots,N$ using 
	(\ref{equ:LA-delta-cost}).
	\FOR{$m=1,2,\dots,M$}
	\STATE 3. Choose $i^* = \argmax_i \Delta_i(t-1)$.
	\STATE 4. Set $\hat{u}_{i^*}(t-1) \leftarrow \hat{u}_{i^*}(t-1)+1$.
	\STATE 5. Re-calculate $\Delta_{i^*}(t-1)$ using 
	(\ref{equ:LA-delta-cost}).
	\ENDFOR
	\STATE 6. Calculate $\lambda_i^{la}(t-1;\vu(t-1))$ with (\ref{eq:la-allocation}).
	\STATE 7. Take measurements $y_i(t)$ as in (\ref{eq:yt}).
	\STATE 8. Update posterior dist. $\vxi(t)$ with (\ref{eq:var-update}), (\ref{eq:mean-update}), (\ref{eq:prob-update}).
\ENDFOR
\end{algorithmic}
\caption{$\set{\vlam^{la}(t-1),\vxi(t)}_{t=1}^{T}=$LocalAdapt$(\Lambda,\vxi(0);T)$}
\label{alg:local-adaptive}
\end{algorithm}
       
\subsection{Local adaptive (LA) policy}
In some cases it may be impossible to deploy a sensor with the agility to assign different resources to every location in the scene at every stage $t$.  Instead we consider the situation where there are $M$ local sensors (e.g. UAVs) that can explore a subset of the locations 
with a fixed resource amount per sensor. 
At each stage $t=0,1,\dots,T-1$, define $u_i(t)\in \set{0,1,\dots,M}$ as the number of sensors allocated to location $i$ for $i\in{\cal{X}}$, $\sum_{i=1}^N u_i(t) = M$, and  ${\bm u}(t)=\set{u_i(t)}_{i=1}^{N}$. Define the local adaptive allocation given $\vu(t)$ as 
\begin{equation}
\label{eq:la-allocation}
\lambda_i^{la}(t;\vu(t))= u_i(t)\left(\frac{\Lambda(t)}{M}\right)
\end{equation}
Then we seek the allocation that minimizes \eqref{eq:myopic-cost}: 
\begin{align}
{\hat{\bm u}}(t) = \argmin_{\bm u(t)} ~& {K(t;\vlam^{la}(t;{\bm u}(t)))}
\label{eq:la-optim-prob}
\end{align}
for $t=0,1,2,\dots,T-1$. 

The solution is given by a series of $M$ steps, where at each step, we allocate a single sensor to the location that provides the greatest decrease in (\ref{eq:myopic-cost}).  First, set $\hat{u}_i(t)=0$ for $i=1,2,\dots,N$.  Then, define the decrease in cost by allocating a single additional sensor to location $i$ as
\begin{equation}\label{equ:LA-delta-cost}
\begin{split}
\Delta_i(t) = &\sum_{c=2}^{|\calC|} \frac{\pic{c}(t) h(c)}{\nu^{2}/\sictsqd{c}{t} 
+\hat{u}_i(t)\Lambda(t)/M}\\
&-\sum_{c=2}^{|\calC|} \frac{\pic{c}(t) h(c)}{\nu^{2}/\sictsqd{c}{t} 
+(\hat{u}_i(t)+1)\Lambda(t)/M}.
\end{split}
\end{equation}
In each of $M$ steps, we set $\hat{u}_{i^*}(t)\leftarrow \hat{u}_{i^*}(t)+1$, where $i^*$ is the index with the largest $\Delta_i(t)$.  Note that multiple sensors are allowed to visit the same location.

The greedy allocation above is optimal for \eqref{eq:la-optim-prob} because of the convexity of the myopic cost with respect to each $\lambda_{i}(t)$. As a consequence, the decreases $\Delta_{i}(t)$ at a particular location diminish as more sensors are assigned to it, and we can be sure that the assignment in each step yields the largest decrease globally.
\ignore{
Consider the following myopic solution, which chooses the . First, note that $K(t;\vlam^{la}(t;{\bm u}(t))) = \sum_{i=1}^N K_i^{la}(t;{\bm u}(t))$, where
\begin{equation}
K_i^{la}(t;{\bm u}(t)) = \begin{cases}
\frac{\zi(t)}{\nu^2/\sigma_i^2(t)}, &i \notin {{\bm u}}(t)\\
\frac{\zi(t)}{\nu^2/\sigma_i^2(t)+\Lambda(t)/M}, &i \in {{\bm u}}(t)
\end{cases}
\end{equation}
so that we can independently minimize each $K_i^{la}(t;{\bm u}(t))$ in the summation.  
}
\ignore{To minimize cost in (\ref{eq:la-optim-prob}), the myopic solution should allocate local sensor to locations $M$ locations with highest increment change. <-- We haven't defined the incremental change yet.  Why not keep it the way it was before?}
\ignore{
Define the incremental change of $K_i^{la}(t;{\bm u}(t))$ resulting from including $i\in {\bm u}(t)$ as
Let $\zeta$ be an ordering operator such that
\begin{equation}
\label{eq:la-sort-zeta}
\Delta_{\zeta(1)}(t) \geq \Delta_{\zeta(2)}(t) \cdots \Delta_{\zeta(N)}(t).
\end{equation}
Then the solution to (\ref{eq:la-optim-prob}) is
\begin{equation}
\label{eq:la-allocation-set}
{\bm u}^{la}(t) = \set{\zeta(1),\zeta(2),\dots,\zeta(M)},
\end{equation}
because this is the set of locations that will most reduce the myopic cost at stage $t$.
}

The LA policy is summarized in Algorithm \ref{alg:local-adaptive}.

\subsection{Global uniform/local adaptive (GU/LA) mixture policy}
With no prior information on the location of targets in the scene, the local policy is likely to perform poorly, because it may take a long time to locate targets if only a small number of positions can be queried in each stage. Thus, we consider a third sensing modality where a global sensor is able to gather low resolution information on target location by uniformly observing the scene.  Subsequently, local sensors can use this information to measure the likely locations with higher signal quality.  In this policy, the optimization is two-fold: (a) the percentage of resources used by the global exploration sensor is optimized; and (b), the locations of the local sensors are optimized.  When $\lvert\mathcal{U}(t)\rvert = M$ 
for all $t$, this optimization problem can be written as
\begin{align}
\vlam^{gula} = &\argmin\limits_{T_s,\set{\bm u(t)}_{t=0}^{T-1}} J_T(\vlam^{gula}(\set{\bm u(t)}_{t=0}^{T-1},T_s)) \\
\mathrm{where} \quad & \lambda_i^{gula}(t;\vu(t),T_s)= \begin{cases} 
			{\Lambda(t)}/{N}, & t\leq T_s\\
			u_{i}(t) {\Lambda(t)}/{M}, & t> T_s
		\end{cases} \notag
\end{align}

Given $T_s$, the optimization problem reduces to finding the sensor allocations 
$\vu(t)$ for $t=T_s+1,\dots,T-1$, which is done myopically using steps (1-5) from the Local Adaptive policy (Algorithm \ref{alg:local-adaptive}).  Optimization over $T_s$ is 
done offline (i.e., before any real measurements are taken) by approximating the expectation in $J_T(\vlam)$ through Monte Carlo samples. 
In general, this requires $\mathcal{O}(T)$ Monte Carlo samples to determine the optimal $T_s$ in a $T$-stage policy.  The GU/LA algorithm is summarized by Algorithms \ref{alg:global-uniform-local-adaptive} and \ref{alg:optimal-ts}.

\begin{algorithm}[t]
\begin{algorithmic}
	\STATE 0. Set $\Lambda(t)=\Lambda/T$ for $t=0,1,\dots,T-1$.
	\STATE 1. Find optimal $T_s$ using Algorithm \ref{alg:optimal-ts}
	\FOR{ $t=1,2,\dots,T$}
		\IF{$t\leq T_s$}
			\STATE 2-a. $\lambda_i(t-1)=\Lambda(t)/N$ for $i=1,2,\dots,N$
		\ELSE 
			\STATE 2-b. $\lambda_i(t-1)$ given by steps (1)-(6) in Alg. \ref{alg:local-adaptive}.
		\ENDIF
	\STATE 3. Take measurements $y_i(t)$ as in (\ref{eq:yt})
	\STATE 4. Update posterior dist. $\vxi(t)$ with (\ref{eq:var-update}), (\ref{eq:mean-update}), (\ref{eq:prob-update})
	\ENDFOR
\end{algorithmic}
\caption{$\set{\vlam^{gula}(t-1),\vxi(t)}_{t=1}^{T}=$GU/LA$(\Lambda,\vxi(0);T)$}
\label{alg:global-uniform-local-adaptive}
\end{algorithm}

\begin{algorithm}[t]
\begin{algorithmic}
	\STATE 0. Set $\Lambda(t)=\Lambda/T$ for $t=0,1,\dots,T-1$.
	\FOR{$T_s\in \set{1,\dots,T}$}
		\FOR{$j=1,2,\dots,N_{\mathrm{MC}}$}
			\STATE 1. Draw Monte Carlo sample of $\set{C_i,X_i}_{i=1}^{N}$ from $\vxi(0)$.
			\STATE 2. Set $\vxi^{(j)}(0) = \vxi(0)$.
			\FOR{ $t=1,2,\dots,T$}
				\IF{$t\leq T_s$}
					\STATE 3-a. $\lambda_i(t-1)=\Lambda(t)/N$ for $i=1,2,\dots,N$
				\ELSE 
					\STATE 3-b. $\lambda_i(t-1)$ given by steps (1)-(6) in Alg. \ref{alg:local-adaptive}.
				\ENDIF
				\STATE 4. Draw $\vy(t)$ from (\ref{eq:yt}) given $\vC$, $\vX$, and $\vlam(t-1)$
				\STATE 5. Update posterior dist. $\vxi^{(j)}(t)$ with (\ref{eq:var-update})-(\ref{eq:prob-update})
			\ENDFOR
			\STATE 6. Calculate cost for $(j)$-th sample, $Q(j,T_s) = K(T; \vlam(T)=\mathbf{0})$ 
			using (\ref{eq:myopic-cost}).
		\ENDFOR
		\STATE 7. Calculate $J_T(T_s) \approx (1/N_{\mathrm{MC}}) \sum_{j=1}^{N_{MC}} Q(j,T_s)$.
	\ENDFOR
	\STATE 8. Return $T_s = \argmin_{T'_{s}} J_T(T'_s)$.
\end{algorithmic}
\caption{$T_s=$findOptimalGULAPoint$(\Lambda,\vxi(0);T)$}
\label{alg:optimal-ts}
\end{algorithm}

\section{Simulation}\label{sec:sim}
This section presents a numerical study for comparing 
the performance of proposed policies --- 
global adaptive (GA), local adaptive (LA), and global uniform/local adaptive mixture (GU/LA) --- and a previously proposed policy ARAP \cite{Bashan08_TSP,Wei13_TSP}, which is designed to only detect targets and not classify them.  The global uniform (GU) and full oracle policies discussed 
in Section \ref{sec:performance-bounds} are used as benchmarks.

Simulation parameters are given in Table \ref{table:simulation-parameters} unless otherwise stated. In this scenario, there are very few targets in the scene (5\% on average), which leads to significant performance gaps between the GU policy and adaptive policies.  Moreover, the condition that $\mu_3<\mu_2$ is imposed to account for the fact that high-importance targets are generally harder to detect than low-importance targets.  We set $T=10$ for the GA policy and $T=30$ for the GU/LA and LA policies since the latter two require additional stages 
to effectively search the entire scene.  Note that once the SNR is fixed, the total budget $\Lambda$ is the same regardless of $T$. 
One could additionally optimize over the 
per-stage budget allocations, $\set{\Lambda(t)}_{t=0,1,\dots,T-1}$ as in \cite{Wei13_TSP}, though this is not done in the current paper. 

This section begins by identifying the regimes in which the GA policy nearly achieves the oracle policy performance as a function of the model parameters, as well as the regimes where there are few performance gains over the baseline GU policy.  The performance of the GU/LA policy is analyzed over the same parameters.  Additionally, this section explores the sensitivity of the GU/LA policy to (a) correct specification of the percentage of stages used for global sensing, $T_s/T$, and (b) the number of local sensors available.  Subsequently, we demonstrate the performance improvement by including the global sensor by analyzing the differences between the GU/LA and LA policies.  The section concludes by comparing the performance of the proposed policies across other performance metrics which are not directly optimized, including mean-squared error and misclassification probability.

\begin{figure}[t]
\centering
\subfloat[$G(\vlam^{u},\vlam^{o})$, $h(3) = 50^2$]{
\includegraphics[height=\hpsnrfigheight]{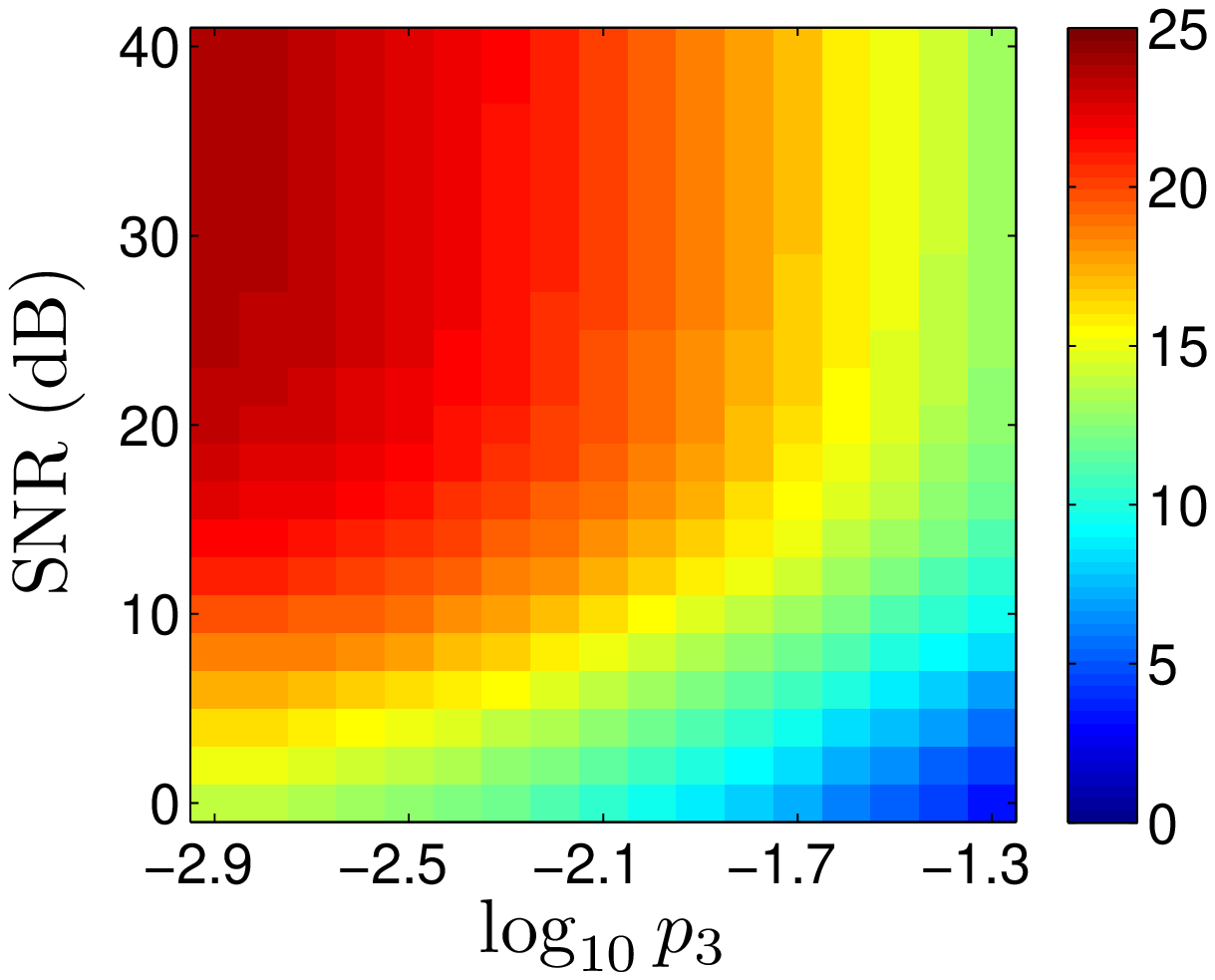}
\label{fig:gain-oracle-snr-p3}
}
\subfloat[$G(\vlam^{u},\vlam^{o})$, SNR $=20$ dB]{
\includegraphics[height=\hpsnrfigheight]{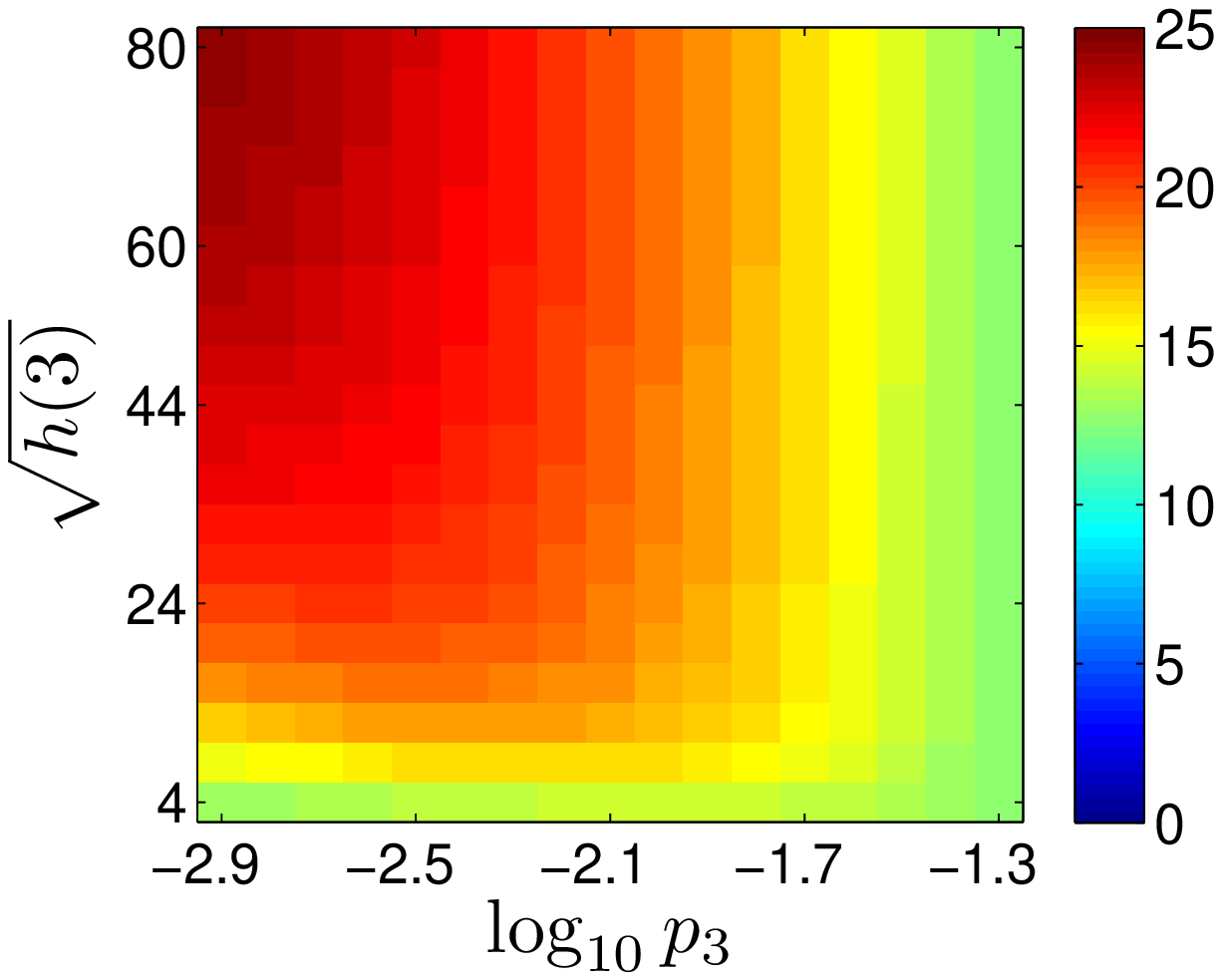}
\label{fig:gain-oracle-h3-p3}
}
\\
\subfloat[$G(\vlam^{u},\vlam^{ga})$, $h(3) = 50^2$]{
\includegraphics[height=\hpsnrfigheight]{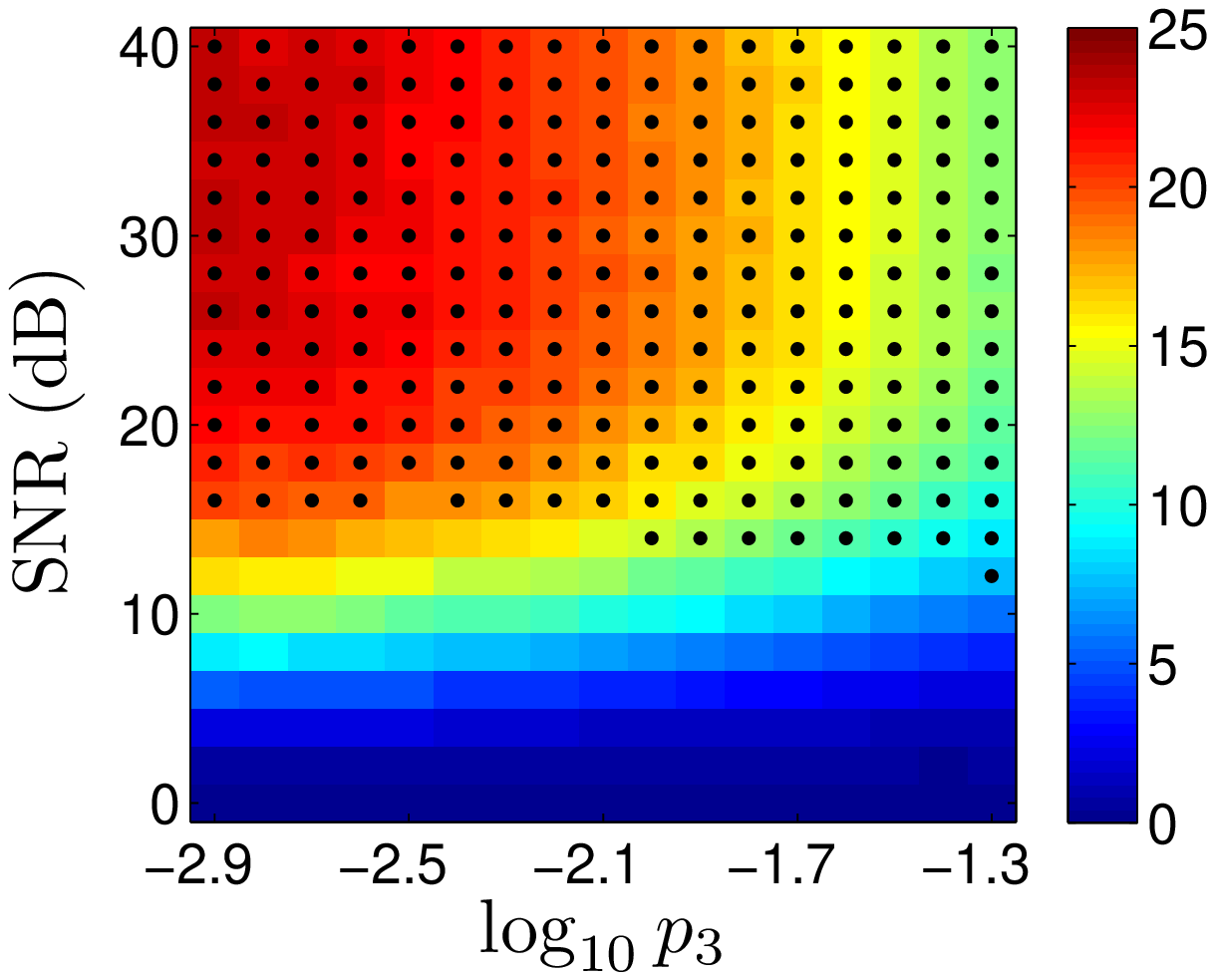}
\label{fig:gain-ga-snr-p3}
}
\subfloat[$G(\vlam^{u},\vlam^{ga})$, SNR $=20$ dB]{
\includegraphics[height=\hpsnrfigheight]{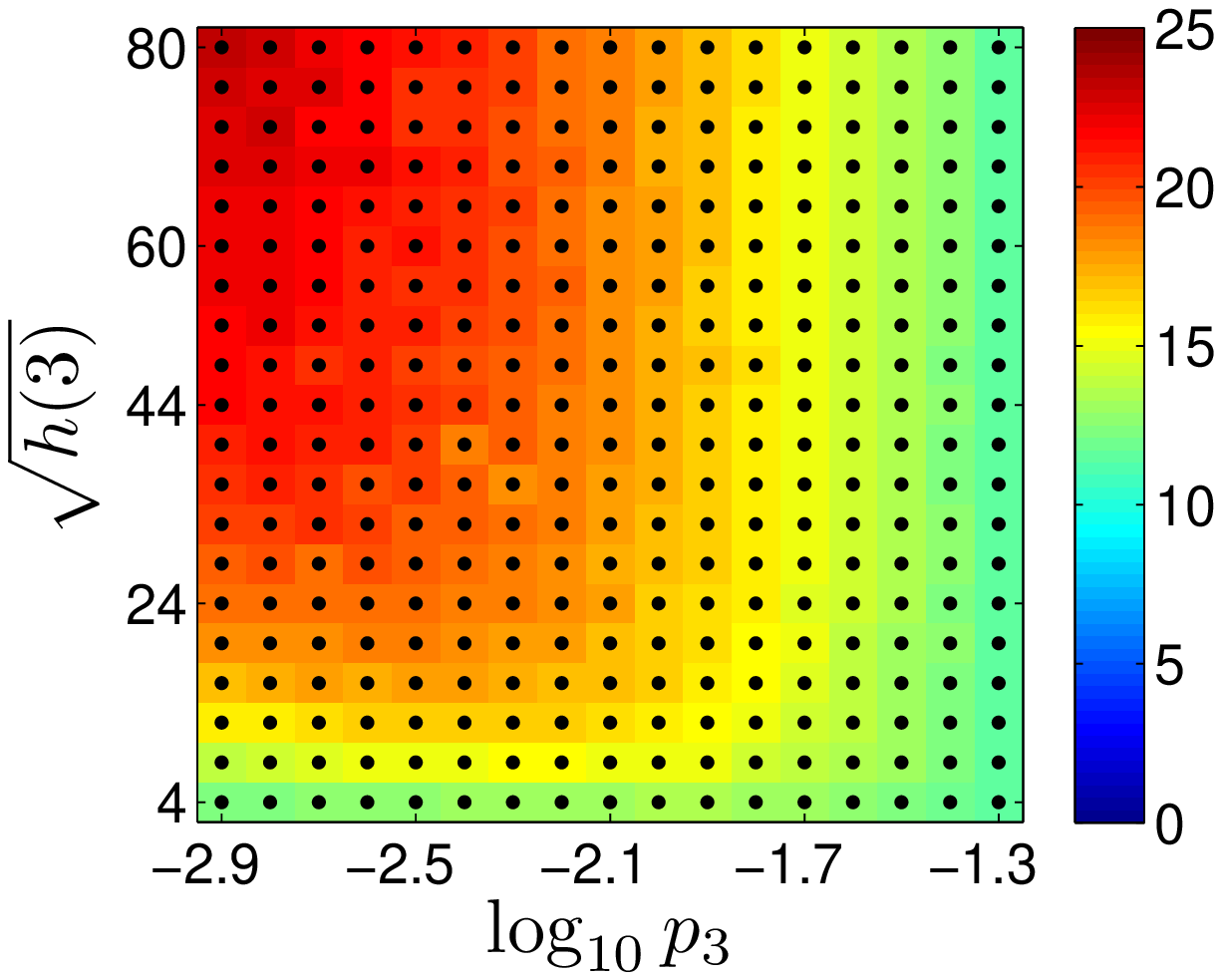}
\label{fig:gain-ga-h3-p3}
}
\\
\subfloat[$G(\vlam^{u},\vlam^{gula})$, $h(3) = 50^2$]{
\includegraphics[height=\hpsnrfigheight]{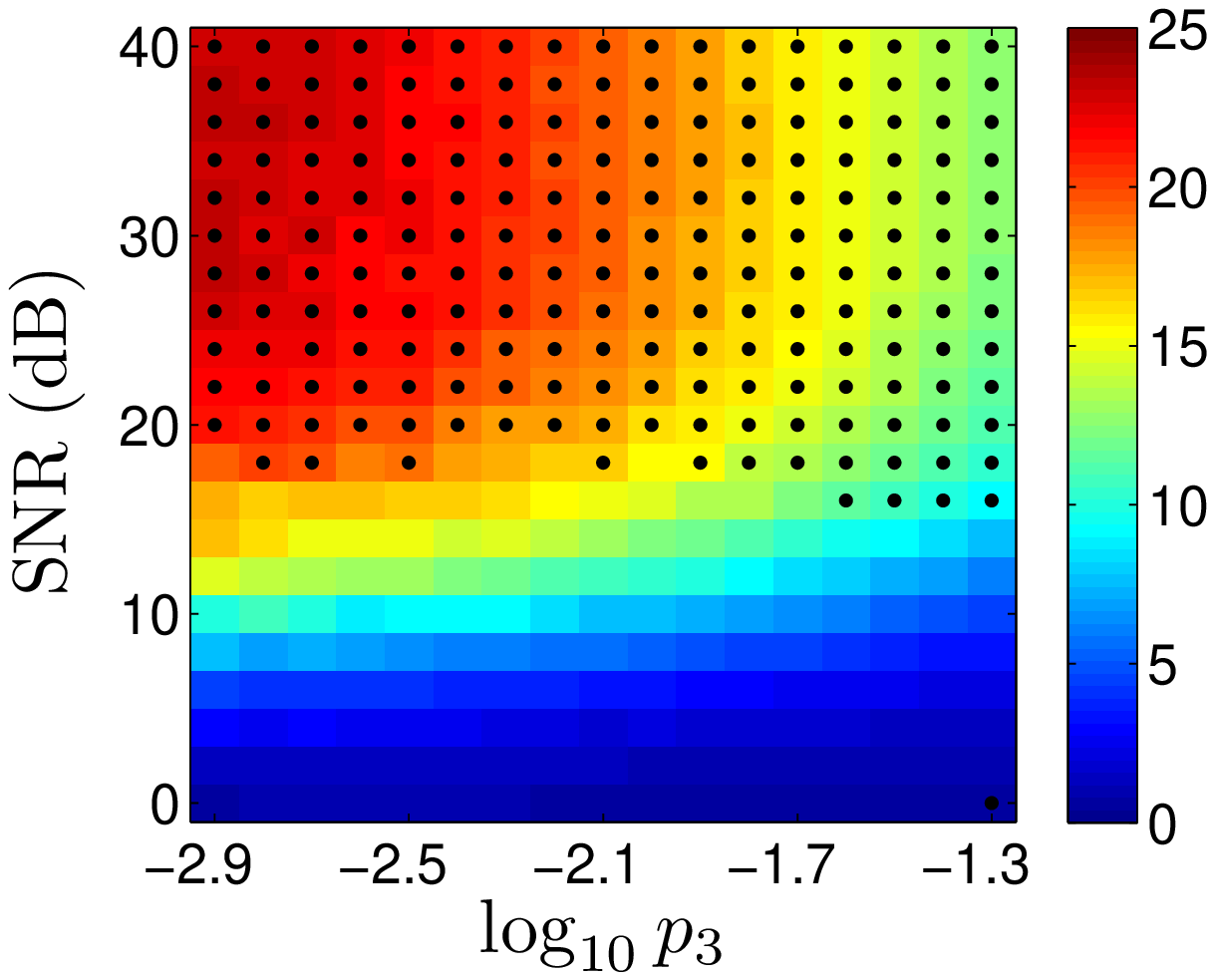}
\label{fig:gula-snr-p3}
}
\subfloat[$G(\vlam^{u},\vlam^{gula})$, 20 dB SNR]{
\includegraphics[height=\hpsnrfigheight]{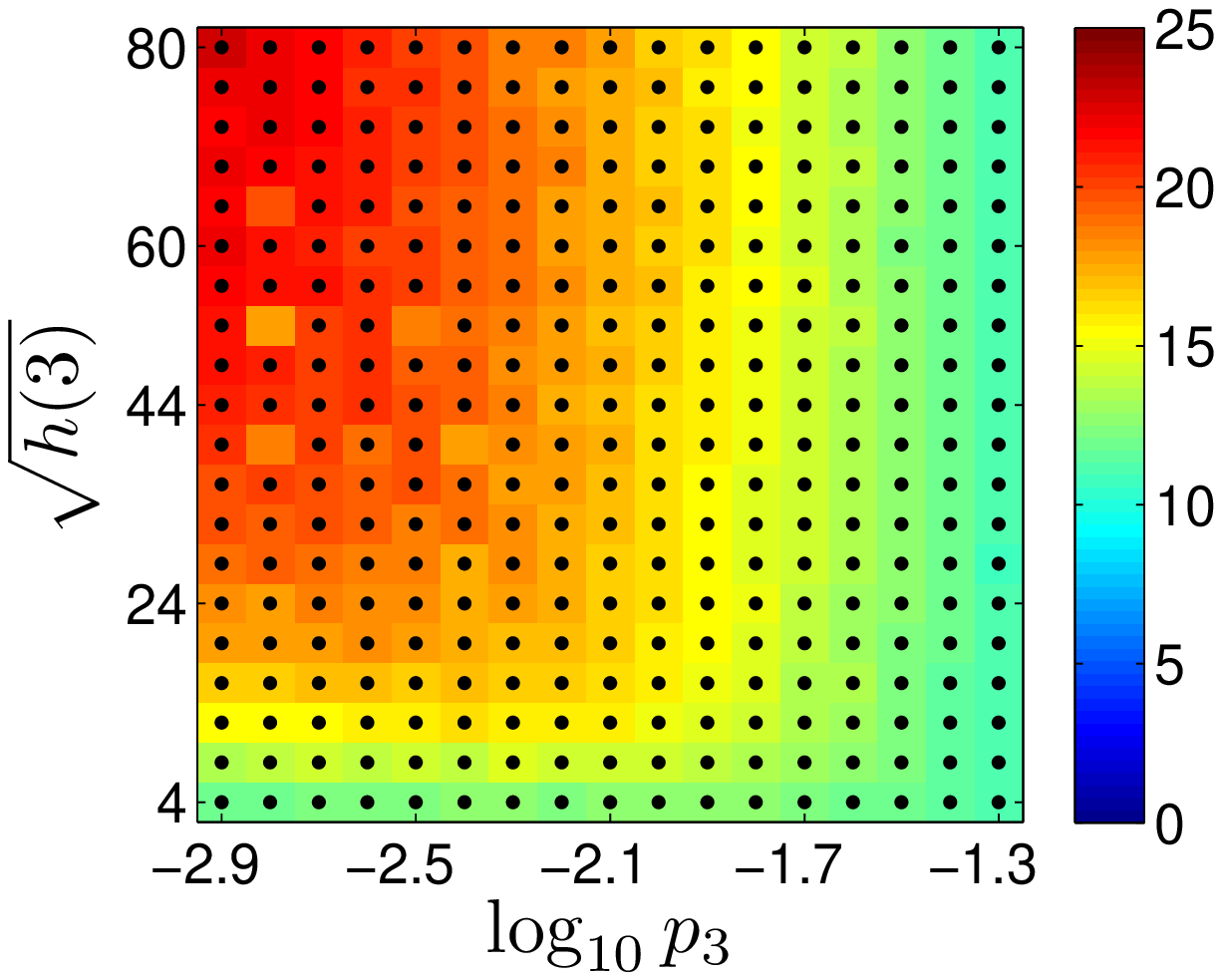}
\label{fig:gula-h3-p3}
}
\caption{Performance gain (in dB) of the GA, GU/LA and full-oracle policies with respect to the GU policy.  Gains are shown for varying SNR, class importance weight $h(3)$ and prior probability $p_3$, with remaining parameters given in Table \ref{table:simulation-parameters}. (a) and (b) show results for the oracle policy, (c) and (d) 
for the GA policy, and (e) and (f) 
for the GU/LA policy.  Black dots indicate that the GA and GU/LA policies are within 3 dB of the oracle policy. For SNR greater than 15 dB, the GA policy achieves within 3 dB of the oracle performance for all values of $h(3)$ and $p_3$.  Similarly, the GU/LA policy comes 
within 3 dB of the oracle when SNR is greater than 20 dB.
}\label{fig:GA-oracle-over-model-params}
\end{figure}

In Fig. \ref{fig:GA-oracle-over-model-params}, the benefits of adaptive sensing are explored by comparing the gain (\ref{eq:gain-definition}) of the GA, GU/LA, and oracle policies to the baseline GU policy.  Figs. \ref{fig:gain-oracle-snr-p3}, \ref{fig:gain-ga-snr-p3}, and \ref{fig:gula-snr-p3} show the gains when varying SNR and the high-value target class probability $p_3$ and fixing $h(3)=2500$.  Meanwhile, Figs. \ref{fig:gain-oracle-h3-p3}, \ref{fig:gain-ga-h3-p3}, and \ref{fig:gula-h3-p3} show the gains when varying $p_3$ and the high-value target mission importance $h(3)$ and fixing SNR to 20 dB.  Black dots indicate that the gains of the GA and GU/LA policies are within 3 dB of the oracle gain.  The GA and GU/LA policies have improved gains when either the SNR increases, the mission importance increases, or the number of high-value targets decreases.  At low SNR levels (about 0 dB), the GA and GU/LA policies have few gains over the GU policy.  However, for sufficiently high SNR levels ($\geq 15$ dB for the GA policy, and $\geq 20$ dB for the GU/LA policy), the policies nearly achieve the performance of the oracle policy over all other model parameters. Note that the GA and GU/LA policies always perform better than the GU policy (i.e., gains are always positive).

\begin{figure}[t]
\centering
\subfloat[$G(\vlam^{u},\vlam^{gula})$ vs. $T_s$ and SNR]{
\includegraphics[height=\hpsnrfigheight]{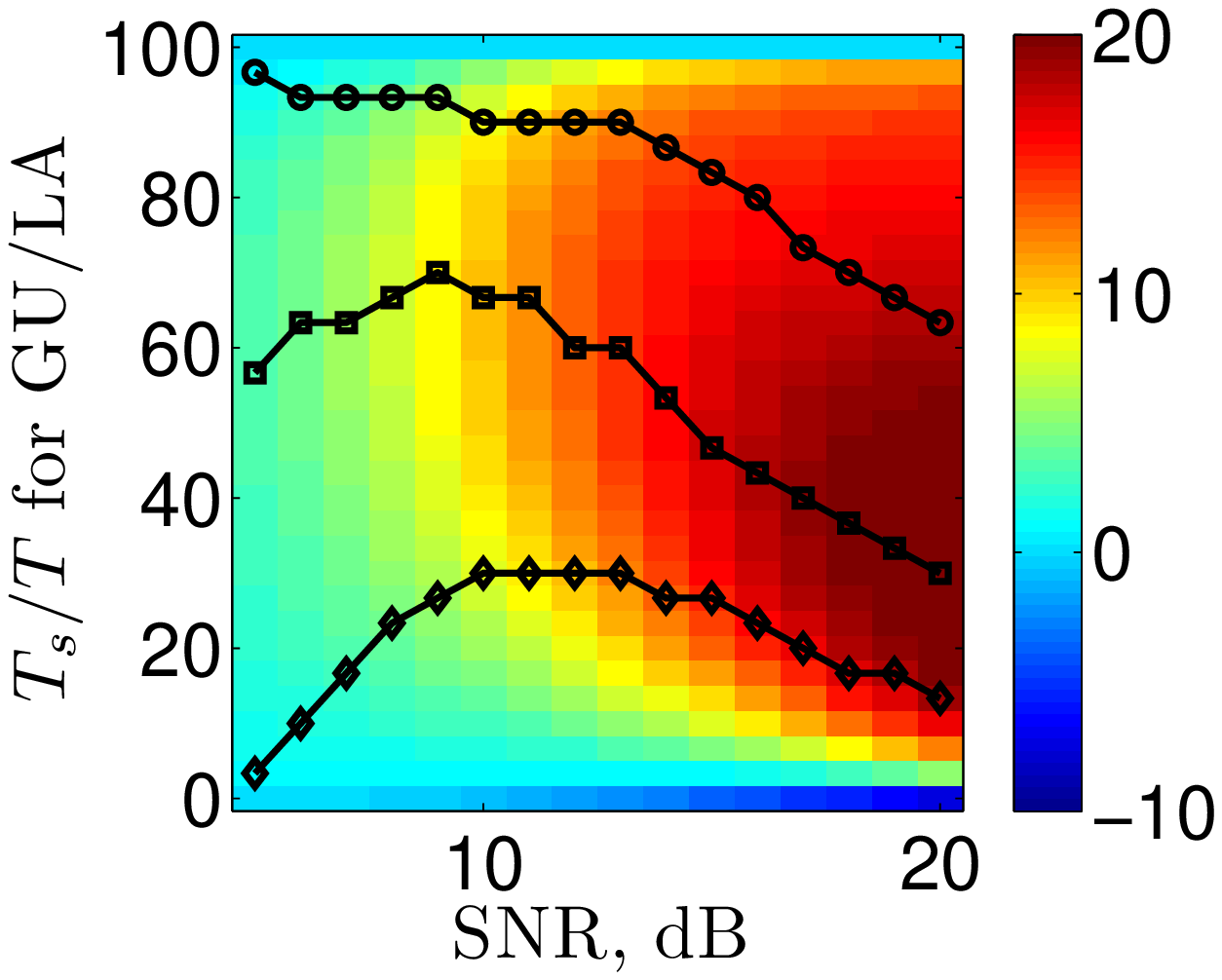}
\label{fig:gula-ts-snr}
}
\subfloat[Optimal $T_s$ vs.~SNR and $p_{3}$]{ 
\includegraphics[height=\hpsnrfigheight]{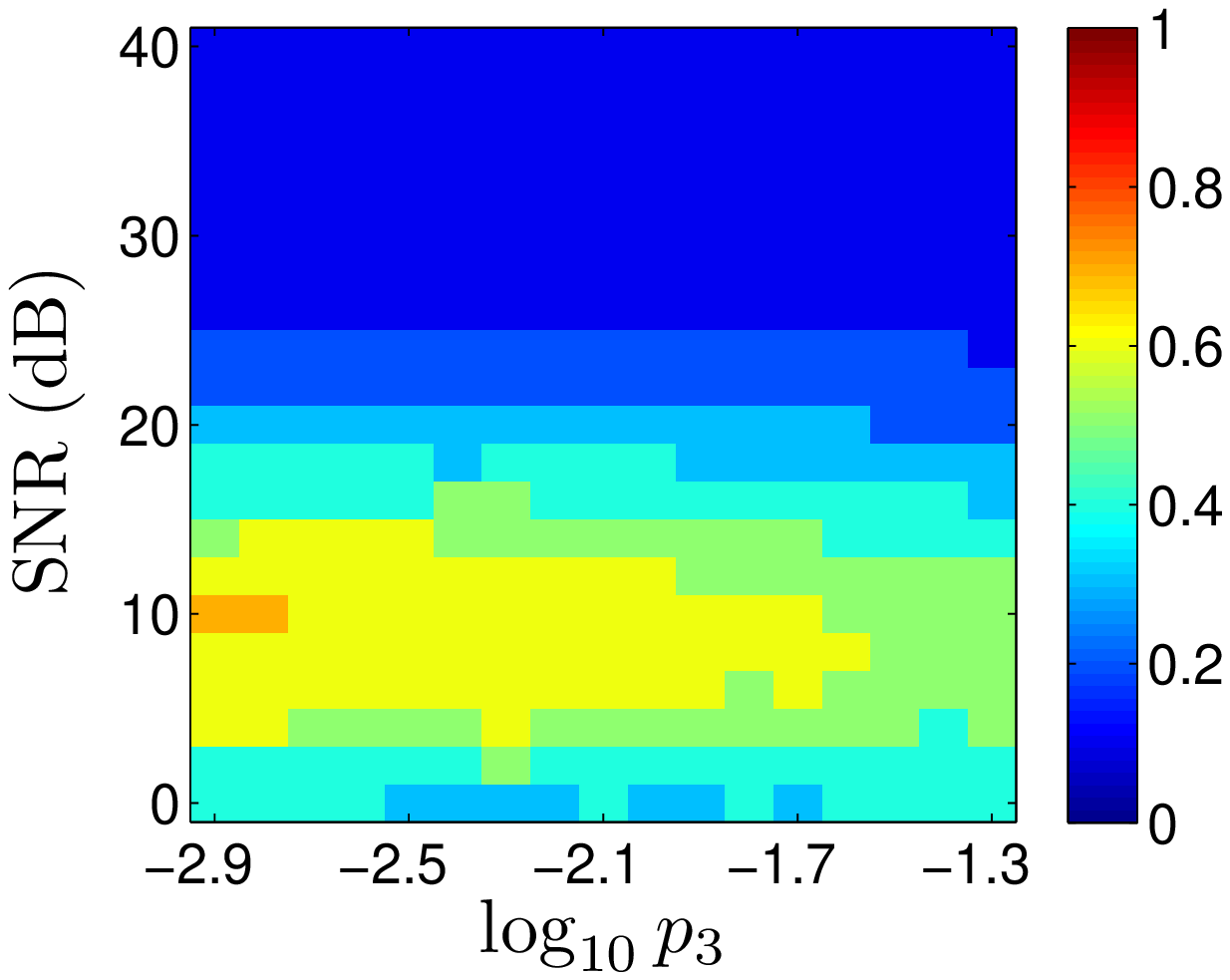}
\label{fig:gula-ts-snr-p3}
}
\caption{
Sensitivity of GU/LA to $T_s/T$. In (a), the gain (in dB) of the GU/LA policy is shown as a function of SNR and the percentage of resources used for the global sensor, $T_s/T$.  
For each SNR value, the optimal percentage as found by Algorithm \ref{alg:optimal-ts} is given by black squares. Circles and diamonds indicate the maximum/minimum $T_s/T$, respectively, where the gain is within 3 dB of the maximum. 
Note that there is significant degradation when $T_s = 0$ (i.e. the LA policy) or $T_s=T$ (i.e., the GU policy).  (b) shows the optimal $T_s/T$ for the GU/LA policy while varying both SNR and $p_3$.
}
\label{fig:gula-over-model-parameters}
\end{figure}

Figs. \ref{fig:gula-ts-snr} and \ref{fig:gula-ts-snr-p3} explore the sensitivity of the GU/LA policy to the percentage of total resources used by the global sensor, $T_s/T$.  In Fig. \ref{fig:gula-ts-snr}, it is seen that there is significant performance degradation when this percentage is close to either extreme.  When $T_s=T$ (i.e., the GU policy), the lack of adaptivity leads to inefficient resource allocation.  Conversely, when $T_s=0$ (i.e., the LA policy), the local sensors have difficulty in finding the locations that contain valuable targets.  Nevertheless, for any fixed SNR, the gain of the GU/LA policy is relatively flat in a region around the optimal $T_s$ value. Circles indicate the maximum $T_s/T$ within 3 dB of the maximum gain, while diamonds indicate the minimum $T_s/T$ within 3 dB.  It is seen that for any SNR, there is a large region where the gain is within 3 dB, which suggests that the GU/LA policy may be robust to small errors in finding the optimal $T_s$.  Fig. \ref{fig:gula-ts-snr-p3} shows the optimal $T_s$ values while varying both SNR and $p_3$.  It is seen that most of the variation occurs when  SNR changes.  Indeed, this was also true when fixing SNR and varying $p_3$ and $h(3)$ (not shown), where it was found that the optimal $T_s$ was nearly constant over all values.  

\ignore{analyzes the performance of the global uniform/local adaptive mixture policy (GU/LA).  The gain over the GU policy $G(\vlam^u,\vlam^{gu/la})$ is given in Fig. \ref{fig:global-local-percentage} (a) and (b) with the same parameters as in Fig.\ref{fig:GA-oracle-over-model-params} (b) and (d). The direct comparison indicates that  $G(\vlam^u,\vlam^{gu/la})$ performs closely to the full adaptive GA policy, except a few cases when $p_3< 10^{-2.5}$ and $\sqrt{h(3)}>25$. Fig.\ref{fig:global-local-percentage} (c) illustrates the GU/LA cost as a function of SNR and the percentage of total resources used by the global uniform sensor $T_s/T$.  Notice $GU/LA$ degenerates to $LA$ when $T_s=0$ and $GU$ when $T_s=T$. There is significant degradation in performance when the percentage $T_s$ is either too high or too low. When $T_s$ is too low, the local sensors cannot pick the right locations because of very poor priors on the locations of targets. Conversely, when $T_s$ is too high, the local sensors do not have enough resources in locations containing targets and thus suffers in estimation accuracy. However, the cost stays relative flat across middle values of $T_s$. This means that as long as there is significant portion of resource committed to global sensing, the gain will be close to the optimal $T_s$ cost. (d) further shows the optimal $T_s/T$ with varying $p_3$ and SNR values, it stays at about $30\%$ when SNR$>$20dB. Therefore, when it is hard to compute the optimal $T_s$ with off-line Monti-Carlo experiments, fixing $T_s$ at $30\%$ would be a good approximate.}

\begin{figure}[t]
\centering
\subfloat[$G(\vlam^{u},\vlam^{la})$]{
\includegraphics[height=\hpsnrfigheight]{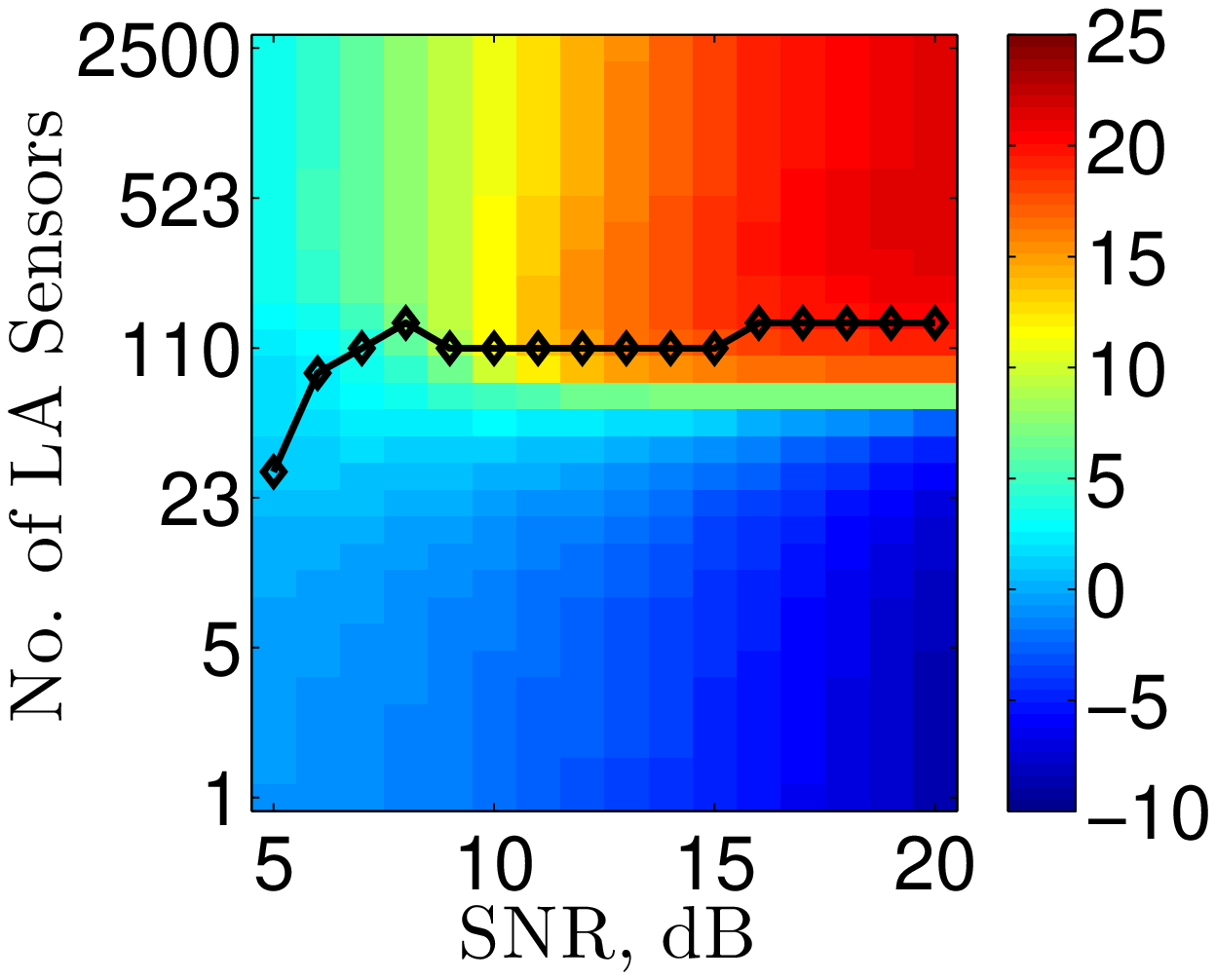} 
\label{fig:la-gain-snr-num-sensors}
}
\subfloat[$G(\vlam^{u},\vlam^{gula})$]{
\includegraphics[height=\hpsnrfigheight]{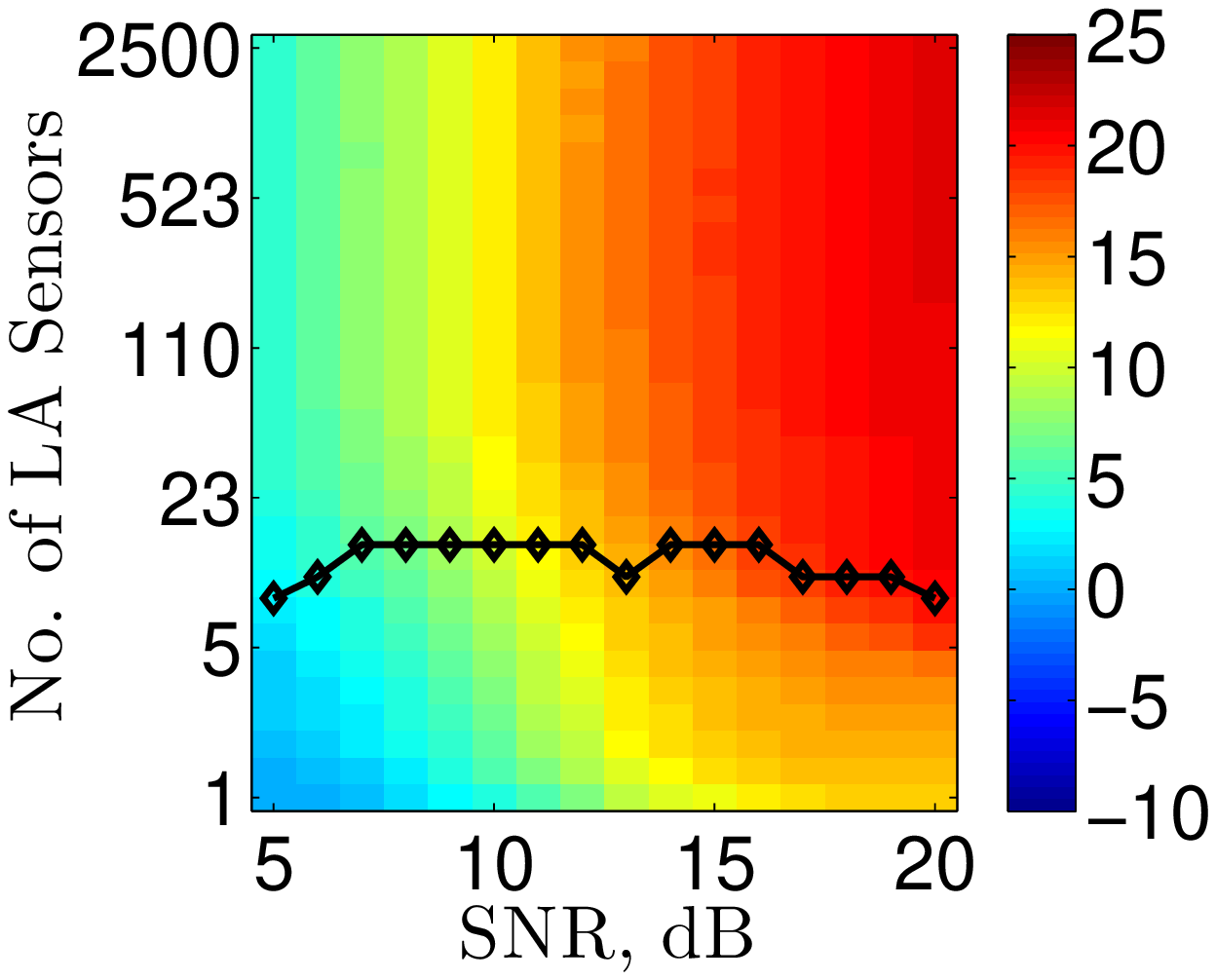}
\label{fig:gula-gain-snr-num-sensors}
}
\ifjournal
\else
\\
\subfloat[$G(\vlam^{u},\vlam)$ for proposed policies]{
\includegraphics[height=\hpsnrfigheight]{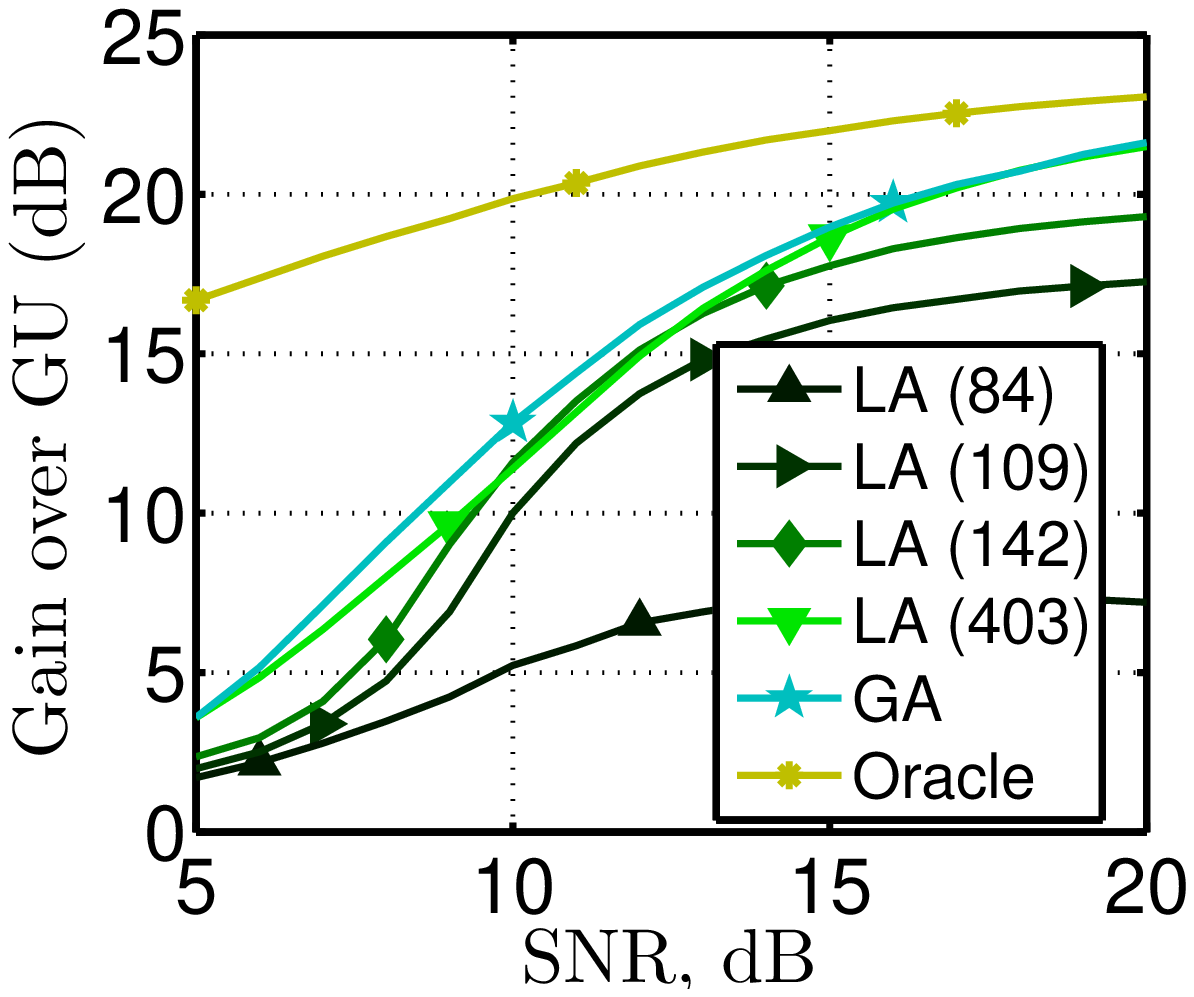}
\label{fig:gains-la-various-sensors-vs-snr}
}
\subfloat[Opt. no of sensors]{
\includegraphics[height=\hpsnrfigheight]{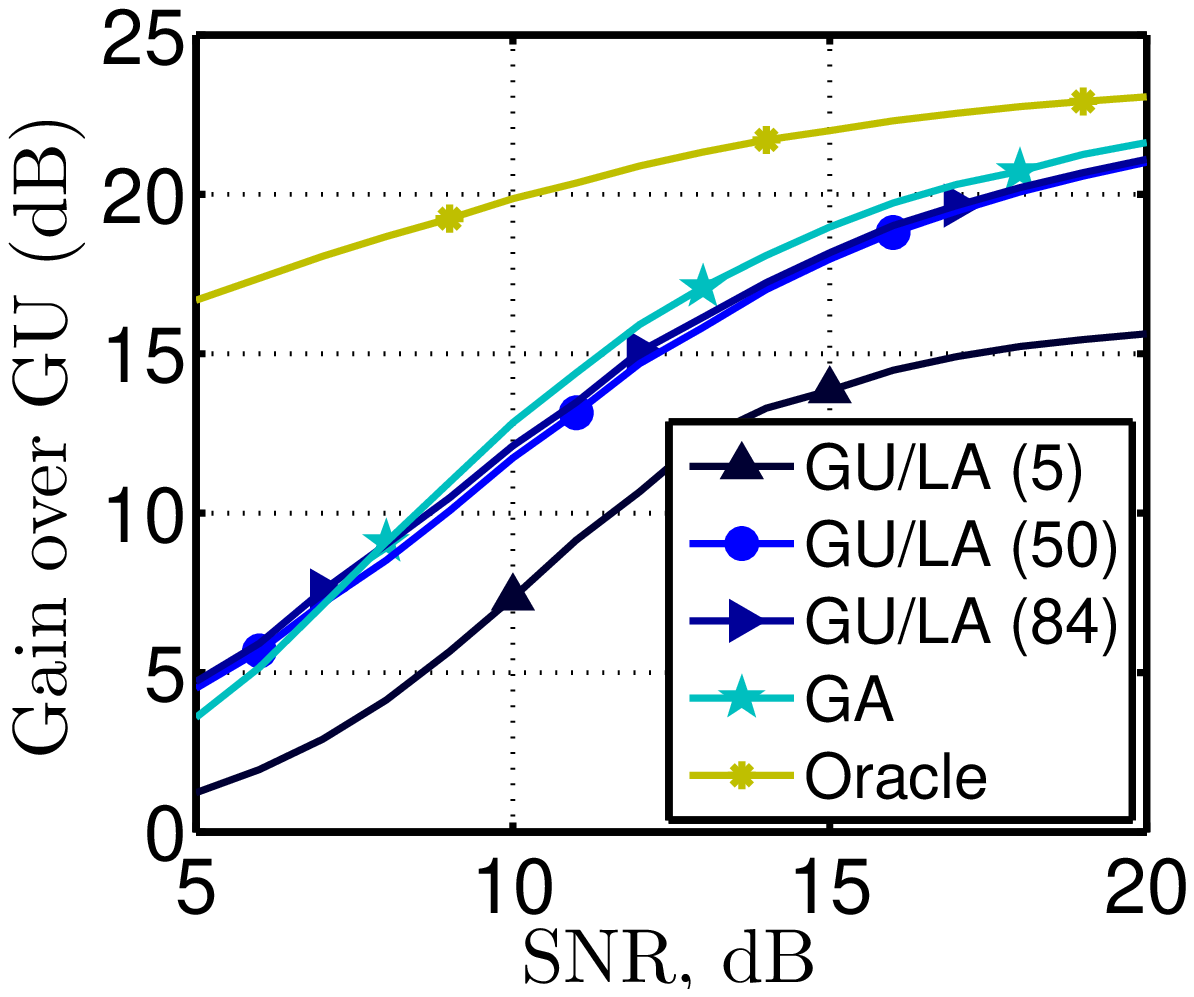}
\label{fig:gains-gula-various-sensors-vs-snr}
}
\fi
\caption{Performance benefit by including a global sensor. (a) and (b) show the gains in dB of the LA and GU/LA policies, respectively, over the GU policy as a function of SNR and the number of local sensors. 
For each SNR value, 
a black diamond indicates the smallest number of sensors needed to achieve gains within 3 dB of the maximum gain. 
Both policies have very good performance when the number of local sensors is large.  However, the LA policy requires at least 100 sensors to acheive within-3dB performance in most cases, and suffers large decreases in performance when this condition is not satisfied.  On the other hand, the GU/LA policy requires many fewer sensors (on the order of 10 sensors) to achieve within-3dB performance.
\ifjournal
\else
In (c) and (d), the performance of the LA and GU/LA policies are plotted in comparison to the GA and oracle policies.  In (c), at least 109 sensors are required to approximate the GA policy.  In (d), only 50 sensors are required, while using 84 sensors only provides marginal improvements.
\fi
}
\label{fig:la-gula-comparisons}
\end{figure}

Fig. \ref{fig:la-gula-comparisons} explores the benefit of including a global sensor by comparing the GU/LA and LA policies. Figs. \ref{fig:la-gain-snr-num-sensors} and \ref{fig:gula-gain-snr-num-sensors} show the gains of the LA and GU/LA policies (\ref{eq:gain-definition}) with respect to the GU policy, while varying SNR and the number of local sensors.  In this simulation, the optimal $T_s$ for the GU/LA policy is given by the values in Fig. \ref{fig:gula-ts-snr}. 
Black diamonds indicate the minimum number of sensors needed to 
achieve within 3 dB of the maximum gain at each SNR.  
Note that the maximum number of sensors is equal to $N=2500$ for both policies.  However, the LA policy requires at least 100 local sensors in almost all cases to attain performance within 3 dB of the maximum, while the GU/LA policy requires an order of magnitude fewer sensors.  Furthermore, a phase transition occurs for the LA policy when fewer than 100 sensors are used, wherein the LA policy actually performs worse than the GU policy.

\ifjournal
\else
Figs. \ref{fig:gains-la-various-sensors-vs-snr} and \ref{fig:gains-gula-various-sensors-vs-snr} show the performance of the LA and GU/LA policies for various number of sensors as compared to the GA and oracle policies.  Similar to above, the LA policy requires many more sensors than the GU/LA policy in order to achieve performance close to the GA policy.
\fi

\begin{figure}[t]
\centering
\subfloat[Class 2 targets]{
\includegraphics[height=\perfheight]{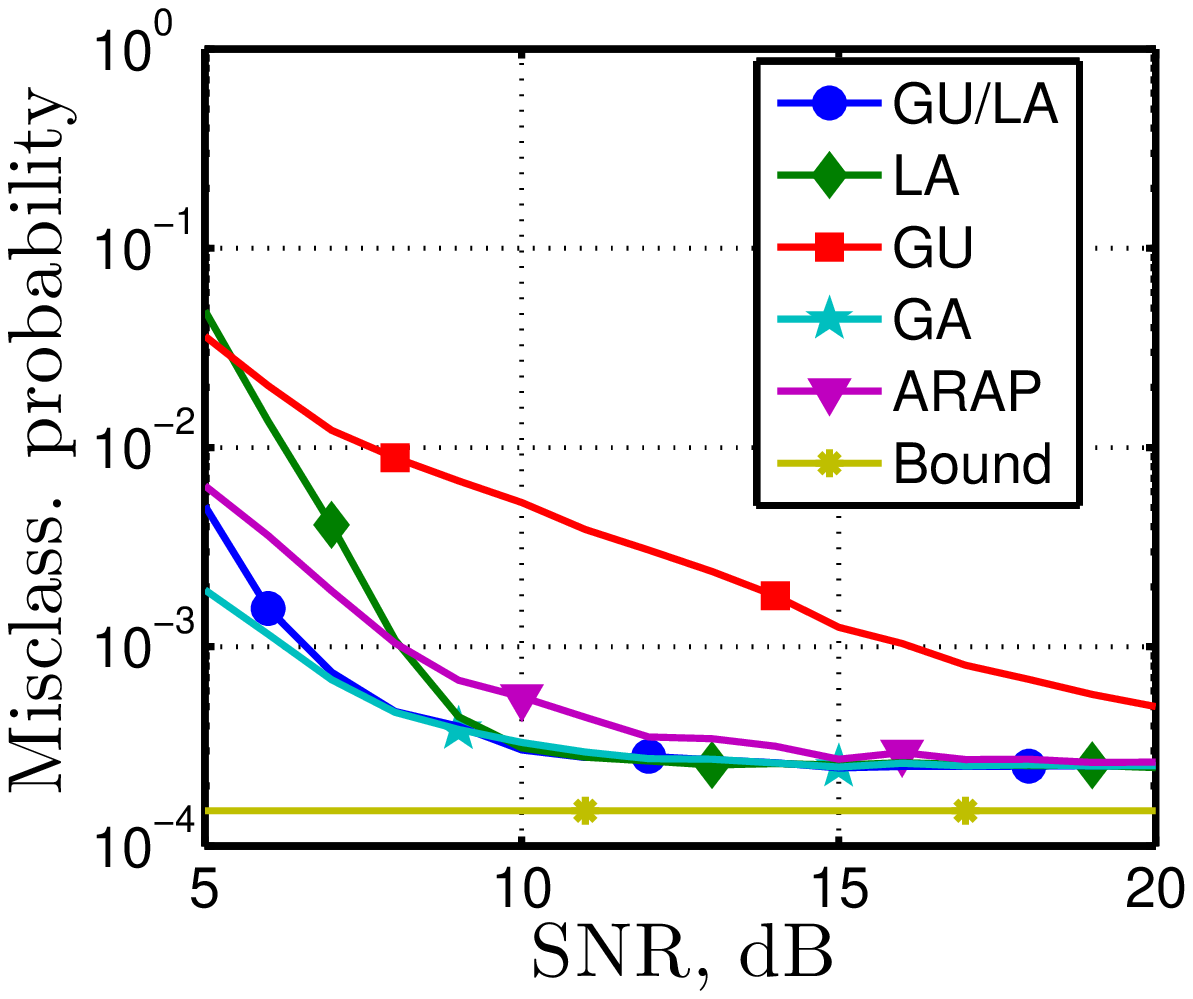}
}
\subfloat[Class 3 targets]{
\includegraphics[height=\perfheight]{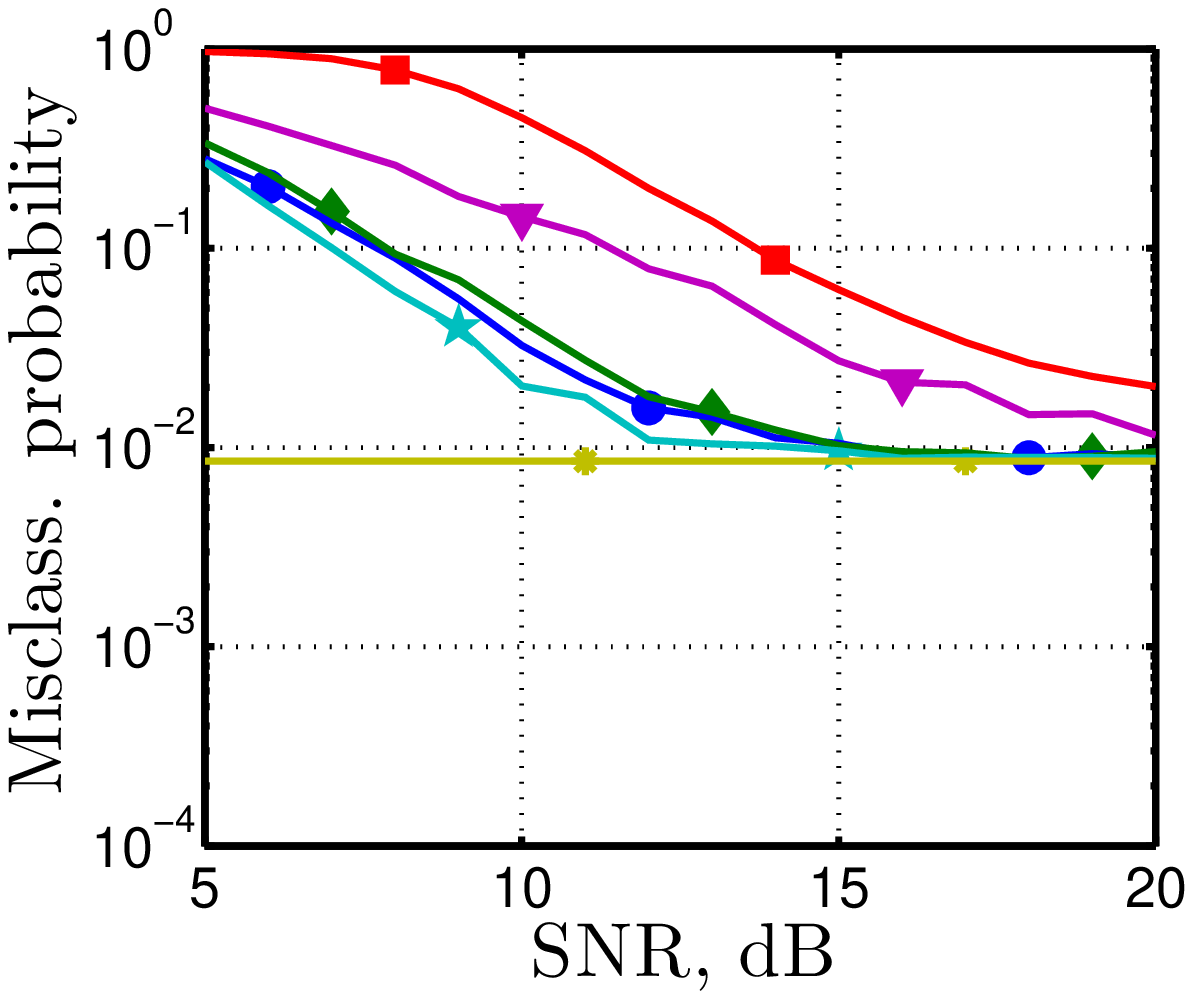}
}
\caption{This figure compares misclassification probability as function of SNR and policy.   All adaptive policies perform significantly better than GU and approach the performance of the oracle as SNR gets large.  Once again LA, GU/LA, and GA perform better than ARAP in reducing misclassification errors for the high-importance targets.}
\label{fig:missProb}
\end{figure}

The proposed policies approximately optimize an objective function that combines estimation and classification errors.  However, the policies also perform well in reducing each of these errors individually as demonstrated in the next figures.  Fig. \ref{fig:Var} plots the posterior variance (i.e., expected estimation error) within the low-value and high-value target classes.  Fig. \ref{fig:missProb} shows the misclassification probabilities within each class.  Fig. \ref{fig:missProb} also includes a bound on these probabilities in the asymptotic case where $\Lambda\rightarrow\infty$.  This bound, 
which is derived in 
\ifjournal
\else
Appendix \ref{app:asymptotic-missclassification-prob},
\fi
depends only on the prior probabilities, 
means, and variance of the nonzero-value target classes.

It is seen that the GA, GU/LA, and LA policies all perform similarly to the oracle policy in all of these metrics as SNR improves.  ARAP, which does not distinguish between high- and low-value targets, performs better for low-value targets and worse for high-value targets. Note that in these plots, the GU/LA policy used only 50 local sensors in comparison to the LA policy which used 400 local sensors.  The GA policy has slightly lower misclassification errors than the GU/LA and LA policies for low SNR values (SNR $<$ 10 dB).

\begin{figure}[t]
\centering
\subfloat[Class 2 targets]{
\includegraphics[height=\perfheight]{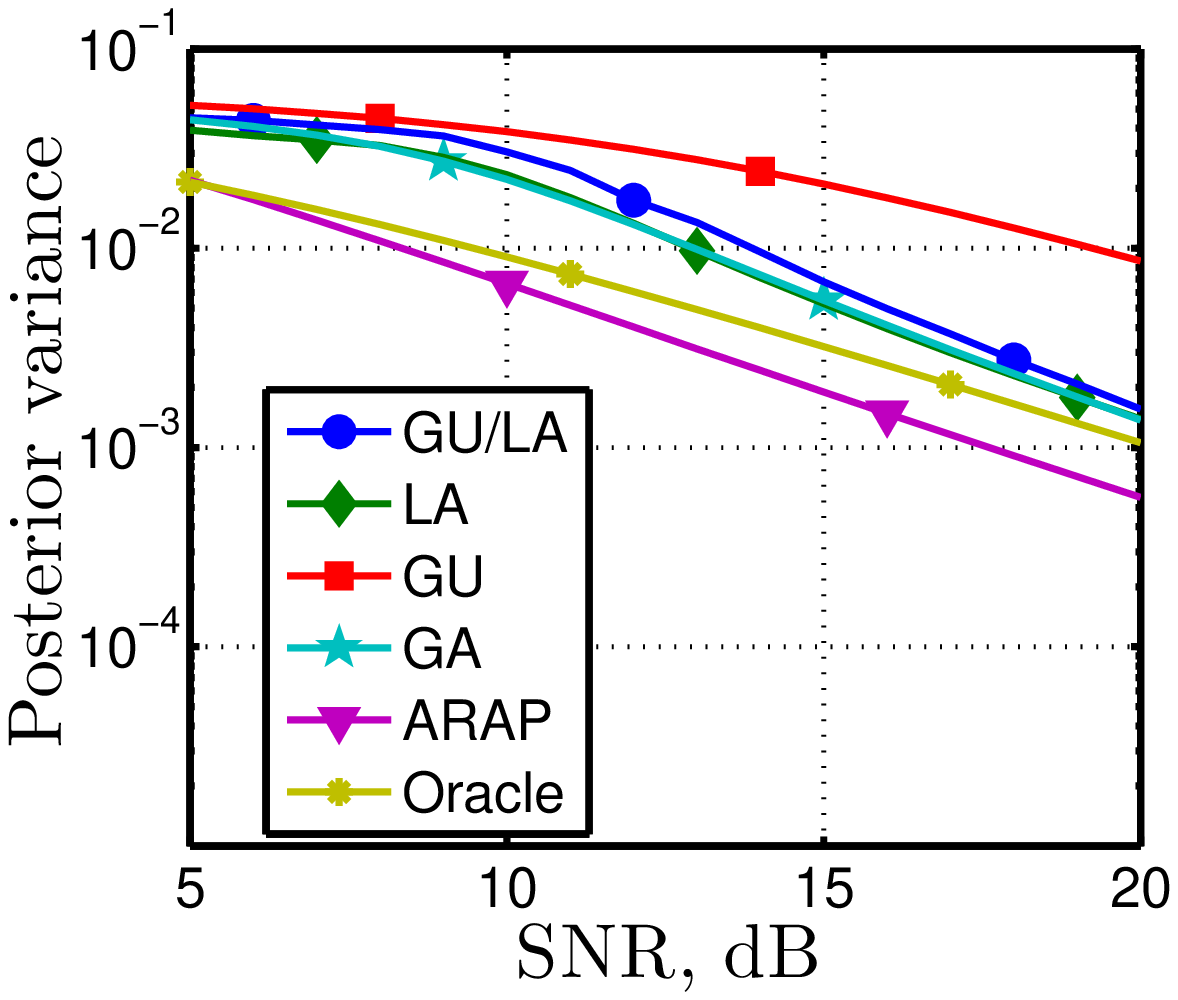}
}
\subfloat[Class 3 targets]{
\includegraphics[height=\perfheight]{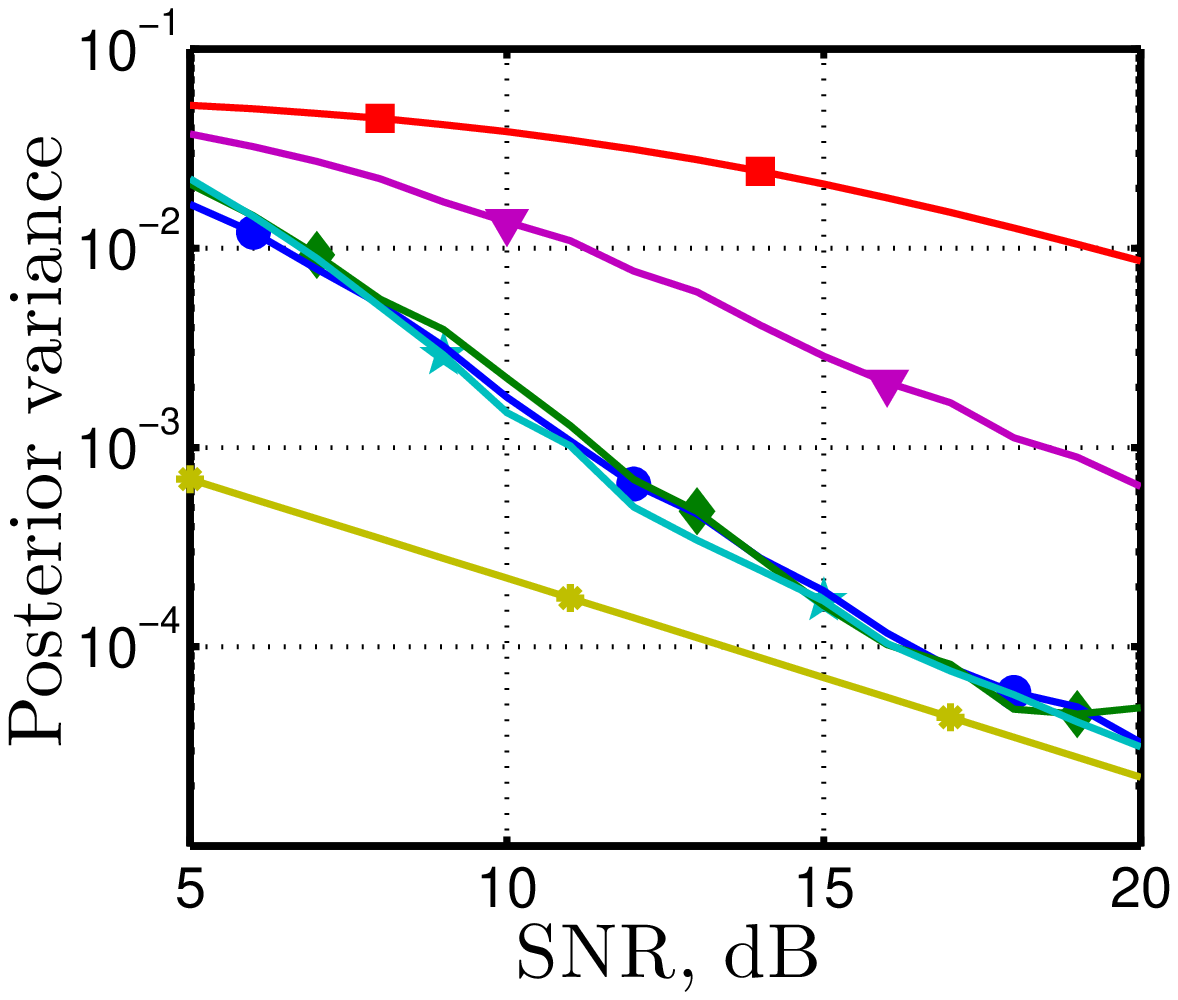}
}
\caption{Posterior variance of the considered policies for both low-importance (class 2) and high-importance (class 3) targets.  GU performs the worst as it does not adapt resource allocation.  The GA, LA, and GU/LA policies perform better than ARAP over high-importance targets, but worse on low-importance targets.  Note that the LA policy uses 400 local sensors, while the GU/LA policy only uses 50 sensors in addition to the GU sensor.}
\label{fig:Var}
\end{figure}

\begin{figure}[t]
\centering
\includegraphics[width=0.65\columnwidth]{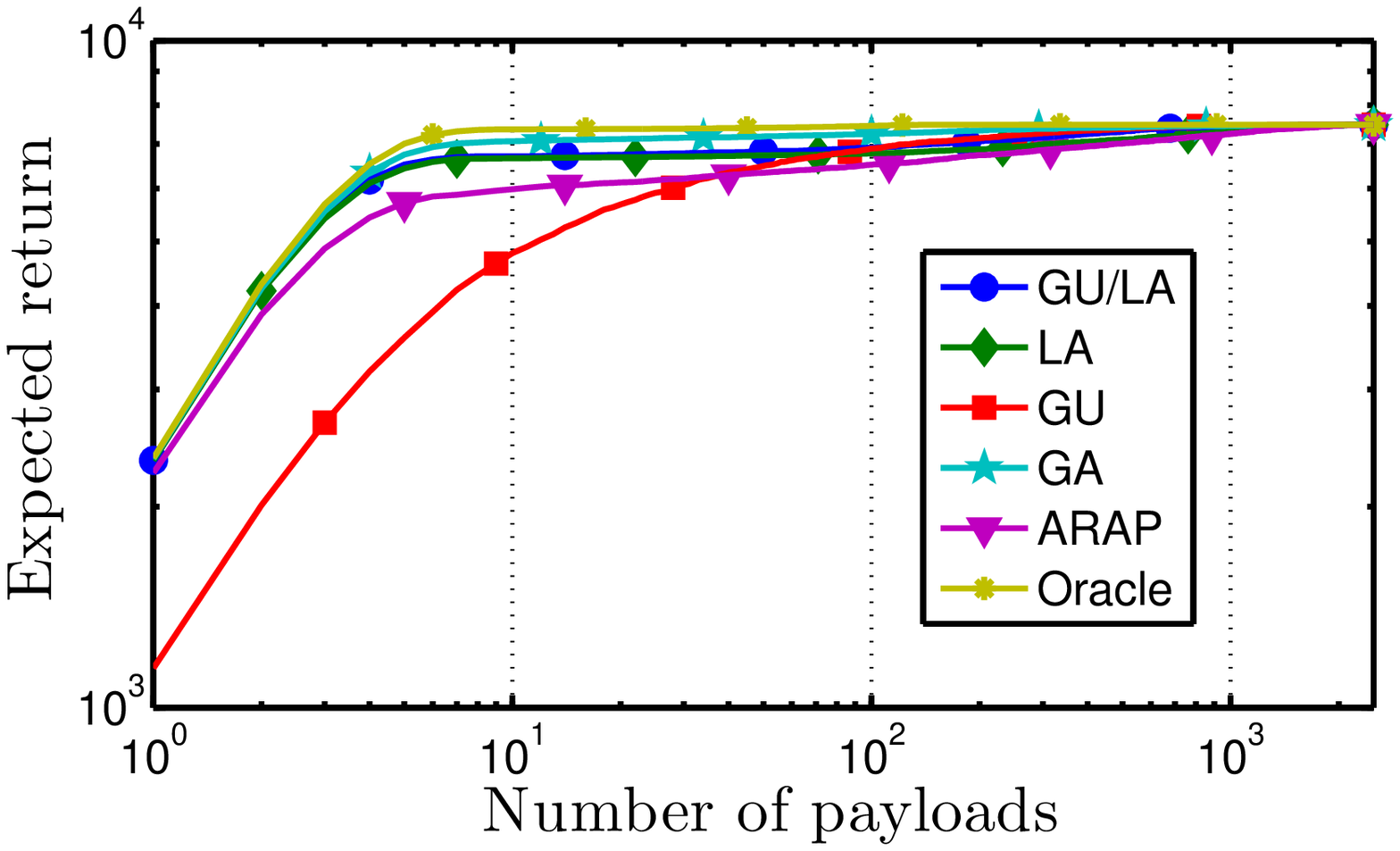}
\caption{The expected return given by (\ref{eq:payload-def}) as function of policy, when there are a limited number of payloads. The GA policy quickly approaches the oracle policy, followed by the GU/LA and LA 
policies. The ARAP 
and GU policies have slower convergence 
to the oracle policy return.}
\label{fig:payloads}
\end{figure}

In some cases, the ultimate goal of the mission is to use the information obtained by the sensors in order to achieve some objective.  For example, this may include dropping payloads for military purposes or for first-aid after natural disasters.  Moreover, it may be the case that there are only a limited number of payloads available.  While the goal of this paper is not to optimize the expected return on these payloads,the performance of the proposed policies as a function of the number of payloads is easily simulated. Define the expected return with $p$ payloads as
\begin{equation}
\label{eq:payload-def}
P_T(p) = \sum\limits_{i=1}^{p} z_{w(i)}(T),
\end{equation}
where $z_i(T)$ is the posterior mean reward at the final stage, as defined by \eqref{equ:Zi}, and $w$ is a sorting operator such that
\begin{equation}
z_{w(1)} \geq \cdots \geq z_{w(N)} 
\end{equation}
Thus payloads are assigned to the $p$ locations with highest expected importance. 
Fig. \ref{fig:payloads} compares $P_T(p)$ as a function of $p$ and policy with SNR = 8 dB.  The GA, GU/LA, and LA policies quickly converge to the oracle policy, followed by ARAP\forinitialsubmission{.  However,}{, but} the LA policy requires 5 times more local sensors  and has larger signal estimation error than the GU/LA policy.

\section{conclusion}\label{sec:conclusion}
The presented policies generalized previous formulations for adaptive target search \cite{Bashan08_TSP,Bashan11_TSP,Wei13_TSP} to include value of information when targets have varying mission importance. New policies are proposed that are able to simultaneously locate, classify and estimate a sparse number of targets embedded in a wide area. These policies approximately optimize an objective function that includes mission importance. We also considered multiple sensor models, which include global uniform sensors, global adaptive sensors, and local adaptive sensors. 

Theoretical upper bounds on the performance of adaptive policies indicate that there are many regimes in which incorporating mission importance leads to significant performance gains.  While the oracle policies are not achievable in practice, the proposed GA policy performs nearly as well as the oracle when SNR is sufficiently high.  In the scenario where a globally adaptive sensor is not available, the LA and GU/LA policies also nearly achieve oracle performance given enough local sensors.  However, the GU/LA policy generally performs better than the LA policy when there are only a few local sensors and it requires an order of magnitude fewer local sensors to achieve within 3 dB of the maximum performance.  Finally, all of the proposed policies are shown to perform well in not only minimizing the objective function, but also in reducing mean-squared estimation error, reducing misclassification probability, and increasing expected return from payloads.


Future research directions will (a) compare and contrast this formulation to one which directly optimizes the expected return rather than the proposed objective function; (b) include time-varying mission-value that can model stale information and dynamic targets; and (c) develop additional analytical results, such as performance bounds that depend directly on the proposed policies rather than the oracle policies, as well as convergence rates as a function of the scene size and number of sensing stages.

\appendices
\section{Proposition 2}\label{app:missclassification-prob}
\begin{proof}
Given the observation history $\vY(t-1)$, the posterior distribution $\pic{c}(t)$ will only depend on $y_i(t)$. $\hat C_i(t)$ (defined in (\ref{equ:estimated-C})) is a function of $\pic{c}(t)$, therefore also only depends on $y_i(t)$. For brevity, $\vY(t-1)$ is left out in the following deductions, but it should be noted that all quantities depend on $\vY(t-1)$.

Define $\text{y}^{(c)}_i(t)$ as the decision region that will lead to $\hat C_i(t)=c$:
\begin{align}\label{equ:y-interval}
\text{y}^{(c)}_i(t)=&\{y_i(t)\in\mathbb{R}|\hat C_i(t)=c\} 
\end{align}

The probability distribution of $y_i(t)$ is given by (\ref{equ:y_t-prob}) and we denote its density as $f(y_{i}(t) \mid C_{i})$. Given Assumption \ref{ass:equal-variance}, the variance of $y_i(t)$ is equal for all $c>1$. Denote the variance for $c=1$ and $c>1$ respectively as:
\begin{align}\label{equ:y(t)}
\Sigma^0_i(t)=&\nu^2/\lambda_i(t-1)\notag \\
\Sigma_i(t)=& \sigma^2_i(t-1)+\nu^2/\lambda_i(t-1)
\end{align}

Denote $m_c = \hatXic{c}(t-1)$, and rank the elements with sequence $\pi(c)$ such that:
\begin{align}\label{eqn:pi}
0 = m_{1} = m_{\pi(1)}<m_{\pi(2)}<m_{\pi(3)}<\cdots
\end{align}
As a consequence of the modification to the MAP classifier in \eqref{equ:estimated-C}, we have the following simplification.

\begin{lemma}
The decision regions $\text{y}^{(c)}_i(t)$ are simple intervals ordered in the same way as in \eqref{eqn:pi}: 
\begin{align}
(-\infty, a_i^{\pi(1)}(t)],~(a_i^{\pi(1)}(t), a_i^{\pi(2)}(t)],\cdots
\end{align}
and the posterior probability at a boundary $a_i^{\pi(c)}(t)$ is the same for the two adjacent classes:
\begin{align}\label{equ:equal-prob}
& \pic{\pi(c)}(t-1)f(a_i^{\pi(c)}|C_i=\pi(c)) \notag \\
=& \pic{\pi(c+1)}(t-1)f(a_i^{\pi(c)}|C_i=\pi(c+1))
\end{align}
\end{lemma}

\ifjournal
\else
	\begin{proof}
First compare the posterior probability when $c\neq 1$:
\begin{align}
& \arg\max_{d\in\mathcal{C},d\neq 1} \pic{d}(t) \notag \\
\intertext{plug in (\ref{eq:prob-update}) and ignore the constant in denominitor}
=& \arg\max_{d\in\mathcal{C},d\neq 1} \pic{d}(t-1) f(y_i(t)|C_i=c) \notag \\
\intertext{plug in (\ref{equ:y_t-prob}) and formula for normal distribution}
=& \arg\max_{d\in\mathcal{C},d\neq 1} \frac{\pic{d}(t-1)}{ \sqrt{\Sigma_i(t)}}
 \exp\left\{ -\frac{(y_i(t)-m_d)^2}{ 2\Sigma_i(t)} \right\} \notag \\
\intertext{take logrithm}
=& \arg\max_{d\in\mathcal{C},d\neq 1} \left(\log\frac{\pic{d}(t-1)}{ \sqrt{\Sigma_i(t)}}
  -\frac{(y_i(t)-m_d)^2}{ 2\Sigma_i(t)} \right) \notag \\
\intertext{change max to min, and ignore constant $\log\sqrt{\Sigma_i(t)}$}
=& \arg\min_{d\in\mathcal{C},d\neq 1} \left( \frac{(y_i(t)-m_d)^2}{2\Sigma_i(t)} -\log \pic{d}(t-1)  \right)
\end{align}
The functions within the minimization are quadratic in $y_i(t)$, and have the same shapes across classes $d$. Therefore, the minimum, i.e. $\text{y}^{(c)}_i(t)$, are simple intervals, and equation (\ref{equ:equal-prob}) holds.

Now consider the case $\hat C_i(t)=1$.
\begin{align}
& \hat C_i(t) = 1 \notag \\
\leftrightarrow~ & 1 = \arg\max_{d\in\mathcal{C}} \pic{d}(t) \notag \\
\leftrightarrow~ &  \pic{1}(t)  \geq \pic{d}(t)\quad \forall d\in\mathcal{C} \notag \\
\leftrightarrow~ & \pic{1}(t-1)f(y_i(t)|C_i=1)\geq \pic{d}(t-1)f(y_i(t)|C_i=d)  \notag \\
\intertext{plug in (\ref{equ:y_t-prob}) }
\leftrightarrow~ & \pic{1}(t-1) \frac{\exp\left\{ -\frac{y_i(t)^2}{ 2\Sigma^0_i(t)}\right\}}{\sqrt{2\pi\Sigma^0_i(t)}} \geq \pic{d}(t-1)\frac{\exp\left\{ -\frac{(y_i(t)-m_d)^2}{ 2\Sigma_i(t)}\right\}}{\sqrt{2\pi\Sigma_i(t)}} \notag \\
\leftrightarrow~ & \frac{\pic{1}(t-1)}{\pic{d}(t-1))}\geq
\sqrt{\frac{\Sigma^0_i(t)}{\Sigma_i(t)}} \exp\left\{ \frac{y_i(t)^2}{ 2\Sigma^0_i(t)} -\frac{(y_i(t)-m_d)^2}{2\Sigma_i(t)}\right\} \notag \\
\intertext{take logrithm}
\leftrightarrow~ & \frac{\sigma_i^2(t)}{2\Sigma^0_i(t)\Sigma_i(t)}y_i(t)^2 +\frac{m_d}{\Sigma_i(t)}y_i(t) -\frac{\mu_d^2}{2\Sigma_i(t)} \notag \\
& \leq \frac{1}{2}\log\frac{\Sigma_i(t)}{\Sigma^0_i(t)} + \log\frac{\pic{1}(t-1)}{\pic{d}(t-1)} \quad \forall d\in \mathcal{C} \notag
\end{align}

Solve above inequality and recall the definition in (\ref{equ:estimated-C}):
\begin{align}
\text{y}^{(1)}_i(t) = (-\infty, a^{\pi(1)}_i(t)]
\end{align}
where
\begin{align}
& a^{\pi(1)}_i(t) =
 \left\{ \begin{array}{ll}
0 & \text{if }  A\leq 0 \\
-\frac{\Sigma^0_i(t)m_{\pi(2)}}{\sigma^2_i(t)}+\sqrt{A} & \text{if $A\geq 0$ } 
\end{array} \right. \notag \\
& A = \frac{\Sigma_i(t)\Sigma_i^0(t)}{\sigma^2_i(t)}
\left(\frac{m^2_{\pi(2)}}{\sigma^2_i(t)} -2\log\frac{\pic{\pi(2)}(t-1)}{\pic{\pi(1)}(t-1)}+\log\frac{\Sigma_i(t)}{\Sigma_i^0(t)}\right) \notag
\end{align}

When $A<0$, the problem becomes misclassification problem over class 1 targets is 0.5, and the problem becomes trivial. Here we consider the case $A\geq 0$, in which case Equation (\ref{equ:equal-prob}) also holds.

To sum up two cases,  Equation (\ref{equ:equal-prob}) holds.
\end{proof}	
\fi
By definition (\ref{equ:missP-def}), 
\begin{align}
 q_i(t) =& \Pr\left(\hat{C}_i(t)\neq C_i|\vY(t-1)\right)\notag \\
	=& 1-\Pr\left(\hat{C}_i(t)= C_i|\vY(t-1)\right)\notag \\
	=& 1- \sum_{c\in\mathcal{C}}\pic{c}(t-1)\int_{\text{y}^{(c)}_i(t)} f(y_i(t)|C_i=c) dy_i(t) \notag \\
	=& 1 - \tilde{q}_i(t)
\end{align}
Where 
\begin{align}\label{equ:correct_class}
\tilde{q}_i(t)= \sum_{c\in\mathcal{C}}\pic{c}(t-1)\int_{\text{y}^{(c)}_i(t)} f(y_i(t)|C_i=c) dy_i(t)
\end{align}
is the probability of correct classification.

Denote $\Phi(\cdot)$ as standard normal cumulative probability function, $\Phi'(\cdot)$ as the standard normal probability density function. Further define normalized boundaries of $\text{y}^{\pi(c)}_i(t)$:
\begin{align}
& b^{\pi(1)}_r = \frac{a^{\pi(1)}_i(t)}{\sqrt{\Sigma^0_i(t)}}, \quad \text{and}\\
& b^{\pi(d)}_l= \frac{a^{\pi(d-1)}_i(t)-m_{\pi(d)}}{\sqrt{\Sigma_i(t)}}, \quad
b^{\pi(d)}_r = \frac{a^{\pi(d)}_i(t)-m_{\pi(d)}}{\sqrt{\Sigma_i(t)}} \notag 
\end{align}
for $ d > 1$. Then
 \begin{align}\label{eqn:qtilde}
&\tilde{q}_i(t)=\sum_{c\in\mathcal{C}}\pic{c}(t-1)\int_{\text{y}^{(c)}_i(t)} f(y_i(t)|C_i=c) dy_i(t) \notag \\
=& \pic{\pi(1)}(t-1) \Phi(b^{\pi(1)}_r) \notag \\
+& \sum_{k> 1}\pic{\pi(k)}(t-1)\left( \Phi(b^{\pi(k)}_r)-\Phi(b^{\pi(k)}_l)\right) \notag \\
\intertext{{\normalsize Splitting the summation and rearranging terms}}
=& \pic{\pi(1)}(t-1)\Phi(b^{\pi(1)}_r) - \pic{\pi(2)}(t-1)\Phi(b^{\pi(2)}_l) \notag \\
+ &\sum_{k> 1}\pic{\pi(k)}(t-1) \Phi(b_r^{\pi(k)})-\pic{\pi(k+1)}(t-1)\Phi(b^{\pi(k+1)}_l) \notag \\
=& M^{1,r}_i(t) + \sum_{k>1} M^k_i(t)
\end{align}

$b^{\pi(k)}_l$, $b^{\pi(k)}_r$ are functions of $a_i^{\pi(c)}(t)$, which are in turn functions of $\Sigma_i(t)$ and $\Sigma^0_i(t)$.  Since $\Sigma_i(t)=\Sigma^0_i(t)+\sigma_i^2(t)$, we can conclude that $M^{1,r}_i(t)$ and $M^k_i(t)$ depend on $\lambda_i(t)$ through $\Sigma^0_i(t)$.

{\small \begin{align}\label{eqn:dM1}
& \frac{d M^{1,r}_i(t)}{d \Sigma_i^0(t)} \notag \\
= &\pic{\pi(1)}(t-1) \Phi'(b^{\pi(1)}_r) \left(-\frac{b^{\pi(1)}_r}{2\Sigma^0_i(t)} +\frac{1}{\sqrt{\Sigma^0_i(t)}}\frac{d a^{\pi(1)}_i(t)}{d\Sigma^0_i(t)}\right) \notag \\
& -  \pic{\pi(2)}(t-1) \Phi'(b^{\pi(2)}_l) \left(-\frac{b^{\pi(2)}_l}{2\Sigma_i(t)}+ \frac{1}{\sqrt{\Sigma_i(t)}}\frac{d a^{\pi(1)}_i(t)}{d\Sigma^0_i(t)}\right) \notag \\
\intertext{\normalsize Plugging in  $\Phi'(b^{\pi(1)}_r)/\sqrt{\Sigma^0_i(t)}= f(a^{\pi(1)}_i(t)|\pi(1))$,  $ \Phi'(b^{\pi(2)}_l)/\sqrt{\Sigma_i(t)}= f(a^{\pi(1)}_i(t)|\pi(2))$ and (\ref{equ:equal-prob}),}
=& -\pic{\pi(1)}(t-1) f(a^{\pi(1)}_i(t)|\pi(1)) \frac{\sqrt{A}}{2\Sigma^0_i(t)\Sigma_i(t)} 
\leq 0
\end{align}}

Similarly, when $k > 1$,
{\small\begin{align}\label{eqn:dMk}
& \frac{dM^k_i(t)}{d\Sigma^0_i(t)} \notag \\
=& \pic{\pi(k)}(t-1)\Phi'(b^{\pi(k)}_r)
\left(-\frac{b^{\pi(k)}_r}{2\Sigma_i(t)}+\frac{1}{\sqrt{\Sigma_i(t)}}\frac{da^{\pi(k)}_i(t)}{d\Sigma^0_i(t)}\right) \notag \\
-& \pic{\pi(k+1)}(t-1)\Phi'(b^{\pi(k+1)}_l)
\left(-\frac{b^{\pi(k+1)}_l}{2\Sigma_i(t)}+\frac{1}{\sqrt{\Sigma_i(t)}}\frac{da^{\pi(k)}_i(t)}{d\Sigma^0_i(t)}\right) \notag\\
= & \pic{\pi(k)}(t-1)f(a^{\pi(k)}_i(t)|\pi(k))\frac{m_{\pi(k)}-m_{\pi(k+1)}}{2\Sigma_i(t)} 
\leq 0
\end{align}}
using \eqref{equ:equal-prob}. Combining \eqref{eqn:dM1} and \eqref{eqn:dMk} with \eqref{eqn:qtilde}, $
\frac{d\tilde{q}_i(t)}{d\Sigma^0_i(t)} \leq 0$.
Since $\frac{d\Sigma^0_i(t)}{d\lambda_i(t)}<0$ from (\ref{equ:y(t)}), we have
\begin{align}
\frac{d\tilde{q}_i(t)}{d\lambda_i(t)}\geq 0, \qquad \frac{dq_i(t)}{d\lambda_i(t)}\leq 0 \notag
\end{align}
\end{proof}

\section{Proof of Proposition \ref{prop:full-oracle-cost-upper-bound}}
\label{app:proof-prop-oracle-cost}

Given Assumptions \ref{ass:equal-variance} and \ref{ass:high-snr-assumption-oracle1}, the number of non-zero allocations for the oracle policy is $N-N_1$, as shown in
\ifjournal
\else
Appendix \ref{app:full-oracle-deriv}.
\fi
Under this condition, we can plug the oracle allocation policy \eqref{eq:oracle-allocation} into the cost function \eqref{eq:oracle-cost-function} to yield
\ifjournal
\else
\begin{equation}
\begin{split}
J_T(\vlam^o)&=\nu^2\expecgen{\vC}{\sum\limits_{i=1}^{N-N_1}\frac{h(C_{\pi(i)})}{c_0 + \dfrac{\left(\Lambda+(N-N_1)c_0\right)\sqrt{h(C_{\pi(i)})}}{\sum_{j=1}^{N}\sqrt{h(C_{\pi(j)})}}-c_0}},\\
&=\nu^2\expecgen{\vC}{\sum\limits_{i=1}^{N-N_1}{\sqrt{h(C_{\pi(i)})}}\left(\dfrac{\Lambda+(N-N_1)c_0}{\sum_{j=1}^{N-N_1}\sqrt{h(C_{\pi(j)})}}\right)^{-1}},\\
&=\nu^2\expecgen{\vC}{\left(\frac{1}{\Lambda+(N-N_1)c_0}\right)\left(\sum\limits_{i=1}^{N-N_1}{\sqrt{h(C_{\pi(i)})}}\right)^2}, 
\end{split}
\end{equation}
\fi
We can further simplify the expression by noting that $h(1)=0$ (i.e. for the zero-value class) so that we can drop the permutation operator:
\ifjournal
\else
\begin{equation}
\label{eq:oracle-cost-expecN}
\begin{split}
J_T(\vlam^o) &= \nu^2\expecgen{\vC}{\left(\frac{1}{\Lambda+(N-N_1)c_0}\right)\left(\sum\limits_{i=1}^{N}{\sqrt{h(C_{i})}}\right)^2},\\
&= \nu^2\expecgen{\vC}{\left(\frac{1}{\Lambda+(N-N_1)c_0}\right)\left(\sum\limits_{c=1}^{|\calC|}{N_c\sqrt{h(c)}}\right)^2},\\
&= \nu^2\expecgen{\vN}{\left(\frac{1}{\Lambda+(N-N_1)c_0}\right)\left(\sum\limits_{c=2}^{|\calC|}{N_c\sqrt{h(c)}}\right)^2},
\end{split}
\end{equation}
\fi
where the last equality once again uses $h(1)=0$ and noting that $J_T(\vlam^o)$ is only random through the number of targets in each class $\vN=\set{N_c}_{c\in\calC}\sim\mathrm{Multinomial}(N,\set{p_c}_{c\in\calC})$.  We note the following properties of Multinomial distributions which will be useful in further simplifying \eqref{eq:oracle-cost-expecN}:
\begin{equation}
\set{N_2,N_3,\dots,N_{|\calC|}} | N_1 \sim \mathrm{Multinomial}\left(N-N_1,\set{\pt_c}_{c=2}^{|\calC|}\right)
\end{equation}
\begin{equation}
\begin{split}
\expec{N_iN_j|N_1} &= \begin{cases}
(N-N_1)\pt_i(1-\pt_i) + (N-N_1)^2\pt_i^2, &i=j \\
-(N-N_1)\pt_i\pt_j + (N-N_1)^2\pt_i\pt_j, &i\neq j
\end{cases}\\
&\forall i,j \in \set{2,3,\dots,|\calC|} 
\end{split}
\end{equation}
where $\pt_c \trieq p_c/(1-p_1)$. Note that $\set{\pt_c}_{c=2}^{|\calC|}$ defines a probability distribution on $C_i$ conditioned on $C_i>1$. Using iterated expectation over (\ref{eq:oracle-cost-expecN}) and the properties above:
{\small \begin{equation}
\begin{split}
\label{eq:iterated-expectation-N}
&J_t(\vlam^o)\\
&=\expecNone{\left(\frac{\nu^2}{\Lambda+(N-N_1)c_0}\right)\expecNN{\left(\sum\limits_{c=2}^{|\calC|}{N_c\sqrt{h(c)}}\right)^2}}
\end{split}
\end{equation}}
where $\vN\setminus N_1=\set{N_2,\dots,N_{|\calC|}}$. 
\ifjournal
\else
Expanding the inner expectation
\begin{equation}
\label{eq:inner-exp-oracle-cost}
\begin{split}
&\expecNN{\left(\sum\limits_{c=2}^{|\calC|}{N_c\sqrt{h(c)}}\right)^2}\\
&\quad=\expecNN{\sum\limits_{c=2}^{|\calC|}{N_c^2h(c)}+2\sum\limits_{c=2}^{|\calC|}\sum\limits_{d=c+1}^{|\calC|}{N_cN_d\sqrt{h(c)h(d)}}}\\
&\quad=\sum\limits_{c=2}^{|\calC|}\left((N-N_1)\pt_c(1-\pt_c) + (N-N_1)^2\pt_c^2\right)h(c)\\
&\qquad+2\sum\limits_{c=2}^{|\calC|}\sum\limits_{d=c+1}^{|\calC|}{(-(N-N_1)\pt_c\pt_d + (N-N_1)^2\pt_c\pt_d)\sqrt{h(c)h(d)}}\\
&\quad= (N-N_1)a_1 + (N-N_1)^2a_2
\end{split}
\end{equation}
where
\begin{align}
a_1 &= \sum_{c=2}^{|\calC|}\pt_c(1-\pt_c)h(c)-2\sum_{c=2}^{|\calC|}\sum_{d=c+1}^{|\calC|}\pt_c\pt_d\sqrt{h(c)h(d)}
\ignore{\\
\nonumber &= \sum_{c=2}^{|\calC|} \pt_c \sqrt{h(c)} \left(\sqrt{h(c)} - \sum_{d=2}^{|\calC|} \pt_d \sqrt{h(d)}\right) = \sum_{c=2}^{|\calC|} \pt_c h(c) - a_2}
\\
 a_2 &= \sum_{c=2}^{|\calC|}\pt_c^2h(c)+2\sum_{c=2}^{|\calC|}\sum_{d=c+1}^{|\calC|}\pt_c\pt_d\sqrt{h(c)h(d)}
\ignore{\\
\nonumber&=\sum_{c=2}^{|\calC|} \pt_c \sqrt{h(c)} \left(\sum_{d=2}^{|\calC|} \pt_d \sqrt{h(d)}\right)\\
\nonumber&=\left(\sum_{c=2}^{|\calC|} \pt_c \sqrt{h(c)}\right)^2}
\end{align}
With some simple algebraic manipulation, it can easily be shown that
\fi
\begin{equation}
\label{eq:a2-app}
a_2 = \left(\sum_{c=2}^{|\calC|} \pt_c \sqrt{h(c)}\right)^2=\firstmomsq
\end{equation}
and
\begin{equation}
\label{eq:a1-app}
a_1 = \sum_{c=2}^{|\calC|} \pt_c h(c) - a_2 = \secondmom-\firstmomsq
\end{equation}
Using (\ref{eq:inner-exp-oracle-cost}) in (\ref{eq:iterated-expectation-N}) yields an expression for the oracle cost that is simply an expectation over a single random parameter $N_1\sim\mathrm{Binomial}(N,p_1)$:
\begin{equation}
\begin{split}
\label{eq:oracle-cost-expecN-one}
J_t(\vlam^o)&=\nu^2\expecNone{\frac{(N-N_1)a_1 + (N-N_1)^2a_2}{\Lambda+(N-N_1)c_0}}
\end{split}
\end{equation}
To complete the proof, we use Lemma \ref{lemma:taylor-series-bounds} below to bound the expectation over $N_1$.  
\ignore{
show that (\ref{eq:oracle-cost-expecN-one}) is a convex function in $N_1$ so that we can apply Jensen's inequality.  In particular, define
\begin{equation}
\expecN{J_t(\vlam^o)} = \expecNone{f(N-N_1)}
\end{equation}
where
\begin{equation}
f(x) = \frac{a_1x + a_2x^2}{\Lambda + xc_0}
\end{equation}
}
\begin{lemma}
\label{lemma:taylor-series-bounds}
Let $f:[0,N]\rightarrow{\mathbf{R}}$ be a function of the form
\begin{equation}
f(x) = (e_1x + e_2x^2)/(\Lambda+xc_0)
\end{equation}
where $(e_1,e_2)$ satisfy $\Lambda e_2 \geq c_0e_1$.  Then for an even integer $d\geq 2$ and a random variable $X\in[0,N]$, we have
\begin{align}
\label{eq:lower-bound-taylor-series}
& \expec{f(X)}\geq  \expec{P_{d-1}^f(x)}\\
\label{eq:upper-bound-taylor-series}
&\expec{f(X)}\leq  \expec{P_{d-1}^f(x)} + \frac{f^{(d)}(0)\expec{(X-\mu_X^{(1)})^d}}{d!}\\
&\expec{P_{d-1}^f(x)} = \sum\limits_{w=0}^{d-1} \frac{f^{(w)}(\mu_X^{(1)})\expec{(X-\mu_X^{(1)})^w}}{w!}\\
&\label{eq:fwx}f^{(w)}(x) = \frac{(-1)^w(\Lambda e_2 - c_0e_1)\Lambda c_0^{w-2}w!}{(\Lambda+c_0x)^{w+1}},\quad w\geq 2
\end{align}
where $f^{(w)}(x)$ is the $w$-th derivative of $f(x)$ and $\mu_X^{(w)}$ is the $w$-th central moment.
\end{lemma}
\begin{IEEEproof}
The proof is given in Appendix \ref{app:taylor-series}.
\end{IEEEproof}

Note that we can rewrite \eqref{eq:oracle-cost-expecN-one} as
\begin{equation}
J_t(\vlam^o)=\nu^2\expec{f(N-N_1)}
\end{equation}
with $e_1=a_1$ and $e_2=a_2$. Using Assumption \ref{ass:high-snr-assumption-oracle1}, we have
\begin{equation}
\begin{split}
\Lambda \geq c_0\Lambda_1 = c_0\left(\frac{\secondmom}{\firstmomsq}-1\right)= c_0a_1/a_2
\end{split}
\end{equation}
so that 
\begin{equation}
\label{eq:a2lambda-a1c0}
a_2\Lambda - a_1c_0 \geq a_2( c_0a_1/a_2) - a_1c_0 = 0
\end{equation}
Therefore, we can apply Lemma \ref{lemma:taylor-series-bounds} with $X=N-N_1$ and $d=2$:
\begin{equation}
\begin{split}
J_t(\vlam^o)&\geq\nu^2\left(f(\mu_X^{(0)}) + f^{(1)}(\mu_X^{(0)})\expec{X-\mu_X^{(0)}}/2\right)\\
&= \nu^2f(N(1-p_1))\\
\label{eq:oracle-cost-jensens}
&=
\nu^2\left(\frac{(N-Np_1)a_1 + (N-Np_1)^2a_2}{\Lambda+(N-Np_1)c_0}\right)
\end{split}
\end{equation}
where we used the fact that $\expec{X-\mu_X^{(0)}}=0$ and $\mu_X^{(0)} = N(1-p_1)$.
\ignore{
\begin{align}
c_2 &= (\Lambda a_2 - c_0a_1)\Lambda/c_0^2\\
\mu_X^{(0)} &= N(1-p_1)\\
\mu_X^{(1)} &= Np_1(1-p_1)\\
\mu_X^{(2)} &= Np_1(1-p_1)(2p_1-1)
\end{align}
Combining (\ref{eq:f-second-deriv}) and (\ref{eq:a2lambda-a1c0}) and noting that  first term in (\ref{eq:f-second-deriv}) is always positive for $x>0$ (since $\Lambda$ and $c_0$ are always positive), we can conclude that $f''(x)\geq 0$ so that $f(x)$ is a convex function.  Since $f(x)$ is convex, then $f(N-N_1)$ is also convex so that we can apply Jensen's inequality:
\begin{equation}
\begin{split}
\label{eq:oracle-cost-jensens}
J_t(\vlam^o)&=\expecNone{f(N-N_1)}\geq f(\expecNone{N-N_1})\\
&= \nu^2\left(\frac{(N-Np_1)a_1 + (N-Np_1)^2a_2}{\Lambda+(N-Np_1)c_0}\right)
\end{split}
\end{equation}
}
The result follows from simple factorization of (\ref{eq:oracle-cost-jensens}) and using the definitions in (\ref{eq:a1-app}) and (\ref{eq:a2-app}).   Note that we can get a tighter bound by applying Lemma \ref{lemma:taylor-series-bounds} with $d>2$.  To derive the upper bound, we apply \eqref{eq:upper-bound-taylor-series} from \ref{lemma:taylor-series-bounds} with $d=2$, where we notice that the first term is just the lower bound and the second term is given by
\begin{equation}
\begin{split}
f^{(2)}(0)&\expec{(X-\mu_X^{(1)})^2}/d!\\
&=\frac{(\Lambda\firstmomsq - c_0 (\secondmom-\firstmomsq))Np_1(1-p_1)}{\Lambda^2}\\
&=\frac{((\Lambda+c_0)\firstmomsq - c_0 \secondmom))Np_1(1-p_1)}{\Lambda^2}
\end{split}
\end{equation}
where the expectation is just the variance of the random variable $N-N_1$, which is $Np_1(1-p_1)$.

\section{Proof of Lemma \ref{lemma:taylor-series-bounds}}
\label{app:taylor-series}
First note that the first two derivatives of $f(x)$ are given by
\begin{align}
f^{(1)}(x) &= \Lambda (e_1 + 2e_2x) + e_2c_0x^2 / (\Lambda + c_0x)^2\\
f^{(2)}(x) &= 2\Lambda (\Lambda e_2 - e_1c_0) / (\Lambda + c_0x)^3
\end{align}
Only the denominator of $f^{(2)}(x)$ depends on $x$, from which the relationship (\ref{eq:fwx}) can easily be derived.  Define the $d$-th order Taylor series approximation to $f$ around point $c$ to be
\begin{equation}
P_{d}^f(x) = \sum\limits_{w=0}^d \frac{f^{(w)}(c)}{w!}(x-c)^w
\end{equation}
Using Taylor's theorem, we have
\begin{equation}
f(x) = P_{d-1}^f(x)+\frac{f^{(d)}(b)}{d!}(x-c)^d
\end{equation}
for some $b\in [0,N]$ (i.e., the domain of $f$).  In particular, when $d$ is even, we have both $f^{(d)}(x)\geq 0$ and $(x-c)^d\geq 0$ for all $x\in[0,N]$.  Therefore
\begin{equation}
\label{eq:fx-lower}
f(x) \geq P_{d-1}^f(x)
\end{equation}
Moreover, using the corollary of Taylor's theorem, we have
\begin{equation}
\left| f(x)-P_{d-1}^f(x)\right|\leq \frac{\max\limits_{b\in [0,N]} |f^{(d)}(b)|}{d!}|x-c|^d
\end{equation}
From (\ref{eq:fwx}), it is clear the maximum of $|f^{(d)}(b)|$ occurs when $b=0$.  Moreover, since $d$ is even $|x-c|^d=(x-c)^d$ and $f^{(d)}(0)\geq 0$ .  Since $f(x)>P_{d-1}^f(x)$, we can conclude that
\begin{equation}
\label{eq:fx-upper}
\begin{split}
& f(x)-P_{d-1}^f(x) \leq \frac{f^{(d)}(0)}{d!}(x-c)^d\\
\Leftrightarrow \qquad & f(x) \leq P_{d-1}^f(x) + \frac{f^{(d)}(0)}{d!}(x-c)^d
\end{split}
\end{equation}
\eqref{eq:lower-bound-taylor-series} and \eqref{eq:upper-bound-taylor-series} result from applying linearity of expectation to \ref{eq:fx-lower} and \ref{eq:fx-upper} with Taylor series expansion around $c=\expec{X-\expec{X}}$.

\ifjournal
\else
	\section{Proof of Proposition \ref{prop:cost-MSE-equivalence}}
\label{app:cost-MSE-equivalence}
We start by using linearity of the summation operators to rewrite the cost (\ref{eq:multiclass-objective-function1}) as
\begin{equation}
J_T(\vlam)=\expec{\sum\limits_{c\in\cal{C}}h(c)\left(\sum\limits_{i=1}^N\Iic{c}(X_i-\hatXic{c}(T))^2\right)},
\end{equation}
To simplify (\ref{eq:multiclass-objective-function1}), it is useful to note that each term of the inner summation is non-zero only when $C_i=c$.  Moreover, using iterated expectation and the linearity of expectation, we have $J_T(\vlam) =$
\begin{equation}
\sum\limits_{c\in\cal{C}}h(c)\expecY{\expec{\sum\limits_{i=1}^N\Iic{c}(X_i-\hatXic{c}(T))^2\Big | C_i=c,\vY(T)}}.
\end{equation}
Given $C_i=c$, $\Iic{c}$ is deterministic and hence independent of the other terms.  Using this property, (\ref{eq:Xi-cme}), and the definition of variance, we have
\begin{equation}
\label{eq:multiclass-objective-function2}
\begin{split}
J_T(\vlam) &= \sum\limits_{c\in\cal{C}}h(c)\expecY{\sum\limits_{i=1}^N\pic{c}(T)(\sic{c}(T))^2}\\
&=\expecY{\sum\limits_{i=1}^N\sum\limits_{c\in\cal{C}}h(c)\pic{c}(T)(\sic{c}(T))^2}
\end{split},
\end{equation}
where $\pic{c}(T)=\Pr(C_i=c|\vY(T))$ and $(\sic{c}(T))^2=\mathrm{var}\left[X_i|C_i=c,\vY(T)\right]$.  

When conditioned on $C_i=c$, we know that $X_i$ is Gaussian by definition.  It can be shown that when $X_i$ is Gaussian and $\vlam(t-1)$ is a deterministic function of previous measurements, then any new measurements $\vy(t)$ are also Gaussian.  This leads to the following simple recursion on the conditional variances:
\begin{equation}
\frac{1}{(\sic{c}(T+1))^2}=\frac{1}{(\sic{c}(T))^2}+\frac{\lambda_i(T)}{\nu^2}=\frac{1}{\sigma_c^2}+\frac{\sum_{t=0}^{T}\lambda_i(t)}{\nu^2},
\end{equation}
so that
\begin{equation}
\label{eq:sic-recursion-tr}
(\sic{c}(T+1))^2 = \nu^2\left[\frac{\nu^2}{\sigma_c^2}+{\sum_{t=0}^{T}\lambda_i(t)}\right]^{-1} = \nu^2\left[\frac{\nu^2}{\sigma_c^2}+\olami\right]^{-1}
\end{equation}
where $\olami = {\sum_{t=0}^{T-1}\lambda_i(t)}$. Plugging this into (\ref{eq:multiclass-objective-function2}), we get
\begin{equation}
\begin{split}
J_T(\vlam)&=\expecY{\sum\limits_{i=1}^N\sum\limits_{c\in\cal{C}}h(c)\pic{c}(T)\si^2(T)}\\
&=\nu^2\expecY{\sum\limits_{i=1}^N\sum\limits_{c\in\cal{C}}\frac{h(c)\pic{c}(T)}{\nu^2/\sigma_c^2+\olami}}
\end{split}
\end{equation}
When Assumption \ref{ass:equal-variance} holds, we can simplify this further, so that
\begin{equation}
\begin{split}
J_T(\vlam)&=\nu^2\expecY{\sum\limits_{i=1}^N\frac{\sum_{c\in\cal{C}}h(c)\pic{c}(T)}{\nu^2/\sigma_0^2+\olami}}\\
&=\nu^2\expecY{\sum\limits_{i=1}^N\frac{z_i(T)}{\nu^2/\sigma_0^2+\olami}}.
\end{split}
\end{equation}
where $\zi(T) = \sum_{c\in\cal{C}}h(c)\pic{c}(T)$.  Using iterated expecation and Lemma \ref{lemma:zic-yt} (below), we get the final result
\begin{equation}
\begin{split}
J_T(\vlam)&=\nu^2\expecYY{\expecy{\sum\limits_{i=1}^N\frac{\zi(T)}{\nu^2/\sigma_0^2+\olami}}},\\
&=\nu^2\expecYY{\sum\limits_{i=1}^N\frac{\expecy{\zi(T)}}{\nu^2/\sigma_0^2+\olami}},\\
&=\nu^2\expecYY{\sum\limits_{i=1}^N\frac{\zi(T-1)}{\nu^2/\sigma_0^2+\olami}}.
\end{split}
\end{equation}
Note that using (\ref{eq:sic-recursion-tr}), we can rewrite an equivalent result in terms of the final-stage allocations $\vlam(T-1)$ as
\begin{equation}
\begin{split}
J_T(\vlam)&=\nu^2\expecYY{\sum\limits_{i=1}^N\frac{\zi(T-1)}{\nu^2/\si^2(T-1)+\lambda_i(T-1)}}.
\end{split}
\end{equation}
The derivation of the final expressions in \eqref{eq:final-cost-form} without Assumption \ref{ass:equal-variance} is entirely analogous.

\begin{lemma}
\label{lemma:zic-yt}
\begin{equation}
\label{eq:zic-yt}
\expecy{\zi(T)}=\zi(T-1)
\end{equation}
\end{lemma}
\begin{IEEEproof}
Note that
\begin{equation}
\begin{split}
\Pr(\vC|\vY(T)) &= \Pr(\vC|\vY(T-1),\vy(T)) \\
&= \frac{\Pr\left(\vy(t)|\vC,\vY(t-1)\right)\Pr\left(\vC|\vY(T-1)\right)}{\Pr\left(\vy(t)|\vY(T-1)\right)}
\end{split}
\end{equation}
Then
\begin{equation}
\label{eq:fc-yt}
\begin{split}
\mathbb{E}&_{\vy(T)|\vY(T-1)}\left[\Pr(\vC|\vY(T))\right]
\\ &=\int\frac{\Pr\left(\vy(t)|\vC,\vY(t-1)\right)\Pr\left(\vC|\vY(T-1)\right)}{\Pr\left(\vy(t)|\vY(T-1)\right)}\\
&\qquad \cdot\Pr\left(\vy(t)|\vY(T-1)\right)d\vy(t)\\
&=\int \Pr\left(\vC|\vY(T-1)\right) \Pr\left(\vy(t)|\vC,\vY(t-1)\right)d\vy(t)\\
&=\Pr\left(\vC|\vY(T-1)\right)\int \Pr\left(\vy(t)|\vC,\vY(t-1)\right)d\vy(t)\\
&=\Pr\left(\vC|\vY(T-1)\right)
\end{split}
\end{equation}
Note that $f\left(\vC|\vY(T)\right)$ is the multivariate extension of $\pic{c}(T)=\Pr(C_i=c|\vY(T))$, from which we can conclude
\begin{equation}
\label{eq:pic-yt}
\expecy{\pic{c}(T)}=\pic{c}(T-1)
\end{equation}
Using (\ref{eq:pic-yt}) in the definition of $z_i(T)$ yields the result.
\end{IEEEproof}

	\section{Proof of Lemma \ref{lemma:cost-determ-lambda}}
\label{app:cost-determ-lambda}
Applying the law of total expectation to (\ref{eq:final-cost-form})
\begin{equation}
\label{eq:cost-determ-lambda1}
\begin{split}
J_T(\vlam) &= 
\nu^2\expecY{\sum\limits_{i=1}^N\sum_{c=2}^{|\calC|}\frac{h(c)\pic{c}(T-1)}{\nu^2/\sigma_c^2+\olami(T-1)}} \\
& = \nu^2\expecgen{\vC}{\expecgen{\vY(T-1) | \vC}{\sum\limits_{i=1}^N\sum_{c=2}^{|\calC|}\frac{h(c)\pic{c}(T-1)}{\nu^2/\sigma_c^2+\olami(T-1)}}}\\
\end{split}
\end{equation}
Since $\lambda_i(t)$ is a function only of $\vC$, the denominator conditioned on $\vC$ is deterministic so that
\begin{equation}
\begin{split}
J_T(\vlam) & = 
\nu^2\expecgen{\vC}{\sum\limits_{i=1}^N\sum_{c=2}^{|\calC|}\frac{\expecgen{\vY(T-1) | \vC}{h(c)\pic{c}(T-1)}}{\nu^2/\sigma_c^2+\olami(T-1)}}
\end{split}
\end{equation}
Moreover, we have
\begin{equation}
\label{eq:cost-determ-lambda2}
\begin{split}
\expecgen{\vY(t) | \vC}{\sum_{c\in\calC} h(c) \pic{c}(t)} &= \expecgen{\vY(t) | \vC}{\sum_{c\in\calC} h(c) \expecgen{C_i}{\Iic{c}(t)}}\\
&= \expecgen{\vY(t) | \vC}{\sum_{c\in\calC} h(c) \Iic{c}(t)} = h(C_i)\\
\end{split}
\end{equation}
Plugging (\ref{eq:cost-determ-lambda2}) into (\ref{eq:cost-determ-lambda1}) yields
\begin{equation}
J_T(\vlam) = \nu^2\expecgen{\vC}{\sum\limits_{i=1}^N\sum_{c=2}^{|\calC|}\frac{h(C_i)}{\nu^2/\sigma_c^2+\olami(T-1)}}.
\end{equation}
Note that Assumption \ref{ass:equal-variance} further simplifies the denominator.

	\section{Derivation of (\ref{eq:oracle-allocation})}
\label{app:full-oracle-deriv}
The solution here follows \cite{Wei13_TSP} and begins by letting $c_i=\nu^2/\sigma_{C_i}^2$. Then, define $g(k)$ to be the monotonically non-decreasing function of $k=0,\dots,N$ with $g(0)=0$, 
\begin{equation}
g(k) = \frac{c_{\pi(k+1)}}{\sqrt{h(C_{\pi(k+1)})}} \sum\limits_{i=1}^k\sqrt{h(C_{\pi(i)})}- \sum\limits_{i=1}^kc_{\pi(i)},
\end{equation}
for $k=1,\dots,N-1$, and $g(N)=\infty$. Define $k^*$ by the interval $(g(k-1),g(k)]$ to which the budget parameter $\Lambda$ belongs.  Since $g(k)$ is monotonic, the mapping from $\Lambda(t)$ to $k^*$ is one-to-one. The optimal full oracle policy is then 
\begin{equation}
\overline{\lambda}_{\pi(i)}^{o}=
\left(\Lambda(t)+\sum\limits_{j=1}^{k^*}
c_{\pi(j)}\right)\dfrac{\sqrt{h(C_{\pi(i)})}}{\sum_{j=1}^{k^*}
\sqrt{h(C_{\pi(j)}}}-c_{\pi(i)},
\end{equation}
when $i\leq k^*$, 
and zero else.

We can simplify this expression when Assumption \ref{ass:equal-variance} holds.  In particular, from (\ref{eq:index-permutation-definition}), we can easily see that
\begin{equation}
\label{eq:h-pi-ordering-oracle}
h(C_{\pi(i)}) = \begin{cases}
h(|\calC|), & i=1,2,\dots,N_{|\calC|}\\
h(|\calC|-1), & i=N_{|\calC|}+1,\dots,N_{|\calC|}+N_{|\calC|-1}\\
\vdots, &\vdots\\
h(2), & i=N-N_1-N_2+1,\dots,N-N_1\\
h(1)=0, & i=N-N_1+1,\dots,N
\end{cases}
\end{equation}
Inspecting \eqref{eq:h-pi-ordering-oracle} further, we note an important property in the case where $C_{\pi(i)}\neq C_{\pi(i+1)}$. Specifically, we have
\begin{equation}
\label{eq:d-pi-def}
\pi(i) = \sum\limits_{c=d}^{|\calC|} N_c
\end{equation}
for some $d\in\set{2,3,\dots,|\calC|}$.  Moreover, if $d$ is chosen to satisfy \eqref{eq:d-pi-def} and $N_c>0$ for all $c\in\calC$, then we have
\begin{equation}
\label{eq:hd-def}
\begin{split}
h(C_{\pi(i)}) &= h(d)\\
h(C_{\pi(i)+1}) &= h(d-1)
\end{split}
\end{equation}
Thus, we have
\begin{equation}
\begin{split}
g(k) &= c_0 \left( \sum\limits_{i=1}^k \frac{\sqrt{h(C_{\pi(i)}})}{\sqrt{h(C_{\pi(k+1)}})} - k\right)
\end{split}
\end{equation}
where $c_0 = \nu^2/\sigma_0^2$. There are two regimes of interest to further simplify this function: (a) when $C_{\pi(k+1)}=C_{\pi(k)}$ and (b) when $C_{\pi(k+1)}\neq C_{\pi(k)}$.  In the former case we note that
\begin{equation}
\label{eq:gk-sameclass}
\begin{split}
g(k) &= c_0\left(\sum\limits_{i=1}^k\frac{\sqrt{h(C_{\pi(i)})}}{\sqrt{h(C_{\pi(k)})}} - k\right)\\
&= c_0\left(\sum\limits_{i=1}^{k-1}\frac{\sqrt{h(C_{\pi(i)})}}{\sqrt{h(C_{\pi(k)})}} - (k-1)\right)= g(k-1)
\end{split}
\end{equation}
Therefore, within intervals where $C_{\pi(k)}=C_{\pi(k+1)}$, we have equal values of $g(k)$.  This implies that when there is sufficient budget to search for a single target of a given class, then we should search for all targets of that class.  When $C_{\pi(k+1)}\neq C_{\pi(k)}$, we note from \eqref{eq:d-pi-def} and \eqref{eq:hd-def} that
\begin{equation}
\label{eq:gk-diffclass}
\begin{split}
g(\pi(k)) &= c_0\left(\frac{1}{\sqrt{h(\tilde{d}(k)-1)}}\sum\limits_{i=1}^k\sqrt{h(C_{\pi(i)})}- \sum\limits_{c=\tilde{d}(k)}^{|\calC|} N_c\right)\\
&= c_0\left(\frac{1}{\sqrt{h(\tilde{d}(k)-1)}}\sum\limits_{c=\tilde{d}(k)}^{|\calC|}{N_c\sqrt{h(c)}}- \sum\limits_{c=\tilde{d}(k)}^{|\calC|} N_c\right)
\end{split}
\end{equation}
where
\begin{equation}
\label{eq:tilde-dk}
\tilde{d}(k) = \set{d\in\set{2,3,\dots,|\calC|}:  \sum\limits_{c=d}^{|\calC|} N_c \leq k < \sum\limits_{c=d-1}^{|\calC|} N_c}
\end{equation}
Using (\ref{eq:gk-sameclass}), (\ref{eq:gk-diffclass}), (\ref{eq:tilde-dk}), and noticing that $g(1)=0$, we have
\begin{equation}
g(k) = \begin{cases}
0, & k<N_{|\calC|},\\
\infty, &k \geq N-N_1\\
c_0\sum\limits_{c=\tilde{d}(k)}^{|\calC|}N_c\left(\dfrac{\sqrt{h(c)}}{\sqrt{h(\tilde{d}(k)-1)}}-1\right), &\mathrm{else}.
\end{cases}
\end{equation}
Note that $\tilde{d}(k)$ is constant for all targets of the same class, and thus so is $g(k)$.  The number of nonzero allocations $k^*$ is given by the interval where $\Lambda\in(g(k^*-1),g(k^*)]$.  We can conclude that when the oracle policy allocates any resources to a target with class $c$, then it must allocate to all targets with class $c$.  The oracle policy is then given by
\begin{equation}
{\overline{\lambda_{\pi(i)}^{o}}}=\begin{cases}
\dfrac{\left(\Lambda+k^*c_0\right)\sqrt{h(C_{\pi(i)})}}{\sum_{j=1}^{k^*}\sqrt{h(C_{\pi(j)})}}-c_0, & i=1,\dots,k^*\\
0, & \mathrm{else}\end{cases}
\end{equation}
We can simplify this further using Assumption \ref{ass:high-snr-assumption-oracle1} where we have
\begin{equation}
\begin{split}
\Lambda \geq c_0 \Lambda_0 &= c_0\sum\limits_{c=2}^{|\calC|} N_c\sqrt{\frac{h(c)}{h(2)}}-(N-N_1)\\
&= g(N-N_1-1)
\end{split}
\end{equation}
Since $g(N-N_1)=\infty$, we have $\Lambda \in (g(N-N_1-1),g(N-N_1)]$ so that $k^*=N-N_1$.  Plugging into \eqref{eq:oracle-allocation}, we have 
\begin{equation}
\begin{split}
\overline{\lambda_{\pi(i)}^o} &= \left(\Lambda+(N-N_1)c_0\right)\left(\dfrac{\sqrt{h(C_{i})}}{\sum_{c=2}^{|\calC|}N_c\sqrt{h(c)}}\right)-c_0
\end{split}
\end{equation}
for $C_i>1$ and zero else.

	\section{Probability of satisfying (\ref{eq:lambda-min-oracle1})}
\label{app:snr-condition-prob}

Assumption \ref{ass:high-snr-assumption-oracle1} requires 
that $\Lambda \geq \Lambda_\mathrm{min} \geq c_0\Lambda_0$, where $\Lambda_0$ depends on random variables $\set{N_c}_{c=1}^{|\calC|}$.  Here we show that $\Lambda_{0}$ is bounded from above by $\Lambda_{0}'$ in \eqref{eqn:Lambda0'} with probability converging exponentially to $1$ as $N$ increases. Hence requiring $\Lambda_{\mathrm{min}} \geq c_{0}\Lambda_{0}'$ implies that $\Lambda_\mathrm{min} \geq c_0\Lambda_0$ with the same probability or greater.

First we recognize $\Lambda_{0}$ as the sum of $N$ i.i.d.\ random variables $G_{1},\ldots,G_{N}$, where 
\begin{equation}
G_{i} = 
\begin{cases}
0, & C_{i} = 1,\\
\left( \sqrt{\frac{h(C_{i})}{h(2)}} - 1 \right), & C_{i} > 1.
\end{cases}
\end{equation}
Given the ordering of classes, $G_{i}$ is bounded above and below by $G_{\mathrm{max}} = \sqrt{h(|\mathcal{C}|) / h(2)} - 1$ and $0$ respectively. The mean of $G_{i}$ is 
\begin{equation}
\mu_{G} = \sum_{c=2}^{|\mathcal{C}|} p_{c} \left( \sqrt{\frac{h(c)}{h(2)}} - 1 \right) = \pbar \left(\frac{\firstmom}{\sqrt{h(2)}} - 1\right).
\end{equation}
We may now apply the Chernoff-Hoeffding bound for independent bounded random variables \cite{Hoeffding63} to the probability that $\Lambda_{0}$ exceeds its mean $N \mu_{G}$ by a multiplicative factor $\alpha > 1$.  Using \eqref{eqn:Lambda0'}, this yields the exponentially decaying bound 
\begin{equation}
\Pr(\Lambda_{0} > \alpha N \mu_{G}) = \Pr(\Lambda_{0} > \Lambda_{0}') \leq \exp\left(-N D(\alpha \hat{p}_{1} \Vert\ \hat{p}_{1}) \right), 
\end{equation}
where 
\begin{equation}
D(p \Vert q) = p \log\left(\frac{p}{q}\right) + (1-p) \log\left(\frac{1-p}{1-q}\right)
\end{equation}
is the Kullback-Leibler divergence between Bernoulli random variables, and 
\begin{equation}
\hat{p}_{1} = \frac{\mu_{G}}{G_{\mathrm{max}}} = \pbar \frac{\firstmom - \sqrt{h(2)}}{\sqrt{h({|\calC|})} - \sqrt{h(2)}}.
\end{equation}

	\section{Proof of Proposition \ref{prop:location-oracle-cost-upper-bound}}
\label{app:proof-prop-location-oracle-cost}
Plugging the location-only oracle allocation policy \eqref{eq:location-oracle-allocation} into the cost function \eqref{eq:oracle-cost-function}, we get
\begin{equation}
\label{eq:cost-location-only-intermed}
\begin{split}
J_T(\vlam^{lo})&=\nu^2\expecgen{\vC}{\sum\limits_{i=1}^{N}\frac{h(C_{i})}{c_0 + \Lambda/(N-N_1)}}\\
&=\nu^2\expecgen{\vN}{\dfrac{\sum_{c=2}^{|\calC|}N_ch(c)}{c_0 + \Lambda/(N-N_1)}}
\end{split}
\end{equation}
Once again using properties of the multinomial distribution conditioned on $N_1$, we have
\begin{equation}
\begin{split}
J_T(\vlam^{lo}) &= \nu^2\expecgen{N_1}{\dfrac{\sum_{c=2}^{|\calC|}(N-N_1)\pt_ch(c)}{c_0 + \Lambda/(N-N_1)}}\\
&=
\left(\dfrac{\nu^2\sum_{c=2}^{|\calC|}p_ch(c)}{1-p_1}\right)\expecgen{N_1}{\left(\dfrac{(N-N_1)^2}{c_0(N-N_1)+\Lambda}\right)}\\
&=
\nu^2\secondmom\expecgen{N_1}{\left(\dfrac{(N-N_1)^2}{c_0(N-N_1)+\Lambda}\right)}
\end{split}
\end{equation}
The expression within the expecation can be evaluated using Lemma \ref{lemma:taylor-series-bounds} with $e_2 = 1$, $e_1 = 0$, $X=N-N_1$, $\mu_X^{(1)}=N(1-p_1)$ and $d=2$ so that
\begin{equation}
\begin{split}
J_T(\vlam^{lo}) &\geq \nu^2\secondmom f(N(1-p_1))\\
&= \nu^2\secondmom\left(\dfrac{(N-Np_1)^2}{c_0(N-Np_1)+\Lambda}\right)\\
&= \nu^2\secondmom\left(\dfrac{N^2(1-p_1)^2}{c_0N(1-p_1)+\Lambda}\right)\\
&= \frac{\nu^2N^2(1-p_1)^2\secondmom}{\Lambda+N(1-p_1)c_0}.
\end{split}
\end{equation}
To derive the upper bound, we apply \eqref{eq:upper-bound-taylor-series} from Lemma \ref{lemma:taylor-series-bounds} to the expecation in \eqref{eq:cost-location-only-intermed} with $d=2$, where we notice that the first term of \eqref{eq:upper-bound-taylor-series} is just the lower bound stated above and the second term is given by
\begin{equation}
\begin{split}
f^{(2)}(0)&\expec{(X-\mu_X^{(1)})^2}/d!\\
&=\frac{\Lambda^2 Np_1(1-p_1)}{\Lambda^3}\\
&=\frac{Np_1(1-p_1)}{\Lambda}
\end{split}
\end{equation}
where the expectation is once again the variance of the random variable $N-N_1$, which is $Np_1(1-p_1)$.

	\section{Proof of Proposition \ref{prop:uniform-cost}}
\label{app:proof-prop-uniform-cost}
Using Lemma \ref{lemma:cost-determ-lambda} and (\ref{eq:uniform-allocation-allstages})
\begin{equation}
\begin{split}
J_T&(\vlam^u)=\nu^2\expec{\sum\limits_{i=1}^N\frac{h(C_i)}{\nu^2/\sigma_0^2+\Lambda/N}}\\
&= \left(\frac{\nu^2}{\nu^2/\sigma_0^2+\Lambda/N}\right)\expec{\sum\limits_{i=1}^Nh(C_i)}\\
&= \left(\frac{\nu^2}{\nu^2/\sigma_0^2+\Lambda/N}\right)\expec{\sum\limits_{c=2}^{|\calC|}N_ch(c)}\\
&= \left(\frac{\nu^2N}{\nu^2/\sigma_0^2+\Lambda/N}\right)\sum\limits_{c=2}^{|\calC|}p_ch(c)=\frac{\nu^2N\pbar\secondmom}{\nu^2/\sigma_0^2+\Lambda/N}
\end{split}
\end{equation}
where the second-to-last equality can be evaluated noting that $\vN\sim\mathrm{Multinomial}(N,\set{p_c}_{c\in\calC})$.

	\section{Full-oracle cost with general prior target variances}
\label{app:proof-prop-oracle-cost-general}

Plugging (\ref{eq:oracle-allocation}) into the cost function \eqref{eq:oracle-cost-function} yields

 \begin{equation}
\begin{split}
J_T(\vlam^o)&=\nu^2\expecgen{\vC}{\sum\limits_{i=1}^{N}\frac{h(C_{\pi(i)})}{\dfrac{\nu^2}{\sigma_{C_{\pi(i)}}^2} + \dfrac{\left(\Lambda+\nu^2\sum\limits_{j=1}^{k^*}\sigma_{C_{\pi(j)}}^{-2}\right)\sqrt{h(C_{\pi(i)})}}{\sum_{j=1}^{k^*}\sqrt{h(C_{\pi(j)})}}-\dfrac{\nu^2}{\sigma_{C_{\pi(i)}}^2}}},\\
&=\nu^2\expecgen{\vC}{\left(\Lambda+\nu^2\sum\limits_{j=1}^{k^*}\sigma_{C_{\pi(j)}}^{-2}\right)^{-1}\left(\sum\limits_{i=1}^{N}{\sqrt{h(C_{\pi(i)})}}\right)^2}, \end{split}
\end{equation}
As in the equal-variance case, given sufficent SNR (similar to Assumption  \ref{ass:high-snr-assumption-oracle1}), the number of non-zero allocations for the oracle policy is $N-N_1$ and we can further simplify the expression by noting that $h(1)=0$ (i.e. for the zero-value class) so that we can drop the permutation operator and replace the summations over the classes and number of targets in each class:
\begin{equation}
\label{eq:oracle-cost-expecN-general}
\begin{split}
J_T(\vlam^o) &= \nu^2\expecgen{\vC}{\left(\Lambda+\nu^2\sum\limits_{c=2}^{|\calC|}N_c\sigma_{c}^{-2}\right)^{-1}\left(\sum\limits_{i=1}^{N}{\sqrt{h(C_{i})}}\right)^2},\\
&= \nu^2\expecgen{\vN}{\left(\Lambda+\nu^2\sum\limits_{c=2}^{|\calC|}N_c\sigma_{c}^{-2}\right)^{-1}\left(\sum\limits_{c=1}^{|\calC|}{N_c\sqrt{h(c)}}\right)^2},\\
\end{split}
\end{equation}
where the last equality once again uses $h(1)=0$ and noting that $J_T(\vlam^o)$ is only random through the number of targets in each class $\vN=\set{N_c}_{c\in\calC}\sim\mathrm{Multinomial}(N,\set{p_c}_{c\in\calC})$.  

As opposed to the previous specialized case of equal prior target variance, there is no simple way to analytically evaluate this expectation.  Nevertheless, it is simple to evaluate the expectation through Monte Carlo approximation by taking samples from $\vN$.  In this way, we can evaluate the bounds to compare to the equal-variance case.  Note that only the first quantity in (\ref{eq:oracle-cost-expecN-general}) is different in comparison to the equal-variance case, and it only depends on the parameters through the prior target variances and the number of targets in each class.

To complete the analysis, we also include the costs of the location-only and uniform policies in the general case.  In particular, following Appendices \ref{app:proof-prop-location-oracle-cost} and \ref{app:proof-prop-uniform-cost} with appropriate modifications, we have

\begin{equation}
\label{eq:cost-location-only-intermed-general}
\begin{split}
J_T(\vlam^{lo})&=\nu^2\expecgen{\vC}{\sum\limits_{i=1}^{N}\frac{h(C_{i})}{\nu^2/\sigma_{C_i}^2 + \Lambda/(N-N_1)}}\\
&=\nu^2\expecgen{\vN}{\sum_{c=2}^{|\calC|}\dfrac{N_ch(c)}{\nu^2/\sigma_c^2 + \Lambda/(N-N_1)}}\\
&=\nu^2\expecgen{N_1}{\sum_{c=2}^{|\calC|}\dfrac{(N-N_1)\pt_ch(c)}{\nu^2/\sigma_c^2 + \Lambda/(N-N_1)}}\\
&=
\left(\dfrac{\nu^2}{1-p_1}\right)\sum_{c=2}^{|\calC|}p_ch(c)\expecgen{N_1}{\left(\dfrac{(N-N_1)^2}{\nu^2(N-N_1)/\sigma_c^2+\Lambda}\right)}
\end{split}
\end{equation}
Once again, this expression differs from the previous ones in the equal-variance case only as a function of the target prior variances and $\vN$. Using a slightly modified version of Lemma \ref{lemma:taylor-series-bounds}, we get
\begin{equation}
\begin{split}
J_T(\vlam^{lo}) &\geq \left(\dfrac{\nu^2}{1-p_1}\right)\sum_{c=2}^{|\calC|}\left(\dfrac{p_ch(c)N^2(1-p_1)^2}{\nu^2 N(1-p_1)/\sigma_c^2+\Lambda}\right)\\
&= \nu^2N^2(1-p_1)\sum_{c=2}^{|\calC|}\dfrac{p_ch(c)}{\Lambda+N(1-p_1)\nu^2/\sigma_c^2}.
\end{split}
\end{equation}
Similarly,
\begin{equation}
\label{eq:cost-uniform-general}
\begin{split}
J_T(\vlam^{u})&=\nu^2\expecgen{\vC}{\sum\limits_{i=1}^{N}\frac{h(C_{i})}{\nu^2/\sigma_{C_i}^2 + \Lambda/N}}\\
&=\nu^2\expecgen{\vN}{\sum_{c=2}^{|\calC|}\dfrac{N_ch(c)}{\nu^2/\sigma_c^2 + \Lambda/N}}\\
&=\nu^2{\sum_{c=2}^{|\calC|}\dfrac{p_ch(c)}{\nu^2/\sigma_c^2 + \Lambda/N}}\\
&=(\nu^2N){\sum_{c=2}^{|\calC|}\dfrac{p_ch(c)}{N\nu^2/\sigma_c^2 + \Lambda}}
\end{split}
\end{equation}

Note that in all cases, when $\sigma_c^2=\sigma_0^2$, these results simplify to the expressions under the equal-variance assumption.

	\section{Derivation of asymptotic bound on misclassification error}\label{app:asymptotic-missclassification-prob}
In the regime where $\Lambda\rightarrow\infty$, we have a simple expression for the misclassification error using the maximum a posteriori (MAP) estimator. Here, we present the bound for $|\calC|=3$ (i.e., 2 non-zero classes), though the bound is easily extended. The problem then reduces to determining decision regions for when to classify $C_i=2$ versus $C_i=3$.  From (\ref{eq:xi_given_ci}), we have 
\begin{align}
\label{eq:y_ci_2}
f(y_i(t)|C_i=2) &\sim \mathrm{Normal}(\mu_2,\sigma_0^2)\\
\label{eq:y_ci_3}
f(y_i(t)|C_i=3) &\sim \mathrm{Normal}(\mu_3,\sigma_0^2)
\end{align}
where we assume that $\mu_2>\mu_3$ and $p_2>p_3$ (i.e., $C_i=2$ is more likely and easier to detect than $C_i=3$).  By the MAP estimator, we have
\begin{equation}
\hat{C}_i = \begin{cases}
2, & f(y_i(t)|C_i=2)p_2\geq f(y_i(t)|C_i=3)p_3,\\
3, & f(y_i(t)|C_i=2)p_2< f(y_i(t)|C_i=3)p_3
\end{cases}
\end{equation}
Note that these in terms of the log-likelihood ratio
\begin{equation}
\begin{split}
\log(LRT) &= \log(f(y_i(t)|C_i=2)p_2) - \log(f(y_i(t)|C_i=3)p_3)\\
&= \log(p_2/p_3) - \frac{1}{2\sigma_0^2}\left((y_i(t)-\mu_2)^2-(y_i(t)-\mu_3)^2\right)\\
&= \log(p_2/p_3) - \frac{1}{2\sigma_0^2}\left(2y_i(t)(\mu_3-\mu_2)+(\mu_2^2-\mu_3^2)\right)
\end{split}
\end{equation}
and
\begin{equation}
\hat{C}_i = \begin{cases}
2, & \log(LRT)\geq 0,\\
3, & \log(LRT)< 0,
\end{cases}
\end{equation}
Then we have
\begin{equation}
\log(LRT)\geq 0 \Leftrightarrow y_i(t) \geq y_0 \trieq \frac{(\mu_2^2-\mu_3^3) -2\sigma_0^2\log(p_2/p_3)}{2(\mu_2-\mu_3)}
\end{equation}
Moreover, the classification error of Class 3 is given by
\begin{equation}
\Pr(\hat{C}_i = 2|C_i=3) = \Pr(y_i(t)\geq y_0 | C_i=3),
\end{equation}
which is simply evaluated using \eqref{eq:y_ci_3}.  Similarly, the misclassification error for class 2 is given by
\begin{equation}
\Pr(\hat{C}_i = 3|C_i=2) = \Pr(y_i(t)< y_0 | C_i=2),
\end{equation}
which is simply evaluated using \eqref{eq:y_ci_2}.

\fi

\bibliographystyle{IEEEtran}
\bibliography{./BIB_all/tempBib,./BIB_all/ACL_all,./BIB_all/ACL_Publications,./BIB_all/ACL_bef2000}
\end{document}